\documentclass[a4paper,twocolumn,10pt,unpublished]{quantumarticle}
\pdfoutput=1

\usepackage[T1]{fontenc}
\usepackage[dvipsnames]{xcolor}
\usepackage[utf8]{inputenc}
\usepackage[english]{babel}
\usepackage{tikz}
\usetikzlibrary{arrows.meta,positioning,patterns,patterns.meta,quantikz2}
\usepackage{pgfplots}
\pgfplotsset{compat=1.18}
\usepgfplotslibrary{groupplots}
\pgfplotsset{
    numerics/.style={
        font=\footnotesize,
        tick label style={font=\scriptsize},
        label style={font=\footnotesize},
        axis lines*=left,
        tick align=outside,
        tick style={black!60},
        axis line style={black!60},
        ymin=0, ymax=1.05,
        ytick={0, 0.2, 0.4, 0.6, 0.8, 1},
        grid=major,
        grid style={black!10, line width=0.3pt},
        legend cell align=left,
        legend style={font=\scriptsize, draw=none, fill=white, fill opacity=0.85, text opacity=1, inner sep=1.5pt, row sep=-1pt},
        every axis plot/.append style={line width=0.9pt, mark size=1.4pt},
        extra y ticks={1},
        extra y tick style={grid=major, grid style={black!45, densely dotted, line width=0.5pt}, tick label style={opacity=0}},
    },
    crqaoa/.style={NavyBlue, mark=*},
    cqaoa/.style={NavyBlue, densely dashed, mark=o, mark options={solid}},
    rqaoa/.style={BrickRed, mark=triangle*},
    qaoa/.style={BrickRed, densely dashed, mark=triangle, mark options={solid}},
}
\usepackage{bm} % bold math
\usepackage{bbm} % black board math
\usepackage{braket}
\usepackage{dsfont}
\usepackage[bb=stix]{mathalpha} % blackboard math
\usepackage{amsmath}
\usepackage{amssymb}
\usepackage{amsthm}
\usepackage{thmtools}
\usepackage{mathtools}
\usepackage{multirow}
\usepackage{hhline}
\usepackage{tabularx}
\usepackage{booktabs}
\usepackage{colortbl}
\usepackage[noend]{algpseudocode}
\makeatletter
\newcounter{algorithm}
\renewcommand*\thealgorithm{\@arabic\c@algorithm}
\def\fps@algorithm{tbp}
\def\ftype@algorithm{4}
\def\ext@algorithm{loa}
\def\fnum@algorithm{Algorithm\nobreakspace\thealgorithm}
\newenvironment{algorithm}[1][tbp]{\@float{algorithm}[#1]\hrule
    \vspace*{3pt}}{
    \vspace*{3pt}\hrule
    \end@float}
\makeatother
\usepackage[numbers,sort&compress]{natbib}
\usepackage{silence}
\usepackage{etoolbox}
\apptocmd{\sloppy}{\hbadness 10000\relax}{}{}

\usepackage{hyperref}
\hypersetup{
    colorlinks  = true,
    citecolor   = PineGreen,
    linkcolor   = MidnightBlue,
    urlcolor    = PineGreen
}

\addto\extrasenglish{

}

\DeclareMathOperator*{\spn}{span} % linear span
\DeclareMathOperator*{\sign}{sign} % sign
\DeclareMathOperator*{\successor}{succ} % successor
\newcommand*{\bit}{\mathbb{b}} % single bit
\newcommand*{\qubit}{\mathbb{q}} % single-qubit system
\newcommand*{\one}{\mathds{1}} % identity operator
\newcommand*{\N}{\mathbb{N}} % natural numbers
\newcommand*{\Z}{\mathbb{Z}} % integers
\newcommand*{\R}{\mathbb{R}} % real numbers
\newcommand*{\C}{\mathbb{C}} % complex numbers
\newcommand*{\lo}{\mathcal{L}} % linear bounded operators
\newcommand*{\abs}[1]{\left\lvert#1\right\rvert} % absolute value / Hamming weight
\newcommand*{\defcolon}{\,:\,} % colon for properties
\newcommand*{\notequiv}{\not{\hspace*{-3pt}\equiv}\hspace*{3pt}} % noot equivalent
\newcommand\scriptin{\raisebox{0.15ex}{$\scriptscriptstyle\in$}} % "element of" symbol for usage in sub- and superscripts

\declaretheorem[style=plain]{theorem}

\declaretheorem[style=plain,sibling=theorem]{corollary}
\declaretheorem[style=plain,sibling=theorem]{lemma}
\declaretheorem[style=plain,sibling=theorem]{proposition}
\declaretheorem[style=definition,sibling=theorem]{definition}

\begin{document}

\title{Recursive Quantum Alternating Operator Ansatz}

\author[1,2]{Lennart~Binkowski}
\email{lennart.binkowski@itp.uni-hannover.de}
\orcid{0009-0001-3723-7699}
\author[1,3]{Stuart~Flannigan}
%\email{stuart@strangeworks.com}
\orcid{0000-0002-3144-5534}
\affil[1]{Strangeworks, Austin, TX 78746, United States}
\affil[2]{Institut f\"{u}r Theoretische Physik, Leibniz Universit\"{a}t Hannover, Hannover, Lower Saxony, Germany}
\affil[3]{Commutator Studios GmbH, 40211 Düsseldorf, Germany}

\begin{abstract}
    The recursive quantum approximate optimisation algorithm~(RQAOA) shrinks an unconstrained combinatorial optimisation problem by eliminating one variable at a time, guided by correlations measured in variational quantum states.
    We extend this recursion to constrained problems by building on the quantum alternating operator ansatz, which restricts the evolution to feasible solutions through a feasible initial state and feasibility-preserving mixers.
    However, every elimination alters the feasibility structure, so that such a mixer for the original problem generally fails to be one for the subproblems.
    Previous extensions to Maximum Independent Set and Minimum Vertex Cover sidestep this with tailored reduction rules that keep every subproblem within the original problem class.
    We resolve this by identifying, for a given constraint structure, a broader problem class that is closed under all eliminations, and by efficiently constructing feasible initial states and feasibility-preserving mixers for every class member.
    We do so for monotone constraints, such as for the knapsack problem, for linear equality constraints with polynomially bounded weights, including Hamming weight constraints, and for combinations of equality constraints with a single inequality constraint, such as the cardinality-constrained knapsack problem.
    Conversely, we show that these restrictions cannot be lifted in general without leaving the realm of efficient constructions.
    Exact simulations on knapsack and vertex-weighted max-bisection instances demonstrate that our recursive scheme can outperform both its non-recursive counterpart and penalty-based QAOA and RQAOA.
\end{abstract}

\maketitle

\section{\label{section:Introduction}Introduction}

The quantum approximate optimisation algorithm~\cite{Farhi2014AQuantumApproximateOptimizationAlgorithm} and its generalisation, the quantum alternating operator ansatz~\cite{Hadfield2019FromTheQuantumApproximateOptimizationAlgorithmToAQuantumAlternatingOperatorAnsatz}, are among the most studied candidates for tackling combinatorial optimisation problems on digital quantum computers.
Both alternate between a phase separator, encoding the objective function, and a mixer, driving transitions between candidate solutions.
The quantum alternating operator ansatz additionally handles hard constraints intrinsically: a feasible initial state and feasibility-preserving mixers confine the entire evolution to the feasible subspace, so that neither penalty terms nor post-selection are required.
Designing such mixers for a given constraint structure is an intricate task, and Hadfield~\textit{et~al.}~\cite{Hadfield2019FromTheQuantumApproximateOptimizationAlgorithmToAQuantumAlternatingOperatorAnsatz} already compiled a catalogue of mixer families for a broad range of problem classes.
The alternative of treating constraints softly, i.e., adding a penalty term to the objective function, retains the standard mixer, but introduces several issues.
Penalty weights large enough to rule out infeasible optima dominate the energy landscape and have to be tuned carefully, inequality constraints require additional slack qubits~\cite{Heuvel2026ResourceEfficientFormulationsOfTheKnapsackProblemForQuantumComputing}, and a substantial fraction of the prepared state may still be supported on infeasible solutions.

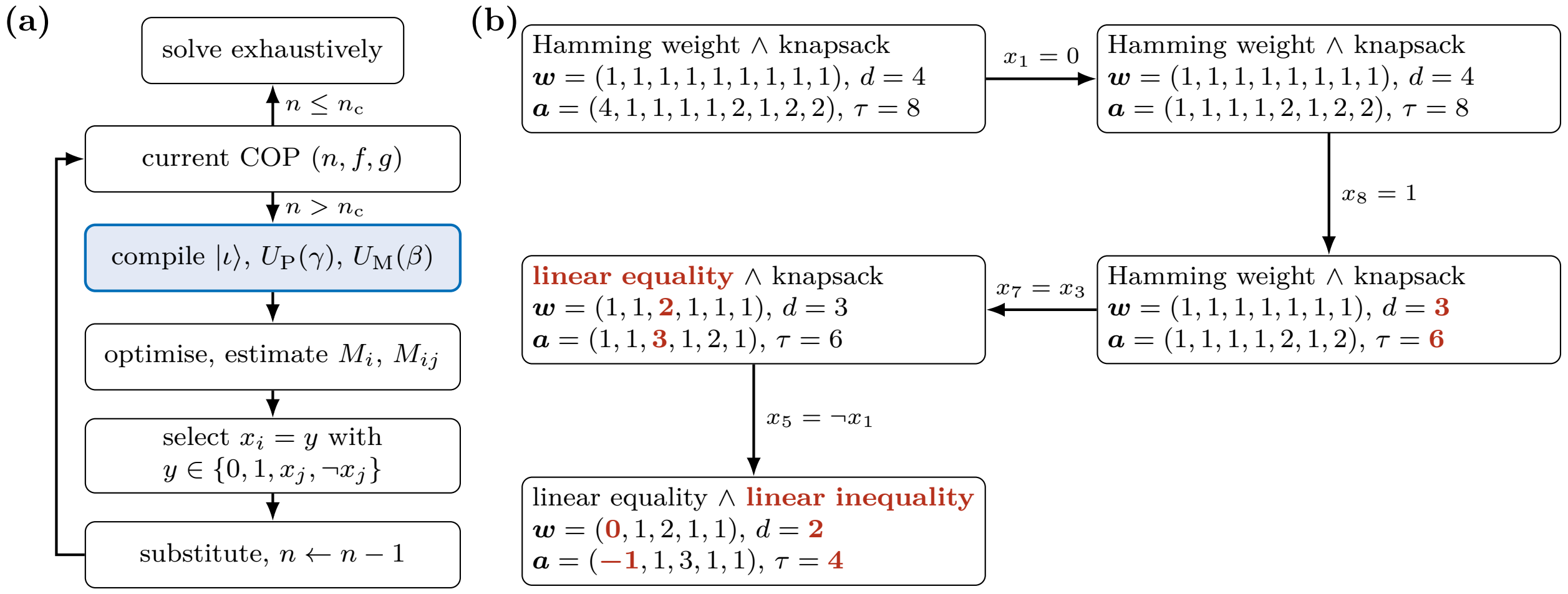
\begin{figure*}[t]
    \centering
    \tikzset{
        loopstep/.style={draw, rounded corners=3pt, align=center, font=\footnotesize, text width=3.8cm, minimum height=0.7cm, inner sep=2.5pt, fill=white},
        thiswork/.style={loopstep, fill=NavyBlue!10, draw=NavyBlue, thick},
        flow/.style={-{Latex[length=2mm]}, thick},
        cop/.style={draw, rounded corners=3pt, font=\footnotesize, inner sep=3pt, align=left, fill=white},
        sublabel/.style={font=\scriptsize, align=left},
    }
    \newcommand*{\chg}[1]{\textcolor{BrickRed}{\bm{#1}}}
    \begin{tikzpicture}
        % panel (a): RQAOA loop
        \node[font=\bfseries] at (-2.6, 1.45) {(a)};
        \node[loopstep, text width=2.6cm] (exh) at (0, 1.15) {solve exhaustively};
        \node[loopstep] (cop) at (0, 0) {current COP $(n, f, g)$};
        \node[thiswork] (compile) at (0, -1.05) {compile $\ket{\iota}$, $U_{\mathrm{P}}(\gamma)$, $U_{\mathrm{M}}(\beta)$};
        \node[loopstep] (opt) at (0, -2.1) {optimise, estimate $M_{i}$, $M_{i j}$};
        \node[loopstep] (select) at (0, -3.15) {\mbox{select $x_{i} = y$ with} \\
            \mbox{$y \in \{0, 1, x_{j}, \neg x_{j}\}$}};
        \node[loopstep] (subst) at (0, -4.2) {\mbox{substitute, $n \gets n - 1$}};
        \draw[flow] (cop) -- node[right, font=\scriptsize] {$n \leq n_{\mathrm{c}}$} (exh);
        \draw[flow] (cop) -- node[right, font=\scriptsize] {$n > n_{\mathrm{c}}$} (compile);
        \draw[flow] (compile) -- (opt);
        \draw[flow] (opt) -- (select);
        \draw[flow] (select) -- (subst);
        \draw[flow] (subst.west) -- ++(-0.3, 0) |- (cop.west);

        % panel (b): one substitution of each type
        \begin{scope}[xshift=5.1cm, cop/.append style={text width=4.7cm}]
            \node[font=\bfseries] at (-2.75, 1.45) {(b)};
            \node[cop] (s0) at (0, 0.85) {Hamming weight $\land$ knapsack \\
                $\bm{w} = (1, 1, 1, 1, 1, 1, 1, 1, 1)$, $d = 4$ \\
                $\bm{a} = (4, 1, 1, 1, 1, 2, 1, 2, 2)$, $\tau = 8$};
            \node[cop] (s1) at (6.1, 0.85) {Hamming weight $\land$ knapsack \\
                $\bm{w} = (1, 1, 1, 1, 1, 1, 1, 1)$, $d = 4$ \\
                $\bm{a} = (1, 1, 1, 1, 2, 1, 2, 2)$, $\tau = 8$};
            \node[cop] (s2) at (6.1, -1.6) {Hamming weight $\land$ knapsack \\
                $\bm{w} = (1, 1, 1, 1, 1, 1, 1)$, $d = \chg{3}$ \\
                $\bm{a} = (1, 1, 1, 1, 2, 1, 2)$, $\tau = \chg{6}$};
            \node[cop] (s3) at (0, -1.6) {\textcolor{BrickRed}{\textbf{linear equality}} $\land$ knapsack \\
                $\bm{w} = (1, 1, \chg{2}, 1, 1, 1)$, $d = 3$ \\
                $\bm{a} = (1, 1, \chg{3}, 1, 2, 1)$, $\tau = 6$};
            \node[cop] (s4) at (0, -3.95) {linear equality $\land$ \textcolor{BrickRed}{\textbf{linear inequality}} \\
                $\bm{w} = (\chg{0}, 1, 2, 1, 1)$, $d = \chg{2}$ \\
                $\bm{a} = (\chg{-1}, 1, 3, 1, 1)$, $\tau = \chg{4}$};
            \draw[flow] (s0) -- node[above, sublabel] {$x_{1} = 0$} (s1);
            \draw[flow] (s1) -- node[right, sublabel] {$x_{8} = 1$} (s2);
            \draw[flow] (s2) -- node[above, sublabel] {$x_{7} = x_{3}$} (s3);
            \draw[flow] (s3) -- node[right, sublabel] {$x_{5} = \neg x_{1}$} (s4);
        \end{scope}
    \end{tikzpicture}
    \caption{\label{figure:Overview}
        \textbf{(a)} The (generalised) RQAOA loop.
        The subproblems need not belong to the root problem's class, so that the compilation of feasible initial states and feasibility-preserving mixers for every subproblem becomes the central task of this work.
        \textbf{(b)} Four consecutive substitutions, one of each type, applied to a cardinality-constrained knapsack problem with $n = 9$ items, i.e., a Hamming weight constraint $\bm{w} \cdot \bm{x} = d$ with $\bm{w} = \bm{1}$ and a capacity constraint $\bm{a} \cdot \bm{x} \leq \tau$.
        Indices refer to the current subproblem, whose remaining variables are re-indexed after every substitution; changes are highlighted.
        Fixing variables preserves the problem class, whereas merging variables via $x_{i} = x_{j}$ produces non-uniform weights, and $x_{i} = \neg x_{j}$ produces a vanishing weight and a negative coefficient, so that neither the Hamming weight nor the knapsack structure survives.
    }
\end{figure*}

For Hamming weight and one-hot constraints, \texttt{XY} mixers are well established~\cite{Wang2020MixersAnalyticalAndNumericalResultsForTheQuantumAlternatingOperatorAnsatz}, and Fuchs~\textit{et~al.}~\cite{Fuchs2022ConstraintPreservingMixersForTheQuantumApproximateOptimizationAlgorithm} construct constraint-preserving mixers for arbitrary feasible sets, at a cost scaling with the number of feasible solutions.
Grover mixers~\cite{Bartschi2020GroverMixersForQaoaShiftingComplexityFromMixerDesignToStatePreparation} shift the complexity from mixer design entirely to the preparation of a superposition of all feasible solutions, which, for instance, the quantum tree generator~\cite{Wilkening2025AQuantumAlgorithmForSolving01KnapsackProblems} accomplishes for the knapsack problem, also within QAOA~\cite{Christiansen2025QuantumTreeGeneratorImprovesQaoaStateOfTheArtForTheKnapsackProblem}.
In all cases, however, the circuits are designed for a fixed problem instance or class.
Some restriction of the constraint class is unavoidable, since already the preparation of a single feasible state is NP-hard for general linear equality constraints, which encode the subset sum problem.

Shallow-depth instances of these hybrid algorithms are, however, subject to fundamental limitations.
Bravyi~\textit{et~al.}~\cite{Bravyi2020ObstaclesToVariationalQuantumOptimizationFromSymmetryProtection} showed that locality and symmetry of the quantum approximate optimisation algorithm limit its performance on certain MaxCut instances at low circuit depths, and proposed the recursive quantum approximate optimisation algorithm~(RQAOA) as a non-local remedy.
RQAOA repeatedly prepares a variational state, estimates one- and two-point correlations, and uses the strongest one to eliminate a variable by fixing it or by identifying it with (the negation of) another variable, thereby producing a sequence of ever smaller subproblems.
Fin\v{z}gar~\textit{et~al.}~\cite{Finzgar2024QuantumInformedRecursiveOptimizationAlgorithms} abstracted this idea into the quantum-informed recursive optimisation~(QIRO) framework, which admits general state preparation routines, problem-specific simplification rules, and backtracking.

The benefit of such a recursion may not seem obvious at first sight, since the very first variational state already comprises all $n$ qubits, so that RQAOA does not extend the size of tractable problems beyond what the underlying quantum device can handle.
Instead, its advantages lie elsewhere.
First, every substitution propagates the information of a single correlation to the entire problem, since all subsequent variational states are prepared for a modified objective function, which circumvents the locality- and symmetry-based obstructions of low-depth circuits~\cite{Bravyi2020ObstaclesToVariationalQuantumOptimizationFromSymmetryProtection}.
Second, the quantum device is merely queried for the sign of the strongest correlation, rather than for a state with large overlap with an optimal solution.
This considerably weaker requirement is expected to be more robust against imperfect parameter optimisation, hardware noise, and finite sampling, as long as these do not alter the dominant correlation.
Third, every subproblem has one variable less than its predecessor, so that its circuit acts on fewer qubits and is optimised over a smaller instance, possibly warm-started from the previous parameters.
Fourth, the recursion terminates with a residual problem small enough to be solved classically, and thus always returns a definite solution rather than a probability distribution over candidate solutions.
These advantages come at the expense of a number of circuit optimisations growing linearly in $n$.
RQAOA has been shown to outperform QAOA of the same depth on MaxCut, e.g., for cycle graphs~\cite{Bravyi2020ObstaclesToVariationalQuantumOptimizationFromSymmetryProtection} and complete graphs~\cite{Bae2024RecursiveQaoaOutperformsTheOriginalQaoaForTheMaxCutProblemOnCompleteGraphs}.

While the RQAOA looks promising for unconstrained problems, its penalty-based adaptation to constrained problems bears several risks.
Most importantly, every substitution is derived from correlations of the variational state, so that substantial weight on infeasible solutions may well suggest assignments which no feasible solution obeys.
Without backtracking, all subsequent subproblems are then infeasible, which can only be detected, and repaired, by additional classical checks.
Hard constraints systematically rule out this failure mode.
If every variational state is supported on feasible solutions only, then every correlation-driven substitution leads to a subproblem which still admits a feasible solution.
In addition, the feasible subspace may be exponentially smaller than the full Hilbert space, so that every recursion step explores a smaller search space.

Extending recursive schemes to constrained problems turns out to be considerably more delicate than in the unconstrained case, though.
Every elimination step alters the feasibility structure, so that a mixer which is feasibility-preserving and transition-providing for the current problem generally fails to be so for the subproblem.
Existing extensions to constrained problems, namely to Maximum Independent Set~\cite{Finzgar2024QuantumInformedRecursiveOptimizationAlgorithms} and Minimum Vertex Cover~\cite{RamosRuiz2026OnTheUseOfTheQuantumAlternatingOperatorAnsatzInQuantumInformedRecursiveOptimizationACaseStudyOnTheMinimumVertexCover}, circumvent this issue by fine-tuned reduction rules which guarantee that every subproblem belongs to the original problem class.
Only very recently, Ramos-Ruiz~\textit{et~al.}~\cite{RamosRuiz2026OnTheUseOfTheQuantumAlternatingOperatorAnsatzInQuantumInformedRecursiveOptimizationACaseStudyOnTheMinimumVertexCover} combined the quantum alternating operator ansatz with QIRO, again tailored to a single problem class.
For most constraint structures, however, insisting on the original problem class is not an option: already a single substitution $x_{i} = \neg x_{j}$ turns a Hamming weight constraint into a weighted linear equality constraint, and a knapsack constraint into a linear inequality constraint with coefficients of mixed signs.
\autoref{figure:Overview} illustrates this drift for a cardinality-constrained knapsack problem.
A systematic account of how to equip recursive schemes with feasible initial state preparation and feasibility-preserving circuits for all subproblems has thus been missing.

In this work, we identify broader classes of constraint structures which are closed under the substitutions performed by RQAOA.
For these classes, we design mixers and initial states, together with efficient rules for constructing them for every subproblem.
The guiding idea is simple:
Instead of asking whether a subproblem still belongs to the root problem's class, we choose a class which contains the root problem and is closed under all four substitutions, and design circuits for this entire class.
For a knapsack constraint, this is the class of single linear inequality constraints with arbitrary integer coefficients; for a Hamming weight constraint, it is the class of linear equality constraints with integer weights whose absolute values sum up to at most the root problem's number of variables.
Two mechanisms then carry the constructions.
Constraints which are monotone in every variable admit single-qubit rotations controlled on the feasibility of the flipped bitstring, which connect every feasible solution to a fixed feasible solution.
Linear equality constraints admit exchanges of two blocks of qubits of equal total weight, and a classical dynamic program over attainable subset sums selects polynomially many such exchanges, which jointly connect all feasible solutions, and moreover yields a feasible initial state as well as all variables forced by the constraint.
Crucially, all these constructions are carried out classically and efficiently anew for every subproblem, so that the quantum device never has to resolve feasibility on its own.
After reviewing combinatorial optimisation problems, the quantum alternating operator ansatz, and existing instantiations of RQAOA and QIRO in \autoref{section:Preliminaries}, we present our main contributions and numerical experiments, before concluding with a summary and open problems; \autoref{table:Catalogue} summarises the resulting circuit families.

\textit{Design criteria.}
In \autoref{section:GeneralisedRecursiveQAOA}, we formulate design criteria for a recursive quantum alternating operator ansatz, extending Hadfield~\textit{et~al.}'s requirements from a single problem to all subproblems reachable by RQAOA-driven substitutions, and show that each such substitution yields a feasible subproblem.

\textit{Unate constraints.}
In \autoref{subsection:UnateFeasibilityPredicates}, we generalise the controlled bit flip mixer for Maximum Independent Set to unate feasibility predicates, derive sufficient conditions under which substitutions preserve unateness, and give an efficient rule for eliminating all variables forced by a substitution at once, which generalises the multiple-variable elimination of Fin\v{z}gar~\textit{et~al.} and Ramos-Ruiz~\textit{et~al.}

\textit{Single linear inequality constraints.}
In \autoref{subsection:SingleLinearInequalityConstraints}, we show that single linear inequality constraints are closed under all four substitutions, so that the unate mixer applies without any restriction, and that it constitutes a variational generalisation of the quantum tree generator for the knapsack problem~\cite{Wilkening2025AQuantumAlgorithmForSolving01KnapsackProblems}.

\textit{Linear equality constraints.}
In \autoref{subsection:HammingWeightConstraints}, we introduce our main contribution: generalised \texttt{XY} mixers exchanging uniform blocks of qubits for Hamming weight constraints and, more generally, linear equality constraints, alongside a characterisation of their mixing properties via the notion of fibre graphs.
For polynomially bounded weights, a dynamic program over attainable subset sums constructs, for every subproblem, a polynomial-sized family of block exchanges providing transitions between all feasible solutions, together with a feasible initial state and elimination of forced variables, in polynomial time.
In the appendices, we show that for general weights such families may be exponentially large or NP-hard to construct, and that inferring a subproblem's family from the current problem's family may incur an exponential overhead even for plain Hamming weight constraints.

\textit{Constraints on linear forms.}
In \autoref{subsection:LinearFormConstraints}, we extend these results to constraints on several linear forms with an arbitrary set of admissible values, covering systems of equality constraints, interval and parity constraints, and their conjunctions, by complementing the block exchanges with a star of controlled moves between the fibres.

\textit{Equality and inequality constraints.}
In \autoref{subsection:EqualityAndInequalityConstraints}, we construct connecting families for a system of linear equality constraints together with a single linear inequality constraint, whose size and construction time do not depend on the inequality's coefficients, and which subsume both previous mixer constructions.
In particular, the cardinality-constrained knapsack problem of \autoref{figure:Overview} belongs to this class, together with all of its subproblems, so that one construction serves the entire recursion.

\textit{Limitations.}
We delimit the scope of efficient constructions: unless $\mathrm{NP} \subseteq \mathrm{BQP}$, no efficient feasible initial state preparation exists for linear equality constraints with general weights, for systems of polynomially many equality constraints, or for polynomially many inequality constraints, even when feasibility of every subproblem is promised.

\textit{Numerical experiments.}
In \autoref{section:NumericalExperiments}, we demonstrate our approach in exact simulations on knapsack and vertex-weighted max-bisection instances with up to $24$ variables, at circuit depth $p = 2$ and with equal budgets of circuit optimisations.
Combining hard constraints with recursion outperforms the non-recursive quantum alternating operator ansatz as well as penalty-based QAOA and RQAOA on every considered instance.
Moreover, on max-bisection, the subproblems leave the root problem's class, so that our general constructions are required in practice.

\section{\label{section:Preliminaries}Preliminaries}

In this section, we review multinomial and parity basis representations of objective functions and constraints, the quantum alternating operator ansatz with all of its building blocks, and the recursive quantum approximate optimisation algorithm with its existing extensions.
Readers familiar with these concepts may skip the respective subsections and only consult \autoref{table:SubstitutionRules} for constraint processing, \autoref{table:RQAOAStateOfTheArt} for an overview of existing extensions to the RQAOA, and \autoref{table:Catalogue} for an overview of our proposal.

Throughout this article, we use the notational short-hands $[n] \coloneqq \{1, \ldots, n\}$ for $n \in \N$, $\bit \coloneqq \{0, 1\}$ for the bit, as well as $\qubit \coloneqq \C^{2}$ for the qubit.
For a bitstring $\bm{x} \in \bit^{n}$, $\abs{\bm{x}}$ denotes its Hamming weight, i.e., the number of bits in $\bm{x}$ set to $1$, and, also given some $i \in [n]$, let $\bm{x}[i \leftarrow b] \coloneqq x_{1} \cdots x_{i - 1} b x_{i + 1} \cdots x_{n} \in \bit^{n}$ denote the bitstring, where the $i$th bit has been replaced by some $b \in \bit$.
Similarly, for a vector $\bm{a} \in \Z^{n}$, $i, j \in [n]$, and $v \in \Z$, let $\bm{a}_{(i)} \in \Z^{n - 1}$ denote the vector with its $i$th entry removed, and $\bm{a}[j \leftarrow\hspace*{-10pt}+ v]$ the vector with $v$ added to its $j$th entry.
For $K \subseteq [n]$ with elements $k_{1} < \cdots < k_{\abs{K}}$, we write $\bm{a}_{K} \coloneqq (a_{k_{1}}, \ldots, a_{k_{\abs{K}}})$ and $\bm{a}_{(K)} \coloneqq \bm{a}_{[n] \setminus K}$; apart from $\bm{a}[j \leftarrow\hspace*{-10pt}+ v]$, these notations also apply to bitstrings.

\subsection{\label{subsection:CombinatorialOptimisationProblems}Combinatorial optimisation problems}

A \emph{combinatorial~optimisation~problem}~(COP) over $\bit^{n}$ is defined by its objective function $f : \bit^{n} \rightarrow \R$ and its feasible set $S \subseteq \bit^{n}$.
Instead of $S$, we will more often work with its indicator function $g : \bit^{n} \rightarrow \bit$, defined via $g(\bm{x}) = 1$ if and only if $\bm{x} \in S$, and also called feasibility predicate.
W.l.o.g., we may assume the COP to be a minimisation problem and state it as
\begin{equation}\label{equation:COP}
    \begin{split}
        \min_{\bm{x} \scriptin \bit^{n}} &f(\bm{x}) \\
        \text{s.t. } &g(\bm{x}) = 1.
    \end{split}
\end{equation}

A COP is called unconstrained if $S = \bit^{n}$ or, equivalently, $g \equiv 1$.
In contrast, the feasibility structures of constrained problems can, in principle, be arbitrary.
However, certain patterns recur across many (practically motivated) problems.
One such example is a Hamming weight constraint $g(\bm{x}) = 1$ if and only if $\abs{\bm{x}} = d$ for some $d \in [n] \cup \{0\}$.
Other constraints that will be of utmost importance are the two unary variable enforcements $g_{i = 0}(\bm{x}) = \neg x_{i}$ and $g_{i = 1}(\bm{x}) = x_{i}$ for some $i \in [n]$, as well as the two binary correlation enforcements $g_{i = j}(\bm{x}) = x_{i} x_{j} \lor \neg x_{i} \neg x_{j}$ and $g_{i\neq j}(\bm{x}) = x_{i} \neg x_{j} \lor \neg x_{i} x_{j}$ for some $i, j \in [n]$, $i \neq j$.
The two unary constraints $g_{i = 0 / 1}$ can be immediately resolved by setting $x_{i}$ accordingly, thus eliminating it as a free variable.
The two binary constraints $g_{i = / \neq j}$ may also be used for reducing the number of variables.
If we choose to keep $x_{j}$ as a free variable, then symbolic substitution of $x_{i}$ with $x_{j}$ when given $g_{i = j}$, and with $\neg x_{j}$ when given $g_{i \neq j}$, in the objective function and remaining constraints effectively eliminates $x_{i}$ as a free variable.
These substitution rules are also summarised on the left-hand side of \autoref{table:SubstitutionRules}.

In order to make potential variable reductions easier, let us consider the multi-linear polynomial representation of objective function $f$, given by
\begin{equation}\label{equation:MultiLinearRepresentation}
    f(\bm{x}) = \sum_{I \subseteq [n]} a_{I} \prod_{i \scriptin I} x_{i},
\end{equation}
where $a_{I} \in \R$ and $\prod_{i \scriptin \emptyset} x_{i} \coloneqq 1$.
The degree of $f$ is defined as $\deg(f) \coloneqq \max\{\abs{I} \defcolon a_{I} \neq 0, I \subseteq [n]\}$.
We will especially focus on objective functions of degree $2$, i.e., on those with bi-linear polynomial representation.
Their terms are covered by the substitution rules shown on the left-hand side of \autoref{table:SubstitutionRules}.

\begin{table*}[th]
    \centering
    \newlength{\colA}\newlength{\substcolw}%
    \settowidth{\colA}{$x_{i} x_{j} \vee \neg x_{i} \neg x_{j}$}%
    \setlength{\substcolw}{\dimexpr(\columnwidth-\colA+6\arrayrulewidth-7\tabcolsep)/3\relax}%
    \begin{tabular}{|c|c|c|c|c|c|c|c|}
        \hline
        \multicolumn{4}{|c|}{bit variables} & \multicolumn{4}{c|}{spin variables} \\
        \hline
        \multirow{2}{*}{\makebox[\colA]{constraint}}
        & \multicolumn{3}{c|}{substitution}
        & \multirow{2}{*}{\makebox[\colA]{constraint}}
        & \multicolumn{3}{c|}{substitution} \\
        \cline{2-4}\cline{6-8}
        & \makebox[\substcolw]{$x_{i}$} & \makebox[\substcolw]{$x_{i} x_{j}$} & \makebox[\substcolw]{$x_{i} x_{k}$} & & \makebox[\substcolw]{$z_{i}$} & \makebox[\substcolw]{$z_{i} z_{j}$} & \makebox[\substcolw]{$z_{i} z_{k}$} \\
        \hhline{|=|=|=|=|=|=|=|=|}
        $x_{i} = 0$ & $0$ & $0$ & $0$ & $z_{i} = +1$ & $+1$ & $z_{j}$ & $z_{k}$ \\
        \hline
        $x_{i} = 1$ & $1$ & $x_{j}$ & $x_{k}$ & $z_{i} = -1$ & $-1$ & $-z_{j}$ & $-z_{k}$ \\
        \hline
        $x_{i} = x_{j}$ & $x_{j}$ & $x_{j}$ & $x_{j} x_{k}$ & $z_{i} z_{j} = +1$ & $z_{j}$ & $+1$ & $z_{j} z_{k}$ \\
        \hline
        $x_{i} = \neg x_{j}$ & $\neg x_{j}$ & $0$ & $\neg x_{j} x_{k}$ & $z_{i} z_{j} = -1$ & $-z_{j}$ & $-1$ & $-z_{j} z_{k}$ \\
        \hline
    \end{tabular}
    \caption{\label{table:SubstitutionRules}
        Substitution rules for implicitly embedding the two unary and two binary constraints.
        The left-hand side shows the constraints and substitution rules expressed for bit variables, while the right-hand side expresses the same concepts in terms of spin variables.
    }
\end{table*}

From an abstract point of view, the expansion~\eqref{equation:MultiLinearRepresentation} corresponds to a concrete basis representation of $f$ in the multinomial basis $\{[\bm{x} \mapsto \prod_{i \scriptin I} x_{i}] \defcolon I \subseteq [n]\}$.
Another commonly used basis is given by the set of parity functions $\{\chi_{I} \coloneqq [\bm{x} \mapsto \prod_{i \scriptin I} (-1)^{x_{i}}] \defcolon I \subseteq [n]\}$ with the convention that $\chi_{\emptyset} \coloneqq 1$.
The respective representation of $f$ in the parity basis reads
\begin{equation*}
    f(\bm{x}) = \hspace*{-1pt} \sum_{I \subseteq [n]} \tilde{a}_{I} \chi_{I}(\bm{x}),\, \text{where } \tilde{a}_{I} = 2^{-n} \hspace*{-2pt} \sum_{\bm{y} \scriptin \bit^{n}} f(\bm{y}) \chi_{I}(\bm{y}).
\end{equation*}

There is a simple translation between the multinomial and the parity basis, given by the two identities $z_{i} \coloneqq (-1)^{x_{i}} = 1 - 2 x_{i}$ and $x_{i} = (1 - z_{i}) / 2$.
From these, it follows that $\deg(f) = \max\{\abs{I} \defcolon \tilde{a}_{I} \neq 0, I \subseteq [n]\}$.
Especially, objective functions with bi-linear polynomial expansion remain of order $2$ in the parity basis.
Let us also reformulate the unary constraints $g_{i = 0 / 1}$ and two binary constraints $g_{i = / \neq j}$ in the parity basis.
Since $(-1)^{0} = +1$, demanding that $x_{i} = 0$ corresponds to enforcing $z_{i} = +1$; analogously, $x_{i} = 1$ if and only if $z_{i} = -1$.
The binary constraint $x_{i} = x_{j}$ has the simple product form $z_{i} z_{j} = +1$ in the parity basis, and $x_{i} = \neg x_{j}$ becomes $z_{i} z_{j} = -1$.
The parity representation of these constraints and their implications are shown on the right-hand side of \autoref{table:SubstitutionRules}.

\subsection{\label{subsection:QuantumAlternatingOperatorAnsatz}Quantum alternating operator ansatz}

The standard procedure of mapping COPs to a digital quantum computer is to replace every bit with a qubit, introducing the $n$-qubit space $\qubit^{\otimes n}$ with the computational basis $\{\ket{\bm{x}} \defcolon \bm{x} \in \bit^{n}\}$ as reference basis.
The COP's objective function $f$ is represented by the diagonal objective Hamiltonian $H_{f} \in \lo(\qubit^{\otimes n})$, defined via $H_{f} \ket{\bm{x}} = f(\bm{x}) \ket{\bm{x}}$.
By construction, the ground states of $H_{f}$ correspond to (superpositions of) minimisers of $f$.
For unconstrained problems, this reformulates the optimisation task as a ground state search over the whole $\qubit^{\otimes n}$.
However, this search has to be restricted to the feasible subspace $\mathcal{S} \coloneqq \spn\{\ket{\bm{x}} \defcolon \bm{x} \in S\} \subsetneq \qubit^{\otimes n}$ if the COP is constrained.\footnote{
    Alternatively, one can use soft constraints to penalise infeasible solutions, rendering the restriction to the feasible subspace obsolete.
    However, this approach is less structurally insightful and introduces several issues, such as sensitivity to hyperparameter fine-tuning and---when there are many more infeasible than feasible solutions as is often the case---concentrating most amplitude mass in the infeasible subspace.
}
The main innovation of the \emph{quantum alternating operator ansatz}~(QAOA) is to ensure restriction to $\mathcal{S}$ via a feasible initial state preparation $\ket{\bm{0}} \mapsto \ket{\iota} \in \mathcal{S}$ and feasibility-preserving parameterised quantum circuits $V(\bm{\theta}) \in \lo(\qubit^{\otimes n})$, $V(\bm{\theta})(\mathcal{S}) \subseteq \mathcal{S}$ for all parameter values $\bm{\theta}$, used to construct trial states $\ket{\bm{\theta}} \coloneqq V(\bm{\theta}) \ket{\iota} \in \mathcal{S}$.
Thereby, constraints never have to be checked explicitly throughout the routine.

The variational circuit $V(\bm{\theta})$ is typically comprised of $p$ alternating applications of an objective function-dependent phase separator $U_{\mathrm{P}}(\gamma)$ and a constraint-dependent mixer $U_{\mathrm{M}}(\beta)$.
Hadfield~\textit{et~al.}~\cite{Hadfield2019FromTheQuantumApproximateOptimizationAlgorithmToAQuantumAlternatingOperatorAnsatz} require the phase separator to be diagonal in the computational basis.
This readily implies that $U_{\mathrm{P}}(\gamma)(\mathcal{S}) \subseteq \mathcal{S}$, i.e., $U_{\mathrm{P}}(\gamma)$ is feasibility-preserving.\footnote{
    One can relax this requirement to diagonality only on the feasible subspace without sacrificing feasibility preservation~\cite{Binkowski2025SymmetryBasedQuantumAlgorithmsForOpenShopSchedulingWithHardConstraints}.
}
Typically, one uses $U_{\mathrm{P}}(\gamma) = e^{-i \gamma H_{f}}$, although also other candidates such as threshold functions are being studied~\cite{Golden2021ThresholdBasedQuantumOptimization}.
Throughout this paper, we will restrict to the more common case of using the exponentiated objective Hamiltonian.
The Pauli expansion of $H_{f}$ is directly related to the parity basis expansion of $f$~\cite{Hadfield2021OnTheRepresentationOfBooleanAndRealFunctionsAsHamiltoniansForQuantumComputing}:
\begin{equation*}
    H_{f} = \sum_{I \subseteq [n]} \tilde{a}_{I} \prod_{i \scriptin I} \texttt{Z}_{i},
\end{equation*}
since $\texttt{Z} \ket{x} = (-1)^{x} \ket{x} = z \ket{x}$.
Since all Pauli strings consist only of identities and Pauli-$\texttt{Z}$ gates, they mutually commute, allowing for the implementation of the phase separator $U_{\mathrm{P}}(\gamma)$ as
\begin{equation*}
    \exp\Bigg(\hspace*{-3pt} -i \gamma \sum_{I \subseteq [n]} \tilde{a}_{I} \prod_{i \scriptin I} \texttt{Z}_{i}\Bigg) = \prod_{I \subseteq [n]} \exp\Bigg(\hspace*{-3pt} -i \gamma \tilde{a}_{I} \prod_{i \scriptin I} \texttt{Z}_{i}\Bigg).
\end{equation*}
Each individually exponentiated Pauli string can be efficiently implemented using a single $\texttt{RZ}$-gate and two enclosing cascades of \texttt{CNOT}-gates~\cite[Figure 4.19]{Nielsen2012QuantumComputationAndQuantumInformation}.
In summary, precomputing the parity basis representation of $f$ is advisable whenever working with the exponentiated objective Hamiltonian $H_{f}$.

Unlike the phase separator, the mixer $U_{\mathrm{M}}(\beta)$ is required to provide transitions between all pairs of feasible computational basis states.
The concrete criterion imposed by Hadfield~\textit{et~al.}~\cite{Hadfield2019FromTheQuantumApproximateOptimizationAlgorithmToAQuantumAlternatingOperatorAnsatz} is: for every $\bm{x}, \bm{y} \in S$, there should exist some value $\beta$ and some positive integer $r$ such that $\abs{\braket{\bm{x} | U_{\mathrm{M}}^{r}(\beta) | \bm{y}}} > 0$.
This property ensures that a QAOA circuit of depth $p = r$ can produce non-zero overlap with the feasible state $\ket{\bm{x}}$, when applied to the initial state $\ket{\iota} = \ket{\bm{y}}$.
However, note that this mixing property is generally not sufficient to ensure exact reachability of any optimum, neither with a fixed circuit size, nor asymptotically for $p \to \infty$.
Requiring the former leads to the notion of exhaustively parameterised quantum circuits~\cite{Schwiering2026ExhaustiveAndFeasibleParametrisationWithApplicationsToTheTravellingSalespersonProblem}.
The latter can be established---by virtue of adiabatic theorems---for mixers of the form $U_{\mathrm{M}}(\beta) = e^{-i \beta H_{\mathrm{M}}}$ by demanding $\ket{\iota}$ to be the ground state of $H_{\mathrm{M}}$, $H_{\mathrm{M}}$ to be feasibility-preserving and component-wise non-positive, as well as irreducible in the computational basis~\cite{Binkowski2024ElementaryProofOfQaoaConvergence}.

\subsection{\label{subsection:RecursiveQAOAStateOfTheArt}Recursive QAOA: state of the art}

The original proposal of the RQAOA~\cite{Bravyi2020ObstaclesToVariationalQuantumOptimizationFromSymmetryProtection} was for the unconstrained MaxCut problem, thus conceptually extending the quantum approximate optimisation algorithm~\cite{Farhi2014AQuantumApproximateOptimizationAlgorithm} rather than its generalisation, the quantum alternating operator ansatz.
That is, the employed mixer is given by
\begin{equation}\label{equation:StandardQAOAMixer}
    U_{\mathrm{M}}(\beta) = e^{-i \beta B},\text{ where } B = \sum_{i = 1}^{n} \texttt{X}_{i},
\end{equation}
where $n$ equals the number of vertices of the considered weighted, undirected graph $G = (V, E, \omega)$.
W.l.o.g., we may identify $G$'s vertex set $V$ with $[n]$ so that $E \subseteq \binom{[n]}{2}$.
Viewing MaxCut as a minimisation problem, the objective function is given by
\begin{align*}
    f_{\text{MC}}(\bm{x}) &\coloneqq -\sum_{\{i, j\} \scriptin E} \omega_{i j} \big(x_{i} (1 - x_{j}) + (1 - x_{i}) x_{j}\big) \\
    &= -\tfrac{1}{2} \sum_{\{i, j\} \scriptin E} \omega_{i j} + \tfrac{1}{2} \sum_{\{i, j\} \scriptin E} \omega_{i j} z_{i} z_{j}.
\end{align*}
For sole optimisation purposes, the constant term and the prefactor of $1 / 2$ may be omitted, yielding a simple objective Hamiltonian with only quadratic terms:
\begin{equation*}
    H_{f_{\text{MC}}} = \sum_{\{i, j\} \scriptin E} \omega_{i j} \texttt{Z}_{i} \texttt{Z}_{j}.
\end{equation*}

The RQAOA protocol for the MaxCut problem consists of the following loop.
First, determine parameter values $(\bm{\beta}^{*}, \bm{\gamma}^{*}) \in \R^{2 p}$, minimising the parameterised expectation value $\braket{\bm{\beta}, \bm{\gamma} | H_{f_{\text{MC}}} | \bm{\beta}, \bm{\gamma}}$.
Second, identify an edge $\{i, j\} \in E$ which contributes the largest term $\abs{M_{i j}} \coloneqq \abs{\braket{\bm{\beta}^{*}, \bm{\gamma}^{*} | \texttt{Z}_{i} \texttt{Z}_{j} | \bm{\beta}^{*}, \bm{\gamma}^{*}}}$.
Third, impose $z_{i} z_{j} = \sign(M_{i j})$, i.e., either $x_{i} = x_{j}$ (for $M_{i j} > 0$) or $x_{i} = \neg x_{j}$ (for $M_{i j} < 0$), and substitute $z_{i}$ by $\pm z_{j}$ in the objective function and Hamiltonian.
For the resulting problem with one less variable, invoke the first step again, and so on.
Following the substitution rules for $z_{i} z_{j} = \pm 1$ in \autoref{table:SubstitutionRules}, the term $\omega_{i j} z_{i} z_{j}$ becomes a constant $\pm \omega_{i j}$, which may be simply omitted, while any term $\omega_{i k} z_{i} z_{k}$, $k \neq j$ becomes $\pm \omega_{i k} z_{j} z_{k}$.
Therefore, the reduced problem is again a MaxCut instance on a graph where vertices $i$ and $j$ have been merged.
If a vertex $k$ was connected to both $i$ and $j$ in the original graph, the weight of the edge connecting the merged vertex $j$ and $k$ is given by $\sign(M_{i j}) \omega_{i k} + \omega_{j k}$.
Defining $\omega$ on the whole set $\binom{[n]}{2}$ with $\omega_{k \ell} = 0$ whenever $\{k, \ell\} \notin E$, this update formula also holds if $k$ was only connected to one or none of the two merged vertices.
Overall, the RQAOA reductions never leave the MaxCut problem class.

The broadest extension of the RQAOA framework is due to Fin\v{z}gar~\textit{et~al.}~\cite{Finzgar2024QuantumInformedRecursiveOptimizationAlgorithms}, who introduced the quantum-informed recursive optimisation~(QIRO) framework.
It keeps RQAOA's core of iteratively eliminating variables by enforcing (anti-)correlations, inferred from measurements of prepared quantum states, and introduces the unary substitution rules $z_{i} = \pm 1$ on top.
Furthermore, it abstracts away from the recursive usage of the quantum approximate optimisation algorithm, allowing for general problem-specific state preparation routines and simplification steps to form subproblems.
Most importantly, QIRO embeds the variable elimination into a backtracking framework, allowing to reconsider reduction choices at prior subproblems if the current path is not promising.
Ultimately, this elevates QIRO from a quantum-assisted heuristic to a quantum-assisted exact solver.
However, as we specifically seek to generalise the RQAOA to a feasibility-preserving setup, we only discuss those enhancements of the QIRO framework which directly influence reduction rules or the construction of QAOA components for subproblems.

Fin\v{z}gar~\textit{et~al.}~\cite{Finzgar2024QuantumInformedRecursiveOptimizationAlgorithms} discuss two concrete problems: Maximum Independent Set~(MIS) and Max-$2$-Sat.
As an unconstrained problem, the latter is conceptually closer to MaxCut.
However, while the MaxCut problem has a particularly elegant formulation in terms of spin variables, the Max-$2$-Sat objective function is more naturally expressed in ordinary bit variables.
Let us already state the problem as a minimisation problem by negating the standard formulation:
One is given $m$ clauses, each having the form $\ell_{j, 1} \land \ell_{j, 2}$, where $j \in [m]$, and $\ell_{j, 1}, \ell_{j, 2} \in \{x_{1}, \ldots, x_{n}, \neg x_{1}, \ldots, \neg x_{n}\}$.
The goal is to satisfy as few clauses as possible, which amounts to minimising
\begin{equation*}
    f_{\text{M2S}}(\bm{x}) \coloneqq \sum_{j = 1}^{m} \ell_{j, 1} \ell_{j, 2}.
\end{equation*}
In spin variables, $f_{\text{M2S}}$ will generally have constant, linear, and quadratic terms.
While a constant offset may again be removed for optimisation purposes, omitting any linear term would fundamentally deviate from the original task.
Accordingly, this problem is a natural candidate for the additional unary substitution rules.
Instead of only recording correlation terms $M_{i j}$, one also stores $M_{i} \coloneqq \braket{\bm{\beta}^{*}, \bm{\gamma}^{*} | \texttt{Z}_{i} | \bm{\beta}^{*}, \bm{\gamma}^{*}}$, and picks the substitution rule corresponding to the maximum over all $\abs{M_{i}}$ and $\abs{M_{i j}}$.
In principle, all four substitution rules can lead to unit clauses, i.e., clauses with only one literal.
However, as we can always write the binary variable $x_{j}$ as $x_{j} x_{j}$, this technically does not lead outside the Max-$2$-Sat problem class.

\begin{table*}[t]
    \centering
    \begin{tabular*}{\textwidth}{@{\extracolsep{\fill}}|c|c|c|c|c|c|}
        \hline
        problem & reference & constrained & objective function & reduction rules & class-closed \\
        \hline
        MaxCut &~\cite{Bravyi2020ObstaclesToVariationalQuantumOptimizationFromSymmetryProtection} & no & quadratic & $g_{i = / \neq j}$ & yes \\
        \hline
        Max-$2$-Sat &~\cite{Finzgar2024QuantumInformedRecursiveOptimizationAlgorithms} & no & linear + quadratic & $g_{i = 0 / 1}$, $g_{i = / \neq j}$ & yes$^{\dagger}$ \\
        \hline
        Maximum Independent Set &~\cite{Finzgar2024QuantumInformedRecursiveOptimizationAlgorithms} & yes & Hamming weight & $g_{i = 0 / 1}$, $g_{i = / \neq j}$ & yes \\
        \hline
        Minimum Vertex Cover &~\cite{RamosRuiz2026OnTheUseOfTheQuantumAlternatingOperatorAnsatzInQuantumInformedRecursiveOptimizationACaseStudyOnTheMinimumVertexCover} & yes & Hamming weight & $g_{i = 0 / 1}$, $g_{i = / \neq j}$ & yes \\
        \hline
    \end{tabular*}
    \caption{\label{table:RQAOAStateOfTheArt}
        Summary of existing RQAOA/QIRO instantiations discussed in \autoref{subsection:RecursiveQAOAStateOfTheArt}.
        The column ``class-closed'' indicates whether every subproblem produced by the listed reductions stays within the original problem class.
        $^{\dagger}$ Requires rewriting $x_{j}$ as $x_{j} x_{j}$ to absorb unary substitutions into the Max-$2$-Sat form.
    }
\end{table*}

The second considered problem, MIS, is a constrained problem.
Given an (unweighted) undirected graph $G = (V, E)$, the goal is to find a largest set of vertices $U \subseteq V$ so that for all $i, j \in U$ it is $\{i, j\} \notin E$.
Using $\abs{V} = n$ bits to encode all possible vertex subsets, the objective function simply becomes the (negative) Hamming weight, i.e., $f(\bm{x}) \coloneqq -\abs{\bm{x}}$.
The independence constraint may be described by $\abs{E}$ many indicator functions $g_{\{\overline{i, j}\}}(\bm{x}) \coloneqq 1 - x_{i} x_{j}$.
Each of these ensures that at most one endpoint of an edge $\{i, j\}$ is included in the encoded vertex subset.
If one cannot ensure full feasibility during the state preparation, these constraints have to be incorporated as penalty terms into the objective function, yielding
\begin{equation*}
    f_{\text{MIS}}(\bm{x}) \coloneqq - \sum_{i = 1}^{n} x_{i} + \lambda \sum_{\{i, j\} \scriptin E} x_{i} x_{j},
\end{equation*}
as described in~\cite{Finzgar2024QuantumInformedRecursiveOptimizationAlgorithms}.
In contrast, if we can guarantee restriction to the feasible subspace, e.g., via a feasibility-preserving QAOA mixer or via the adiabatic protocol for neutral atom computers presented in~\cite{Finzgar2024QuantumInformedRecursiveOptimizationAlgorithms}, the quadratic terms would again vanish.
Let us first consider the two unary constraints $g_{i = 0 / 1}$.
Since the problem is constrained, assigning a value to a variable might also enforce other variables to take on concrete values.
Namely, setting $x_{i} = 1$, i.e., including $i$ into the vertex subset, enforces $x_{j} = 0$ for all $\{i, j\} \in E$, i.e., all neighbouring vertices have to be excluded to maintain the independence constraints.
Conversely, setting $x_{i} = 0$ does not enforce any further variable assignments.
In addition to the unary constraints, Fin\v{z}gar~\textit{et~al.}~\cite{Finzgar2024QuantumInformedRecursiveOptimizationAlgorithms} proposed to also consider the two binary constraints $g_{i = / \neq j}$ whenever $\{i, j\} \in E$.
They observed that, if $M_{i j} > 0$ for some $\{i, j\} \in E$, $x_{i} = x_{j} = 0$ must hold to not violate the constraint $g_{\{\overline{i, j}\}}$, resulting in assigning and thus removing both variables from the problem description.
The case $M_{i j} < 0$, in turn, implies that exactly one of the two vertices $i$ and $j$ is included while the other one is excluded from the vertex subset.
However, as it is not clear at this point which is included, only variables representing vertices neighbouring both $i$ and $j$ may be safely set to $0$.
Especially, neither $x_{i}$ nor $x_{j}$ is removed from the problem description.
Following these rules again guarantees that all subproblems stay within the MIS problem class, but the four rules can differ significantly in the magnitude of variable reduction.

Recent work by Ramos-Ruiz~\textit{et~al.}~\cite{RamosRuiz2026OnTheUseOfTheQuantumAlternatingOperatorAnsatzInQuantumInformedRecursiveOptimizationACaseStudyOnTheMinimumVertexCover} studies the combination of the quantum alternating operator ansatz with QIRO for the Minimum Vertex Cover~(MVC) problem.
One is again given an unweighted undirected graph $G = (V, E)$ and the task to find a smallest set of vertices $U \subseteq V$ so that for all $\{i, j\} \in E$ it is $i \in U$ or $j \in U$.
Encoding vertex subsets as bitstrings with $x_{i} = 1$ if and only if $i \in U$, the objective function simply evaluates the Hamming weight, and the cover constraints are given by $\abs{E}$ indicator functions $g_{\{i, j\}}(\bm{x}) \coloneqq x_{i} \lor x_{j}$.
Furthermore, a vertex subset $U$ is a minimum vertex cover for $G$ if and only if its complement $V \setminus U$ is a maximum independent set for $G$.
Accordingly, all the aforementioned simplification rules could be straightforwardly applied to the MVC as well.
In fact, three out of four simplification rules proposed by Ramos-Ruiz~\textit{et~al.} coincide exactly with the corresponding MIS rules:
If $x_{i} = 0$ is enforced, all variables representing neighbouring vertices have to be set to $1$ to maintain the cover constraints---this corresponds to the MIS rule for $x_{i} = 1$.
If $x_{i} = 1$, the corresponding vertex and all edges to which it is incident can be safely removed from the graph, resembling the MIS rule for $x_{i} = 0$.
In case $M_{i j} > 0$ for some $\{i, j\} \in E$ is selected, both vertices have to be included into the subset $U$ as $x_{i} = x_{j} = 0$ would leave the edge $\{i, j\}$ uncovered; for the MIS one had to set both variables to $0$ instead.
Lastly, for the case of $M_{i j} < 0$, they incorporate the heuristic of adding the vertex with the higher degree to $U$ and setting the other variable to $0$.
Due to the added heuristic, this scenario also always reduces the number of variables.
Note that one could easily add an analogous heuristic to the MIS construction.
Crucially, due to the presented simplification rules, all subproblems belong again to the MVC problem class.

\section{\label{section:GeneralisedRecursiveQAOA}Generalised recursive QAOA}

In this section, we motivate and lay out our design criteria for the recursive quantum alternating operator ansatz.
As it strictly generalises the recursive quantum approximate optimisation algorithm, we will henceforth call it RQAOA as well.
In essence, the (generalised) RQAOA should combine Hadfield~\textit{et~al.}'s design criteria for the quantum alternating operator ansatz with efficient compilation rules for transforming valid QAOA circuits to valid QAOA circuits for all subproblems that can arise by applying one of the four substitutions $x_{i} \in \{0, 1, x_{j}, \neg x_{j}\}$.
In contrast to all examples in the existing literature, we explicitly do not require the subproblems to belong to the same problem class, only that a given QAOA circuit for the subproblem can be efficiently generated from a description of the current problem or its QAOA circuit.
This includes all aforementioned examples which do not leave the problem class; their compilation rules merely exchange the problem parameters, such as number of variables or graph structure, in the description of the current QAOA circuit.

Before stating our concrete design criteria, let us first formalise RQAOA's subproblem generation.
There will always be some root \hyperref[equation:COP]{COP}, represented by its number of variables $n_{0} \in \N$, its objective function $f_{0} : \bit^{n_{0}} \rightarrow \R$, and its feasibility indicator function $g_{0}: \bit^{n_{0}} \rightarrow \bit$, from which subproblems emerge in an iterative/recursive manner.
Concretely, given the current COP $(n, f, g)$---not necessarily the root COP---we select one out of four substitutions $x_{i} \in \{0, 1, x_{j}, \neg x_{j}\}$ for some $i, j \in [n]$, $i \neq j$.
In any case, the variable $x_{i}$ is eliminated after performing the substitution in both, $f$ and $g$, resulting in a subproblem with $n - 1$ variables.
Accordingly, we re-index all variables $x_{k} \mapsto x_{k - 1}$ afterwards, where $k > i$.
Let us denote the derived functions by $f\!\restriction_{x_{i} = y}$ and $g\!\restriction_{x_{i} =y}$, where $y \in \{0, 1, x_{j}, \neg x_{j}\}$.
Crucially, if we use a feasible initial state $\ket{\iota} \in \mathcal{S}$ and a feasibility-preserving QAOA circuit $V(\bm{\beta}, \bm{\gamma})$, and determine the substitution rule corresponding to the maximum over all $\abs{M_{i}} \coloneqq \abs{\braket{\bm{\beta}, \bm{\gamma} \vert \texttt{Z}_{i} \vert \bm{\beta}, \bm{\gamma}}}$ and $\abs{M_{i j}} \coloneqq \abs{\braket{\bm{\beta}, \bm{\gamma} \vert \texttt{Z}_{i} \texttt{Z}_{j} \vert \bm{\beta}, \bm{\gamma}}}$, where $\ket{\bm{\beta}, \bm{\gamma}} \coloneqq V(\bm{\beta}, \bm{\gamma}) \ket{\iota}$, the constructed subproblem will again be solvable.
This is true also for non-variational state preparation and solely relies on the fact that we remain within the feasible subspace.

\begin{lemma}\label{lemma:LocalConsistency}
    Let $(n, f, g)$ be a COP with $g \not{\hspace*{-4pt}\equiv}\hspace*{4pt} 0$, and let $\ket{\psi} \in \spn\{\ket{\bm{x}} : g(\bm{x}) = 1\}$ with $\braket{\psi \vert \psi} = 1$.
    For every $i \in [n]$ and every $j \in [n] \setminus \{i\}$, the following hold true:
    \begin{itemize}
        \item[i)] If $\braket{\psi \vert \texttt{Z}_{i} \vert \psi} \neq -1$, then $g\!\restriction_{x_{i} = 0} \notequiv 0$.
        \item[ii)] If $\braket{\psi \vert \texttt{Z}_{i} \vert \psi} \neq 1$, then $g\!\restriction_{x_{i} = 1} \notequiv 0$.
        \item[iii)] If $\braket{\psi \vert \texttt{Z}_{i} \texttt{Z}_{j} \vert \psi} \neq -1$, then $g\!\restriction_{x_{i} = x_{j}} \notequiv 0$.
        \item[iv)] If $\braket{\psi \vert \texttt{Z}_{i} \texttt{Z}_{j} \vert \psi} \neq 1$, then $g\!\restriction_{x_{i} = \neg x_{j}} \notequiv 0$.
    \end{itemize}
\end{lemma}

\begin{proof}
    We prove only i); the other three statements follow analogously.
    To argue that $g\!\restriction_{x_{i} = 0} \notequiv 0$, we have to show that $[g = 1] \cap [g_{i = 0} = 1] \neq \emptyset$.
    We write $[i = 0] \coloneqq [g_{i = 0} = 1]$ and $[i = 1] \coloneqq [g_{i = 1} = 1]$, so that, evidently, $\bit^{n} = [i = 0]\, \dot{\cup}\, [i = 1]$.
    Setting $\psi_{\bm{x}} \coloneqq \braket{\bm{x} \vert \psi}$, $\bm{x} \in \bit^{n}$, and using that $\braket{\bm{x} \vert \bm{y}} = \delta_{\bm{x} \bm{y}}$ for all $\bm{x}, \bm{y} \in \bit^{n}$, we obtain
    \begin{align*}
        \braket{\psi \vert \texttt{Z}_{i} \vert \psi} &=\hspace*{-3pt} \sum_{\bm{x} \scriptin [i = 0]} \abs{\psi_{\bm{x}}}^{2}\hspace*{-1pt} \braket{\bm{x} \vert \texttt{Z}_{i} \vert \bm{x}} +\hspace*{-3pt} \sum_{\bm{x} \scriptin [i = 1]} \abs{\psi_{\bm{x}}}^{2}\hspace*{-1pt} \braket{\bm{x} \vert \texttt{Z}_{i} \vert \bm{x}} \\
        &= \hspace*{-3pt}\sum_{\bm{x} \scriptin [i = 0]} \abs{\psi_{\bm{x}}}^{2} -\hspace*{-3pt} \sum_{\bm{x} \scriptin [i = 1]} \abs{\psi_{\bm{x}}}^{2},
    \end{align*}
    so that $\braket{\psi \vert \texttt{Z}_{i} \vert \psi} \neq -1$ implies the existence of an $\bm{x} \in [i = 0]$ with $\abs{\psi_{\bm{x}}}^{2} \neq 0$.
    Since $\ket{\psi}$ lies in the feasible subspace, $\psi_{\bm{x}} \neq 0$ implies $\bm{x} \in [g = 1]$, thus showing that indeed $[g = 1] \cap [i = 0] \neq \emptyset$.
\end{proof}

\subsection{\label{subsection:InitialStatePreparation}Initial state preparation}

We require the initial state preparations for the root problem $(n_{0}, f_{0}, g_{0})$ and all feasible subproblems to be trivial to implement and to produce a state strictly within the respective feasible subspace.
This directly extends Hadfield~\textit{et~al.}'s requirements from one problem instance to an entire family of problem instances.
In general, each subproblem will have an independent state preparation routine.
However, if the initial state for the current COP $(n, f, g)$ completely obeys the unary or binary constraint to be enforced for the next variable reduction, then it may be simply reused for the derived problem.
For example, assume a state preparation routine $\ket{\bm{0}} \mapsto \ket{\iota}$ with $\ket{\iota} \in \spn\{\ket{\bm{x}} : g(\bm{x}) = g_{i = 1}(\bm{x}) = 1\}$ and assume that $g_{i = 1}$ is enforced in the next step.
Then one can clearly reuse $\ket{\iota}$ (and hence its preparation routine) for the derived subproblem, simply by ignoring the $i$th qubit (or even omitting it if it stays disentangled from the other qubits) and decrementing the indices of all qubits with index $k > i$.
This is especially convenient and often applicable when $\ket{\iota}$ is a single feasible computational basis state $\ket{\bm{x}}$, since checking whether $\bm{x}$ obeys the enforced constraint and, if so, simply removing the $i$th qubit are two trivial tasks.

\subsection{\label{subsection:PhaseSeparator}Phase separator}

The phase separators for the root problem and all feasible subproblems should be diagonal in the computational basis and depend on the respective problem's objective function.
As for the (non-recursive) quantum alternating operator ansatz, the phase separator will usually be given by $U_{\mathrm{P}}(\gamma) = e^{-i \gamma H_{f}}$.
This automatically ensures that the phase separator for a subproblem can be straightforwardly constructed from the current problem's phase separator.
The substitutions $x_{i} \mapsto 0, 1, x_{j}, \neg x_{j}$ directly correspond to replacing $\texttt{Z}_{i}$ in $H_{f}$ by $\one, -\one, \texttt{Z}_{j}, -\texttt{Z}_{j}$, respectively.
Hence, these mechanical replacements (and subsequent qubit re-indexing) yield $H_{f\!\restriction_{x_{i} = y}}$, where $y \in \{0, 1, x_{j}, \neg x_{j}\}$, and with it the phase separator for each possible derived subproblem.

\subsection{\label{subsection:Mixer}Mixer}

As for the (non-recursive) quantum alternating operator ansatz, the mixer design for the RQAOA is the most intricate and structural component.
First of all, we extend Hadfield~\textit{et~al.}'s criteria from a single problem to a family of derived subproblems:
The mixers for the root problem and all feasible subproblems should preserve and explore the respective feasible subspaces.
That is, for every $(n, f, g)$---whether root or derived problem---and every $\beta \in \R$, the respective mixer $U_{\mathrm{M}}(\beta)$ should fulfil $U_{\mathrm{M}}(\beta)(\mathcal{S}) \subseteq \mathcal{S}$.
Furthermore, for every $\bm{x}, \bm{y} \in \bit^{n}$ with $g(\bm{x}) = g(\bm{y}) = 1$, there should exist some parameter value $\beta$ and some positive integer $r$ such that $\abs{\braket{\bm{x} \vert U_{\mathrm{M}}^{r}(\beta) \vert \bm{y}}} > 0$.
On top of the problem-wise mixing properties, we also require that the mixer for a subproblem can be efficiently built/compiled from a mixer for the current problem.
That is, given some description for the mixer $U_{\mathrm{M}}(\beta)$ for a COP $(n, f, g)$, there should be efficient mechanical rules for constructing a mixer $\tilde{U}_{\mathrm{M}}(\beta)$ for $(n - 1, f\!\restriction_{x_{i} = y}, g\!\restriction_{x_{i} = y})$, where $y \in \{0, 1, x_{j}, \neg x_{j}\}$ and $i, j \in [n]$ with $i \neq j$, whenever $g\!\restriction_{x_{i} = y} \notequiv 0$.
This criterion admits two strategies.
The subproblem's mixer may either be inferred from the current problem's mixer, e.g., by discarding or combining its constituents, or be fabricated anew from a description of the subproblem alone.
Inference is considerably more restrictive, since a mixer tailored to the current feasible set need not provide the building blocks for the subproblem's feasible set.
We will encounter both strategies in the next section, and \autoref{appendix:InferenceFromParent} shows that inference can be exponentially more expensive than fabrication, even for Hamming weight constraints.

\section{\label{section:CatalogueOfRQAOACircuitFamilies}Catalogue of RQAOA circuit families}

\begin{figure*}[t]
    \centering
    \colorlet{subsok}{ForestGreen!80!black}
    \colorlet{subscond}{Orange!90!black}
    \colorlet{subsno}{black!25}
    % closure under the four substitutions x_i = 0, 1, x_j, not x_j
    \newcommand*{\sbs}[4]{\textcolor{#1}{$0$}\enspace\textcolor{#2}{$1$}\enspace\textcolor{#3}{$x_{j}$}\enspace\textcolor{#4}{$\neg x_{j}$}}
    \newcommand*{\sbsall}{\sbs{subsok}{subsok}{subsok}{subsok}}
    \tikzset{
        cls/.style={draw, rounded corners=3pt, align=center, font=\scriptsize, text width=2.45cm, minimum height=1.9cm, inner sep=3pt, fill=white, execute at begin node={\hyphenpenalty=10000\exhyphenpenalty=10000}},
        sec/.style={label={[font=\tiny, fill=white, inner sep=1pt, anchor=center, xshift=-0.4cm]north east:#1}},
        partial/.style={cls, draw=black!60},
        closed/.style={cls, draw=subsok, thick},
        example/.style={cls, dashed, draw=black!45},
        incl/.style={-{Latex[length=1.8mm]}, black!55, thick},
    }
    \begin{tikzpicture}[x=2.84cm, y=2.35cm]
        % row 0: unconstrained problems and problems from the literature
        \node[closed, sec=\S\ref{subsection:BaseCaseUnconstrainedProblems}] (none) at (0, 0) {\textbf{unconstrained} \\
            \mbox{$g \equiv 1$} \\
            \sbsall \\
            always};
        \node[example, sec=\S\ref{subsection:UnateFeasibilityPredicates}] (mis) at (1, 0) {\textbf{MIS} \\
            \mbox{$x_{i} + x_{j} \leq 1$,} \\
            \mbox{$\{i, j\} \in E$} \\
            \sbs{subsok}{subscond}{subscond}{subsno}};
        \node[example, sec=\S\ref{subsection:SingleLinearInequalityConstraints}] (knap) at (2, 0) {\textbf{knapsack} \\
            \mbox{$\bm{a} \cdot \bm{x} \leq \tau$,} \\
            \mbox{$\bm{a} \geq \bm{0}$} \\
            \sbs{subsok}{subsok}{subsok}{subsno}};
        \node[example, sec=\S\ref{subsection:EqualityAndInequalityConstraints}] (cck) at (3, 0) {\textbf{cardinality-} \\
            \textbf{constr.~knapsack} \\
            \mbox{$\abs{\bm{x}} = d$,} \\
            \mbox{$\bm{a} \cdot \bm{x} \leq \tau$} \\
            \sbs{subsok}{subsok}{subsno}{subsno}};
        \node[example, sec=\S\ref{subsection:LinearFormConstraints}] (onehot) at (4, 0) {\textbf{one-hot encoding} \\
            \mbox{$\abs{\bm{x}_{v}} = 1$,} \\
            \mbox{$v \in V$} \\
            \sbs{subsok}{subscond}{subscond}{subsno}};
        \node[example, sec=\S\ref{subsection:HammingWeightConstraints}] (hw) at (5, 0) {\textbf{Hamming weight} \\
            \mbox{$\abs{\bm{x}} = d$} \\
            \sbs{subsok}{subsok}{subsno}{subsno}};
        % row 1
        \node[closed, sec=\S\ref{subsection:SingleLinearInequalityConstraints}] (ineq) at (2, 1) {\textbf{single linear} \\
            \textbf{inequality} \\
            \mbox{$\bm{a} \cdot \bm{x} \leq \tau$} \\
            \sbsall \\
            always};
        \node[closed, sec=\S\ref{subsection:HammingWeightConstraints}] (eq) at (5, 1) {\textbf{linear equality} \\
            \mbox{$\bm{w} \cdot \bm{x} = d$} \\
            \sbsall \\
            poly.~bounded $\bm{w}$};
        % row 2
        \node[partial, sec=\S\ref{subsection:UnateFeasibilityPredicates}] (unate) at (1, 2) {\textbf{unate} \\
            \mbox{$g$ unate in every $x_{i}$} \\
            \sbs{subsok}{subsok}{subscond}{subscond} \\
            poly.-time $g$};
        \node[closed, sec=\S\ref{subsection:EqualityAndInequalityConstraints}] (eqineq) at (3, 2) {\textbf{equalities and} \\
            \textbf{one inequality} \\
            \mbox{$A \bm{x} = \bm{d}$, $\bm{a} \cdot \bm{x} \leq \tau$} \\
            \sbsall \\
            \mbox{$m = O(1)$,} \\
            poly.~bounded $A$};
        \node[closed, sec=\S\ref{subsection:LinearFormConstraints}] (sys) at (4, 2) {\textbf{linear equalities} \\
            \mbox{$A \bm{x} = \bm{d}$} \\
            \sbsall \\
            \mbox{$m = O(1)$,} \\
            poly.~bounded $A$};
        % row 3
        \node[closed, sec=\S\ref{subsection:LinearFormConstraints}] (forms) at (3, 3) {\textbf{constraints on} \\
            \textbf{linear forms} \\
            \mbox{$A \bm{x} \in D$} \\
            \sbsall \\
            \mbox{$m = O(1)$,} \\
            poly.~bounded $A$, \\
            poly.-time $D$};
        % legends
        \node[font=\scriptsize, align=left, anchor=north west] at (-0.45, 3.4) {$\mathcal{X} \rightarrow \mathcal{Y}$: class $\mathcal{X}$ is contained in class $\mathcal{Y}$ \\
            solid: classes treated in this work \\
            dashed: problems from the literature \\
            last line: condition for efficient construction \\
            \S: section treating the class};
        \node[font=\scriptsize, align=left, anchor=north west] at (3.6, 3.4) {closure under substitution $x_{i} = y$: \\
            \textcolor{subsok}{$y$}: always closed \\
            \textcolor{subscond}{$y$}: closed under additional conditions \\
            \textcolor{subsno}{$y$}: not closed in general};
        % inclusions
        \draw[incl] (none) -- (unate);
        \draw[incl] (mis) -- (unate);
        \draw[incl] (knap) -- (ineq);
        \draw[incl] (ineq) -- (unate);
        \draw[incl] (ineq) -- (eqineq);
        \draw[incl] (cck) -- (eqineq);
        \draw[incl] (onehot) -- (sys);
        \draw[incl] (hw) -- (eq);
        \draw[incl] (eq) -- (sys);
        \draw[incl] (sys) -- (eqineq);
        \draw[incl] (eqineq) -- (forms);
    \end{tikzpicture}
    \caption{\label{figure:ClassInclusion}
        Inclusions between the constraint classes treated in \autoref{section:CatalogueOfRQAOACircuitFamilies}, together with their closure under the four RQAOA-driven substitutions $x_{i} = y$, $y \in \{0, 1, x_{j}, \neg x_{j}\}$, and sufficient conditions for efficiently constructing mixers and initial states for all subproblems.
        Classes closed under all four substitutions (solid green frames) are the target classes of our constructions; problems from the literature (dashed frames) are embedded into them.
        Unate predicates are closed under binary substitutions only if the polarities of $x_{i}$ and $x_{j}$ match as specified in \autoref{proposition:UnateReduction}, which motivates the admissibility restriction in \autoref{algorithm:UnateRQAOA}.
        The MIS and one-hot encodings stay within their class under $x_{i} = 1$ and $x_{i} = x_{j}$ only after eliminating the variables forced to $0$, e.g., the neighbours of $i$ for the MIS, or both variables if they share an edge or a one-hot group, respectively; merging variables of different one-hot groups leaves the class.
        Conjunctions of several linear inequalities, such as the MIS constraints, are also constraints on linear forms, albeit with a number of linear forms growing with the problem size; similarly, the unconstrained case is the special case $m = 0$ of every class on the right.
    }
\end{figure*}
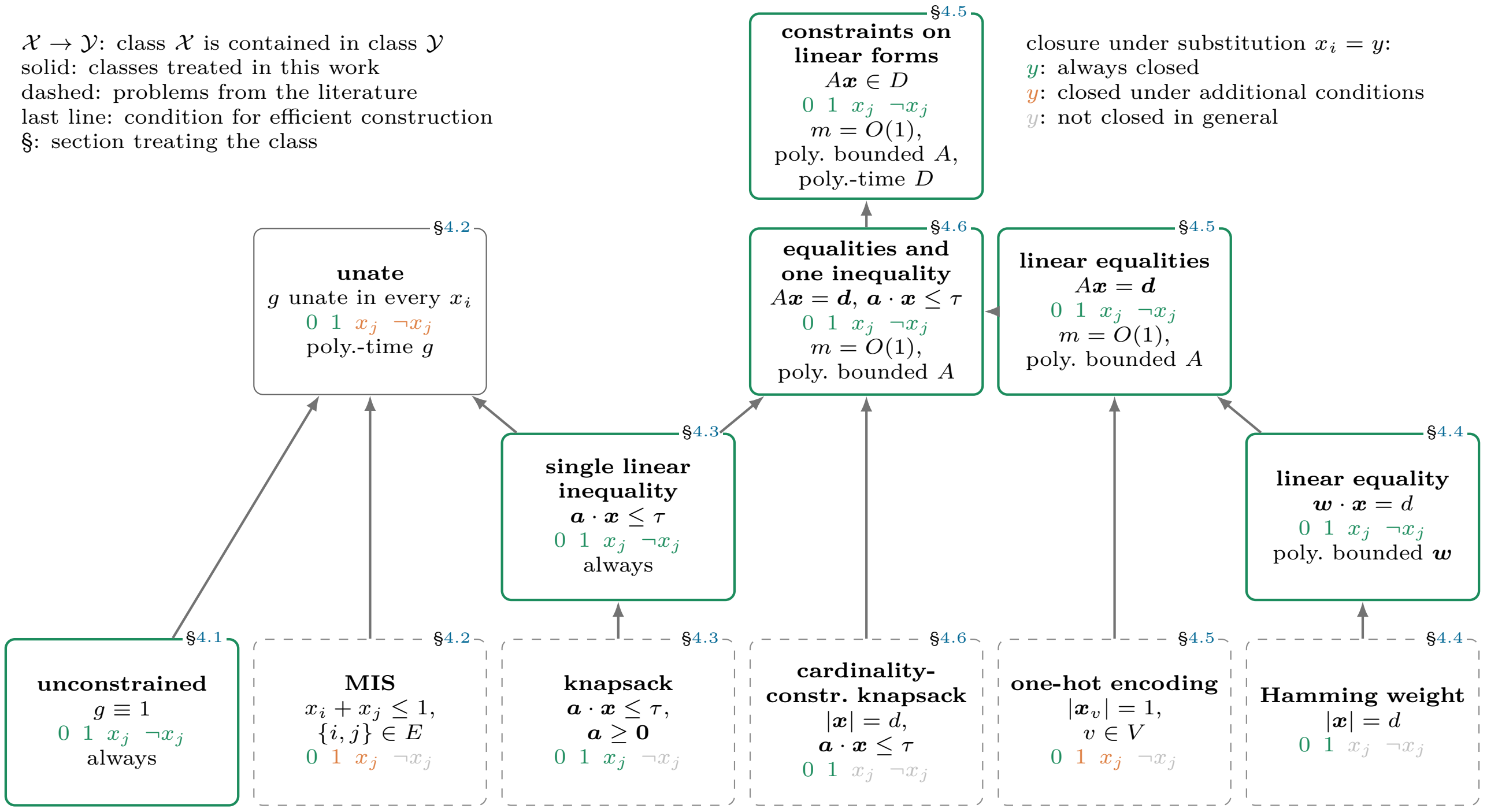

In this section, we give concrete examples of RQAOA circuit families, satisfying the design criteria laid out in the previous section.
We will iterate through several problem classes that have already been addressed by Hadfield~\textit{et~al.}~\cite{Hadfield2019FromTheQuantumApproximateOptimizationAlgorithmToAQuantumAlternatingOperatorAnsatz}, and also introduce novel constructions for less explored problem classes.
Since the root problem's phase separator and the mechanical rules to derive phase separators for every subproblem are, to a certain extent, trivial, we will aggregate problems by their feasibility structure rather than by their objective functions.
\autoref{table:Catalogue} summarises the catalogue, and \autoref{figure:ClassInclusion} shows how the underlying constraint classes relate to each other and which of them are closed under all four substitutions.

\begin{table*}[t]
    \centering
    \small
    \renewcommand{\arraystretch}{1.15}
    % p-columns clash with quantumarticle, hence explicit parboxes with one width per column
    \newcommand*{\catcell}[2]{\parbox[t]{#1}{\raggedright\hyphenpenalty=10000\strut #2\rule[-5pt]{0pt}{0pt}}}
    % column widths as shares (out of 89) of the text width left after padding and rules
    \newcommand*{\catw}[1]{\dimexpr(\textwidth - 8\tabcolsep - 5\arrayrulewidth) * #1 / 89\relax}
    \newcommand*{\catwA}{\catw{19}}
    \newcommand*{\catwB}{\catw{20}}
    \newcommand*{\catwC}{\catw{16}}
    \newcommand*{\catwD}{\catw{34}}
    \begin{tabular}{|l|l|l|l|}
        \hline
        \catcell{\catwA}{feasibility predicate} & \catcell{\catwB}{mixer} & \catcell{\catwC}{initial state} & \catcell{\catwD}{efficient for} \\
        \hline
        \catcell{\catwA}{none} & \catcell{\catwB}{any mixer formulated for a variable number of qubits, e.g.,~\eqref{equation:StandardQAOAMixer}} & \catcell{\catwC}{e.g., $\ket{\bm{+}}$} & \catcell{\catwD}{all instances} \\
        \hline
        \catcell{\catwA}{unate} & \catcell{\catwB}{controlled bit flips} & \catcell{\catwC}{safe bitstring~$\ket{\neg \bm{b}}$} & \catcell{\catwD}{predicates computable in polynomial time} \\
        \hline
        \catcell{\catwA}{linear inequality, e.g., knapsack} & \catcell{\catwB}{controlled bit flips with threshold controls} & \catcell{\catwC}{safe bitstring~$\ket{\neg \bm{b}}$} & \catcell{\catwD}{all instances} \\
        \hline
        \catcell{\catwA}{linear equality, e.g., Hamming weight} & \catcell{\catwB}{generalised \texttt{XY} block exchanges} & \catcell{\catwC}{dynamic program witness} & \catcell{\catwD}{polynomially bounded weights} \\
        \hline
        \catcell{\catwA}{constraints on linear forms} & \catcell{\catwB}{block exchanges and controlled star moves} & \catcell{\catwC}{dynamic program hub} & \catcell{\catwD}{constantly many linear forms with polynomially bounded coefficients, polynomial-time membership test} \\
        \hline
        \catcell{\catwA}{linear equalities and a single linear inequality} & \catcell{\catwB}{block exchanges with threshold controls} & \catcell{\catwC}{$\bm{a}$-minimal witness} & \catcell{\catwD}{constantly many equalities with polynomially bounded coefficients; arbitrary inequality coefficients} \\
        \hline
    \end{tabular}
    \caption{\label{table:Catalogue}
        Overview of the RQAOA circuit families presented in \autoref{section:CatalogueOfRQAOACircuitFamilies}.
        The column ``efficient for'' states sufficient conditions under which the mixers and initial states for all subproblems can be constructed in polynomial time.
    }
\end{table*}

\subsection{\label{subsection:BaseCaseUnconstrainedProblems}Base case: unconstrained problems}

The simplest constraints are no constraints at all.
By enforcing (anti-)correlations through explicit variable substitution, subproblems uncovered by the RQAOA stay unconstrained.
Formally: for an unconstrained COP $(n, f, 1)$ and any $i, j \in [n]$ it is $g\!\restriction_{x_{i} = y} \equiv 1$ for all $y \in \{0, 1, x_{j}, \neg x_{j}\}$.
Hence, any initial state preparation routine and mixer which are formulated for a variable number of qubits can be simply reused for all subproblems.
This is especially true for the standard QAOA mixer~\eqref{equation:StandardQAOAMixer} and the associated initial state preparation $\ket{\bm{0}} \mapsto \texttt{H}^{\otimes n} \ket{\bm{0}} = \ket{\bm{+}}$.
Another admissible construction is the warm-starting QAOA for unconstrained binary optimisation problems~\cite{Egger2021WarmStartingQuantumOptimization}.
If $\deg(f) \leq 2$, then clearly $\deg(f\!\restriction_{x_{i} = y}) \leq 2$ for every $i, j \in [n]$ and every $y \in \{0, 1, x_{j}, \neg x_{j}\}$ so that no possible variable substitution leaves the problem class.
A (more sophisticated) workflow for using warm-starting and RQAOA in concert has already been discussed by Egger~\textit{et~al.}~\cite{Egger2021WarmStartingQuantumOptimization}.

\subsection{\label{subsection:UnateFeasibilityPredicates}Unate feasibility predicates}

Existing extensions of RQAOA to constrained problems are predominantly focused on the MIS and the equivalent MVC problem.
Here we compare current fine-tuned reduction steps to our more general construction.
A concrete feasibility-preserving QAOA mixer for the MIS was already constructed by Hadfield~\textit{et~al.}~\cite{Hadfield2019FromTheQuantumApproximateOptimizationAlgorithmToAQuantumAlternatingOperatorAnsatz} and is given by multiple controlled bit flips:
For a given graph structure $G = ([n], E)$ and vertex $i \in [n]$, let $N(i) \coloneqq \{j \in [n] \defcolon \{i, j\} \in E\}$ be the neighbourhood of $i$ (in $G$).
The independence constraints $g_{\{\overline{i, j}\}}(\bm{x}) = 1 - x_{i} x_{j}$, $\{i, j\} \in E$, may be expressed as $g_{N(i)}(\bm{x}) \coloneqq 1 - x_{i} \land \bigvee_{j \scriptin N(i)} x_{j}$, $i \in [n]$, ensuring that $x_{i}$ is not set as soon as at least one of its neighbours is set.
If $\bm{x}$ is feasible, setting $x_{i} = 0$ always preserves feasibility while setting $x_{i} = 1$ is only allowed if none of (the variables representing) its neighbouring vertices are set.
Hence a bit flip of $x_{i}$ is, in full generality, only feasibility-preserving if conditioned on $\neg \bigvee_{j \scriptin N(i)} x_{j}$.
This classical move can be realised coherently by applying an $\texttt{X}$-gate to the $i$th qubit, negatively controlled on all qubits representing vertices $j \in N(i)$.
Exponentiating the $\texttt{X}$-gate yields a feasibility-preserving, parameterised partial mixer
\begin{equation}\label{equation:PartialMISMixer}
    U_{\text{CM}, i}(\beta) \coloneqq \Lambda_{\text{NOR}(N(i))} \left(e^{-i \beta \texttt{X}_{i}}\right),
\end{equation}
where $\Lambda_{\text{NOR}(N(i))}$ expresses the coherent control pattern.
One can now achieve transitions between all pairs of feasible computational basis states by executing all partial mixers $U_{\text{CM}, i}(\beta)$, $i \in [n]$, in sequence.\footnote{
    Alternatively, one can exponentiate the sum of all suitably controlled \texttt{X}-gates simultaneously.
}

Products of single-qubit operations with classical control patterns, such as~\eqref{equation:PartialMISMixer}, can be straightforwardly adapted to derived subproblems.
However, let us first generalise the MIS mixer design, thereby extending it to a broader family of COPs, to give more general compilation rules for subproblem mixers.
Concretely, we are targeting all COPs with a \emph{unate} feasibility predicate $g$~\cite{Sasao1999SwitchingTheoryForLogicSynthesis}.

\begin{definition}\label{definition:UnateFunction}
    A Boolean function $g : \bit^{n} \rightarrow \bit$ is \emph{positive} (\emph{negative}) \emph{unate in} $x_{i}$ if for all $\bm{x} \in \bit^{n}$ it holds that $g(\bm{x}[i \leftarrow 1]) \geq g(\bm{x})$ ($g(\bm{x}[i \leftarrow 0]) \geq g(\bm{x})$).
    If $g$ is unate in every $x_{i}$, $i \in [n]$, then $g$ is said to be \emph{unate}.
\end{definition}

If we interpret $g : \bit^{n} \rightarrow \bit$ as an indicator function for the set $S \coloneqq [g = 1]$, as we do for COP feasibility predicates, then $g$ being, e.g., positive unate in $x_{i}$ simply means that $\bm{x}[i \leftarrow 1] \in S$ whenever $\bm{x} \in S$.
The MIS clearly has a feasibility predicate which is negative unate in every variable, since removing any vertex from an independent set again yields an independent set.
In order to extend the MIS mixer design to general problems with unate feasibility predicates, we observe that the control pattern in $U_{\text{CM}, i}$, $i \in [n]$, corresponds to checking whether the current bitstring with its $i$th bit replaced by the ``unsafe'' value---$1$ ($0$) if $g$ is negative (positive) unate in $x_{i}$---is again feasible.
Denoting the ``unsafe'' value by $b_{i}$, we may therefore write the control pattern as $\Lambda_{g\!\restriction_{x_{i} = b_{i}}}$.
The following proposition proves that this generalisation always adheres to Hadfield~\textit{et~al.}'s mixer criteria.

\begin{proposition}\label{proposition:BitFlipMixersForUnateFeasibilityPredicates}
    Let $(n, f, g)$ be a COP with unate feasibility predicate $g$ and let $\bm{b} \in \bit^{n}$ so that $b_{i} = 0$ ($b_{i} = 1$) if $g$ is positive (negative) unate in $x_{i}$, $i \in [n]$.
    If $g$ is both, positive and negative unate in $x_{i}$, $b_{i}$ may be chosen arbitrarily.
    The parameterised mixer
    \begin{equation}\label{equation:UnateMixer}
        U_{\texttt{{\upshape CX}}}(\beta) \coloneqq \prod_{i = 1}^{n} U_{\texttt{{\upshape CX}}, i}(\beta) \coloneqq \prod_{i = 1}^{n} \Lambda_{g\restriction_{x_{i} = b_{i}}}\big(e^{-i \beta \texttt{{\upshape X}}_{i}}\big)
    \end{equation}
    is feasibility-preserving and provides transitions between all pairs of feasible computational basis states.
\end{proposition}

\begin{proof}
    First, it suffices to show feasibility preservation of the partial mixers $U_{\texttt{CX}, i}(\beta)$ to prove that $U_{\texttt{CX}}(\beta)$ preserves feasibility.
    Let $\bm{x} \in \bit^{n}$ with $g(\bm{x}) = 1$.
    Since $g$ is unate in $x_{i}$, $g(\bm{x}[i \leftarrow \neg b_{i}]) = 1$.
    Therefore, if $x_{i} = b_{i}$, then $e^{-i \beta \texttt{X}_{i}} \ket{\bm{x}} \in \spn\{\ket{\bm{x}}, \ket{\bm{x}[i \leftarrow \neg b_{i}]}\} \subseteq \mathcal{S}$ so that especially any controlled version maintains feasibility.
    More generally, the control explicitly prevents the application of $e^{-i \beta \texttt{X}_{i}}$, whenever
    \begin{equation*}
        g(\bm{x}[i \leftarrow b_{i}]) = 0 \Leftrightarrow g\hspace*{-3pt}\restriction_{\hspace*{-1pt}x_{i} = b_{i}}\hspace*{-4pt}(x_{1} \cdots x_{i - 1} x_{i + 1} \cdots x_{n}) = 0.
    \end{equation*}
    If now $x_{i} = \neg b_{i}$, then $e^{-i \beta \texttt{X}_{i}}$ is applied if and only if also $g(\bm{x}[i \leftarrow b_{i}]) = 1$, again maintaining feasibility.
    In summary and by linearity, $U_{\texttt{CX}, i}(\beta)$ is feasibility-preserving for every possible $\beta$.
    Second, let $\bm{x}, \bm{y} \in \bit^{n}$ with $g(\bm{x}) = g(\bm{y}) = 1$.
    We will argue that $\braket{\bm{x} \vert U_{\texttt{CX}}^{2}(\beta) \vert \bm{y}} \neq 0$ for all but finitely many $\beta \in [0, 2 \pi)$.
    Each partial mixer reads $U_{\texttt{CX}, i}(\beta) = e^{-i \beta G_{i}}$, where $G_{i} \coloneqq \Lambda_{g\restriction_{x_{i} = b_{i}}}(\texttt{X}_{i})$ is a symmetric matrix with entries in $\bit$ with respect to the computational basis.
    Expanding all $2 n$ exponentials in $U_{\texttt{CX}}^{2}(\beta)$ into their power series yields
    \begin{equation*}
        \braket{\bm{x} \vert U_{\texttt{CX}}^{2}(\beta) \vert \bm{y}} = \sum_{\bm{k} \scriptin \N_{0}^{2 n}} \frac{(-i \beta)^{\abs{\bm{k}}}}{\bm{k}!} \braket{\bm{x} \vert G_{\bm{k}} \vert \bm{y}},
    \end{equation*}
    where $\abs{\bm{k}} \coloneqq \sum_{j} k_{j}$, $\bm{k}! \coloneqq \prod_{j} k_{j}!$, and $G_{\bm{k}}$ denotes the product of the respective powers of the $G_{i}$ in the order of $U_{\texttt{CX}}^{2}(\beta)$.
    All matrix elements $\braket{\bm{x} \vert G_{\bm{k}} \vert \bm{y}}$ are non-negative.
    Hence, if $D$ denotes the smallest $\abs{\bm{k}}$ with $\braket{\bm{x} \vert G_{\bm{k}} \vert \bm{y}} > 0$, then no cancellations occur at leading order, i.e., for some $c > 0$,
    \begin{align*}
        \braket{\bm{x} \vert U_{\texttt{CX}}^{2}(\beta) \vert \bm{y}} &= (-i)^{D} c \beta^{D} + O(\beta^{D + 1}), \\
        e^{-i \beta G_{i}} &= \one + (\cos(\beta) - 1) G_{i}^{2} - i \sin(\beta) G_{i},
    \end{align*}
    where the latter follows from $G_{i}^{3} = G_{i}$.
    Thus, the overlap is a trigonometric polynomial in $\beta$, which, not vanishing identically, has only finitely many zeros in $[0, 2 \pi)$.
    It remains to show that $D$ exists, i.e., that some sequence of enabled bit flips, at most one per partial mixer, leads from $\bm{y}$ to $\bm{x}$ within two sweeps.
    We connect $\bm{x}$ and $\bm{y}$ via the ``safe'' state $\neg \bm{b}$.
    In the first sweep, the $i$th partial mixer flips $y_{i} = b_{i}$ to $\neg b_{i}$ when acting on $(\neg b_{1}) \cdots (\neg b_{i - 1}) y_{i} \cdots y_{n}$, which is feasible by unateness of $g$ and $g(\bm{y}) = 1$, so that the flip is enabled.
    This sweep ends in $\neg \bm{b}$.
    In the second sweep, the $i$th partial mixer flips $\neg b_{i}$ to $x_{i} = b_{i}$ when acting on $x_{1} \cdots x_{i - 1} (\neg b_{i}) \cdots (\neg b_{n})$, which is enabled since $x_{1} \cdots x_{i} (\neg b_{i + 1}) \cdots (\neg b_{n})$ is feasible by unateness of $g$ and $g(\bm{x}) = 1$.
    This sweep ends in $\bm{x}$.
\end{proof}

This mixer design~\eqref{equation:UnateMixer} is valid for all COPs with unate feasibility predicate.
Therefore, if an RQAOA-driven substitution $x_{i} = y \in \{0, 1, x_{j}, \neg x_{j}\}$ produces a subproblem with unate feasibility predicate $g\!\restriction_{x_{i} = y}$, there is again a mixer of the form~\eqref{equation:UnateMixer} for the subproblem.
We describe next the concrete relation between the mixers of the root and the derived problem, and investigate under which circumstances an RQAOA-based substitution yields again a subproblem with unate feasibility predicate.
The mechanical rules for compiling the subproblem mixer from the root problem's one are rather straightforward:
In any case, one variable, say $x_{i}$, is eliminated, so that the partial mixer $U_{\texttt{CX}, i}(\beta)$ can be fully dropped.
Afterwards, we merely have to adapt the control patterns in the remaining partial mixers.
Let us first consider the substitutions $x_{i} = 0 / 1$ and their effect on the $k$th partial mixer $U_{\texttt{CX}, k}(\beta)$, where $k \in [n] \setminus \{i\}$.
Every appearance of (the qubit representing) $x_{i}$ in the control pattern $\Lambda_{g\!\restriction_{x_{k} = b_{k}}}$ has to be replaced with $0 / 1$.
Controlling on constants can evidently be simplified at compile time:
If $x_{i}$ is to be substituted by $1$, every gate controlled on $\neg x_{i}$ will not be executed with certainty, i.e., it can be fully removed from the circuit.
In contrast, controls on $x_{i}$ are redundant; here the control rather than the target is omitted.
Analogously, if $x_{i}$ is to be substituted by $0$, controls on $\neg x_{i}$ vanish and gates controlled on $x_{i}$ can be removed.
The substitutions $x_{i} = x_{j} / \neg x_{j}$ are similar in spirit, but generally lead to less simplified circuits.
If $x_{i}$ is to be substituted by $x_{j}$, all controls on $(\neg) x_{i}$ have to be rewired to $(\neg) x_{j}$.
If instead $x_{i}$ is to be substituted by $\neg x_{j}$, these controls are additionally negated.
Symbolically, the partial mixer for the subproblem becomes
\begin{equation*}
    U_{\texttt{CX}, k}'(\beta) = \Lambda_{(g\restriction_{x_{i} = y})\restriction_{x_{k} = b_{k}}}\big(e^{-i \beta \texttt{X}_{k}}\big),
\end{equation*}
where $y \in \{0, 1, x_{j}, \neg x_{j}\}$.
This compilation rule also holds true for $k = j$, since the substitution $x_{i} = (\neg) x_{j}$ is executed before the substitution $x_{j} = b_{j}$.

When the subproblem's feasibility predicate is not unate, this construction will produce a parameterised circuit that might fail to be feasibility-preserving or to provide transitions between all feasible computational basis states.
This may already happen when a binary substitution identifies two variables of opposite polarities.
However, there is a simple sufficient criterion for an RQAOA-driven substitution to produce again a problem with unate feasibility predicate.

\begin{proposition}\label{proposition:UnateReduction}
    Let $g : \bit^{n} \rightarrow \bit $ be unate.
    For every $i \in [n]$ and every $j \in [n] \setminus \{i\}$, the following hold true:
    \begin{itemize}
        \item[i)] $g\!\restriction_{x_{i} = 0}$ is unate.
        \item[ii)] $g\!\restriction_{x_{i} = 1}$ is unate.
        \item[iii)] $g\!\restriction_{x_{i} = x_{j}}$ is unate if $g$ is either positive unate in both, $x_{i}$ and $x_{j}$, or negative unate in both, $x_{i}$ and $x_{j}$.
        \item[iv)] $g\!\restriction_{x_{i} = \neg x_{j}}$ is unate if $g$ is either positive unate in $x_{i}$ and negative unate in $x_{j}$, or negative unate in $x_{i}$ and positive unate in $x_{j}$.
    \end{itemize}
\end{proposition}

\begin{proof}
    Statements i) and ii) both follow immediately from \autoref{definition:UnateFunction}:
    W.l.o.g., we may consider the case that $i = 1$ and let, for every $k \in [n - 1]$, $e_{k} = 0$ (resp.\ $e_{k} = 1$) if $g$ is negative (resp.\ positive) unate in $x_{k + 1}$.\footnote{
        Note that $e_{k}$ is defined as the ``safe'' value for the respective bit positions; this is the negated version of the bit values $b_{i}$ defined for the parameterised mixer in \autoref{proposition:BitFlipMixersForUnateFeasibilityPredicates}.
    }
    Then, for every $\bm{y} \in \bit^{n - 1}$ and every $k \in [n - 1]$, it holds that
    \begin{align*}
        &g\!\restriction_{x_{i} = 0}(\bm{y}) = g(0 \bm{y}) \\
        &\quad\quad \leq g(0 \bm{y}[k \leftarrow e_{k}]) = g\!\restriction_{x_{i} = 0}(\bm{y}[k \leftarrow e_{k}]),
    \end{align*}
    i.e., $g\!\restriction_{x_{i} = 0}$ is negative (positive) unate in $x_{k}$ if $e_{k} = 0$ ($e_{k} = 1$).
    Since this holds for all $k \in [n - 1]$, $g\!\restriction_{x_{i} = 0}$ is again unate.
    Completely analogously, one argues that $g\!\restriction_{x_{i} = 1}$ is also unate.

    Statements iii) and iv) are similar in spirit and may be proven analogously; we focus on iii).
    Let $\epsilon_{i / j} = 0$ ($\epsilon_{i / j} = 1$) if $g$ is negative (positive) unate in $x_{i} / x_{j}$.
    If $g$ is both, positive and negative unate in one of the variables, prefer setting $\epsilon_{i} = \epsilon_{j}$.
    Overall, this reduces the claim to ``$\epsilon_{i} = \epsilon_{j} \implies g\!\restriction_{x_{i} = x_{j}}$ is unate''.
    W.l.o.g., let $i = 1$ and $j = 2$, and let $e_{k}$, $k \in [n - 1]$, be as before.
    For every $\bm{y} \in \bit^{n - 1}$ and every $k \in [n - 1] \setminus \{1\}$ it is
    \begin{align*}
        &g\!\restriction_{x_{i} = x_{j}}(\bm{y}) = g(y_{1} y_{1} y_{2} y_{3} \cdots y_{k} \cdots y_{n - 1}) \\
        &\quad\ \leq g(y_{1} y_{1} y_{2} \cdots e_{k} \cdots y_{n - 1}) = g\!\restriction_{x_{i} = x_{j}}(\bm{y}[k \leftarrow e_{k}]),
    \end{align*}
    where we have used that $y_{k}$ sits at the $(k + 1)$th bit position in the bitstring $y_{1} y_{1} y_{2} \cdots y_{k} \cdots y_{n - 1}$.
    This shows that $g\!\restriction_{x_{i} = x_{j}}$ is unate in all $x_{k}$, $k \in [n - 1] \setminus \{1\}$.
    Now, if indeed $\epsilon_{1} = \epsilon_{2}$, then also
    \begin{align*}
        &g\hspace*{-3pt}\restriction_{x_{i} = x_{j}}\hspace*{-3pt}(\bm{y}) \leq g(\epsilon_{1} y_{1} y_{2} \cdots y_{n - 1}) \leq g(\epsilon_{1} \epsilon_{2} y_{2} \cdots y_{n - 1}) \\
        &\quad\quad = g(\epsilon_{2} \epsilon_{2} y_{2} \cdots y_{n - 1}) = g\hspace*{-3pt}\restriction_{x_{i} = x_{j}}\hspace*{-3pt}(\bm{y}[1 \leftarrow \epsilon_{2}]),
    \end{align*}
    i.e., $g\hspace*{-3pt}\restriction_{x_{i} = x_{j}}$ is also unate in $x_{1}$ and thus unate overall.
    Replacing $\epsilon_{1}$ with $\neg \epsilon_{1}$, one can prove statement iv) completely analogously.
\end{proof}

Note that iii) and iv) in \autoref{proposition:UnateReduction} cannot be strengthened to necessary conditions in full generality.
For example, writing the MIS' independence constraints as linear inequalities $x_{i} + x_{j} \leq 1$, $\{i, j\} \in E$, let $k \in N(i) \setminus (N(j) \cup \{j\})$ and $\ell \in N(j) \setminus (N(i) \cup \{i\})$ for some edge $\{i, j\} \in E$.
The substitution $x_{i} = \neg x_{j}$ transforms $x_{i} + x_{k} \leq 1$ into $x_{k} \leq x_{j}$, while $x_{j} + x_{\ell} \leq 1$ is left unchanged.
Setting all other variables to $0$, the choice $x_{k} = 0$, $x_{\ell} = 1$ is feasible for $x_{j} = 0$ but not for $x_{j} = 1$, whereas $x_{k} = 1$, $x_{\ell} = 0$ is feasible for $x_{j} = 1$ but not for $x_{j} = 0$.
Hence, the subproblem's feasibility predicate is not unate in $x_{j}$; \autoref{figure:MISCounterexample} illustrates this for a path on four vertices.
This aligns well with the additional rules that both, Fin\v{z}gar~\textit{et~al.}~\cite{Finzgar2024QuantumInformedRecursiveOptimizationAlgorithms} and Ramos-Ruiz~\textit{et~al.}~\cite{RamosRuiz2026OnTheUseOfTheQuantumAlternatingOperatorAnsatzInQuantumInformedRecursiveOptimizationACaseStudyOnTheMinimumVertexCover}, had to impose for the $x_{i} = \neg x_{j}$ substitution, as the MIS feasibility predicate is only negative unate in all variables.

\begin{figure}[t]
    \centering
    \tikzset{
        vtx/.style={circle, draw, thick, minimum size=5.5mm, inner sep=0pt, font=\footnotesize, fill=white},
        elim/.style={vtx, dashed, draw=black!35, text=black!35},
        indep/.style={thick},
        impl/.style={-{Latex[length=2mm]}, thick, NavyBlue},
    }
    \begin{tikzpicture}
        \node[font=\bfseries\footnotesize] at (-0.55, 0.75) {(a)};
        \node[vtx] (k) at (0, 0) {$k$};
        \node[vtx] (i) at (0.8, 0) {$i$};
        \node[vtx] (j) at (1.6, 0) {$j$};
        \node[vtx] (l) at (2.4, 0) {$\ell$};
        \draw[indep] (k) -- (i) -- (j) -- (l);
        \begin{scope}[xshift=3.7cm]
            \node[font=\bfseries\footnotesize] at (-0.55, 0.75) {(b)};
            \node[vtx] (k) at (0, 0) {$k$};
            \node[elim] (i) at (0.8, 0) {$i$};
            \node[vtx] (j) at (1.6, 0) {$j$};
            \node[vtx] (l) at (2.4, 0) {$\ell$};
            \draw[impl] (k) to[bend left=40] node[above, font=\scriptsize] {$x_{k} \leq x_{j}$} (j);
            \draw[indep] (j) -- (l);
            \node[font=\scriptsize] at (2.0, -0.5) {$x_{j} + x_{\ell} \leq 1$};
        \end{scope}
    \end{tikzpicture}

    \vspace*{0.2cm}
    \footnotesize
    \begin{tabular}{c|cc}
        $(x_{k}, x_{\ell})$ & $x_{j} = 0$ & $x_{j} = 1$ \\
        \hline
        $(0, 1)$ & feasible & infeasible \\
        $(1, 0)$ & infeasible & feasible
    \end{tabular}
    \normalsize
    \caption{\label{figure:MISCounterexample}
        Non-unate MIS subproblem.
        \textbf{(a)} Path $k - i - j - \ell$ with independence constraints along its edges, a feasibility predicate which is negative unate in every variable.
        \textbf{(b)} After the substitution $x_{i} = \neg x_{j}$, the constraint $x_{i} + x_{k} \leq 1$ becomes the implication $x_{k} \leq x_{j}$ (arrow), while $x_{j} + x_{\ell} \leq 1$ remains unchanged.
        As shown in the table, raising $x_{j}$ renders $(x_{k}, x_{\ell}) = (0, 1)$ infeasible and lowering it renders $(1, 0)$ infeasible, so that the subproblem's feasibility predicate is neither positive nor negative unate in $x_{j}$.
    }
\end{figure}

Let us now reintroduce Fin\v{z}gar~\textit{et~al.}'s and Ramos-Ruiz~\textit{et~al.}'s fine-tuned rules for multiple-variable elimination.
After a substitution, some of the remaining variables might take the same value in all feasible solutions of the subproblem.
For instance, substituting $x_{i} = 1$ in the MIS forces all variables representing neighbours of $i$ to attain the value $0$.
If a variable $x_{k}$ is fixed like that, then $\abs{M_{k}} = 1$ for every feasible state, so that the selection rule would spend the next RQAOA iterations on eliminating $x_{k}$ (or any other fixed variable).
For unate feasibility predicates, the value enforced upon $x_{k}$ can only ever be the predicate's $k$th ``safe value'' $\neg b_{k}$.
Hence, the partial mixer $U_{\texttt{CX}, k}(\beta)$ of an enforced variable acts as the identity on the feasible subspace, since its control $g\!\restriction_{x_{k} = b_{k}}$ vanishes on every feasible bitstring.
In addition to avoiding wasted RQAOA rounds and redundant control patterns, detecting enforced variables as early as possible saves valuable qubit resources in all subsequent rounds.
We would therefore like to detect and delete such variables in advance; for unate feasibility predicates, this can be done efficiently:
If there existed any feasible solution $\bm{x} \in \bit^{n}$ with $x_{k} = b_{k}$, then replacing the bits $x_{\ell}$, $\ell \in [n] \setminus \{k\}$, one after another by the ``safe'' values $\neg b_{\ell}$ would preserve feasibility by \autoref{definition:UnateFunction}, and would end in $(\neg \bm{b})[k \leftarrow b_{k}]$, which would hence be feasible.
The latter is a simple evaluation test.
If it fails, $x_{k}$ is forced to attain the value $\neg b_{k}$, so that we can execute the unary substitution $x_{k} = \neg b_{k}$ immediately.
A single pass of this test over all variables already suffices to detect all forced variables at once.
Symbolically, we assign each variable $x_{k}$ with $k \in J \coloneqq \{\ell \in [n] \defcolon g((\neg \bm{b})[\ell \leftarrow b_{\ell}]) = 0\}$ its safe value, yielding the still unate subproblem $(n - \abs{J}, f\!\restriction_{\bm{x}_{J} = \neg \bm{b}_{J}}, g\!\restriction_{\bm{x}_{J} = \neg \bm{b}_{J}})$.

\autoref{algorithm:UnateRQAOA} summarises the resulting RQAOA protocol for COPs with unate feasibility predicates, where we use the conventions $\sign(0) \coloneqq +1$ and $\max \emptyset \coloneqq -\infty$.
The initial state $\ket{\iota} = \ket{\neg \bm{b}}$ is a feasible computational basis state, prepared by at most $n$ \texttt{X}-gates; for the MIS, it simply encodes the empty vertex set, while for the MVC, it corresponds to the entire vertex set.
To guarantee unate subproblems, the set $\mathcal{A}$ only contains those pairs for which the substitution selected by the sign of $M_{i j}$ is covered by iii) or iv) of \autoref{proposition:UnateReduction}, i.e., $x_{i} = x_{j}$ requires $b_{i} = b_{j}$ and $x_{i} = \neg x_{j}$ requires $b_{i} \neq b_{j}$.
Under this restriction, the proof of \autoref{proposition:UnateReduction} shows that every remaining variable retains its polarity, so that $\bm{b}_{(i)}$ are again valid unsafe values for the subproblem.
Moreover, by \autoref{lemma:LocalConsistency}, choosing the substitution according to the sign of the selected expectation value guarantees $g\hspace*{-3pt}\restriction_{x_{i} = y} \not{\hspace*{-5pt}\equiv}\hspace*{5pt} 0$, so that the subproblem's initial state, i.e., its safe bitstring, is feasible again.

\subsection{\label{subsection:SingleLinearInequalityConstraints}Single linear inequality constraints}

\begin{algorithm}[t]
    \caption{\label{algorithm:UnateRQAOA}RQAOA for unate feasibility predicates}
    \begin{algorithmic}[1]
        \Require COP $(n, f, g)$ with unate $g$, $\bm{b} \in \bit^{n}$ as in \autoref{proposition:BitFlipMixersForUnateFeasibilityPredicates}, depth $p \in \N$, cut-off $n_{\mathrm{c}} \in \N$
        \While{$n > n_{\mathrm{c}}$}
        \State $\ket{\iota} \gets \ket{\neg \bm{b}}$
        \State $U_{\mathrm{P}}(\gamma) \gets e^{-i \gamma H_{f}}$
        \State $U_{\texttt{CX}}(\beta) \gets \prod_{k = 1}^{n} \Lambda_{g\restriction_{x_{k} = b_{k}}}\big(e^{-i \beta \texttt{X}_{k}}\big)$
        \State $(\bm{\beta}^{*}, \bm{\gamma}^{*}) \gets \arg\min_{\bm{\beta}, \bm{\gamma}} \braket{\bm{\beta}, \bm{\gamma} \vert H_{f} \vert \bm{\beta}, \bm{\gamma}}$
        \State estimate all $M_{i}$ and $M_{i j}$ w.r.t.~$\ket{\bm{\beta}^{*}, \bm{\gamma}^{*}}$
        \State $\mathcal{A} \gets \{(i, j) \defcolon i \neq j,\ \sign(M_{i j}) = (-1)^{b_{i} + b_{j}}\}$
        \If{$\max_{i} \abs{M_{i}} \geq \max_{(i, j) \scriptin \mathcal{A}} \abs{M_{i j}}$}
        \State $i \gets \arg\max_{i} \abs{M_{i}}$
        \State $y \gets 0$ if $M_{i} \geq 0$, else $y \gets 1$
        \Else
        \State $(i, j) \gets \arg\max_{(i, j) \scriptin \mathcal{A}} \abs{M_{i j}}$
        \State $y \gets x_{j}$ if $M_{i j} \geq 0$, else $y \gets \neg x_{j}$
        \EndIf
        \State $(f, g, \bm{b}) \gets (f\!\restriction_{x_{i} = y}, g\!\restriction_{x_{i} = y}, \bm{b}_{(i)})$,\ $n \gets n - 1$
        \State $J \gets \{k \in [n] \defcolon g((\neg \bm{b})[k \leftarrow b_{k}]) = 0\}$
        \State $n \gets n - \abs{J}$
        \State $(f, g, \bm{b}) \gets (f\!\restriction_{\bm{x}_{J} = \neg \bm{b}_{J}}, g\!\restriction_{\bm{x}_{J} = \neg \bm{b}_{J}}, \bm{b}_{(J)})$
        \EndWhile
        \State $\bm{x}^{*} \gets \arg\min\{f(\bm{x}) \defcolon g(\bm{x}) = 1\}$ \Comment{exhaustively}
        \State \Return $\bm{x}^{*}$
    \end{algorithmic}
\end{algorithm}
A prominent family of unate feasibility predicates, for which \autoref{proposition:UnateReduction} does not cover every substitution, but no RQAOA-driven substitution ever produces a non-unate predicate, is given by single 0--1 linear inequality constraints.
For given $n \in \N$, $\tau \in \Z$, and $\bm{a} \in \Z^{n}$, let $t_{\tau, \bm{a}}^{(n)} : \bit^{n} \rightarrow \bit$ be the Boolean function defined by
\begin{equation*}
    t_{\tau, \bm{a}}^{(n)}(\bm{x}) = 1 \text{ if and only if } \sum_{i = 1}^{n} a_{i} x_{i} \leq \tau.
\end{equation*}
Its archetypal instance is the knapsack constraint, where $a_{i} \geq 0$ are the item weights and $\tau$ is the capacity.
Since $\bm{a} \cdot \bm{x}[i \leftarrow 0] \leq \bm{a} \cdot \bm{x}$ if $a_{i} \geq 0$, and $\bm{a} \cdot \bm{x}[i \leftarrow 1] \leq \bm{a} \cdot \bm{x}$ if $a_{i} \leq 0$, the predicate $t_{\tau, \bm{a}}^{(n)}$ is negative unate in every $x_{i}$ with $a_{i} \geq 0$ and positive unate in every $x_{i}$ with $a_{i} \leq 0$; in particular, it is unate.
Especially, the knapsack constraint is negative unate in every variable, since removing an item from a feasible packing again yields a feasible packing.
The safe bitstring $\neg \bm{b}$ minimises $\bm{a} \cdot \bm{x}$, so that the test for forced variables reduces to a comparison: $x_{k}$ is forced if and only if $\abs{a_{k}}$ exceeds the slack $\tau - \sum_{i = 1}^{n} \min\{a_{i}, 0\}$; for the knapsack, these are exactly the items heavier than the capacity, which are usually discarded during preprocessing anyway.
Hence, \autoref{proposition:UnateReduction} guarantees unate subproblems for the substitutions $x_{i} = 0, 1, x_{j}$, but not for $x_{i} = \neg x_{j}$, which merges two items into a single one of possibly negative weight $a_{j} - a_{i}$.
Nonetheless, the class of single linear inequality constraints is closed under all four substitutions, so that the RQAOA never leaves it.
\begin{proposition}\label{proposition:LinearInequalityReduction}
    Let $n \in \N$, $\tau \in \Z$, and $\bm{a} \in \Z^{n}$.
    For every $i \in [n]$ and every $j \in [n] \setminus \{i\}$, it holds that
    \begin{align*}
        t_{\tau, \bm{a}}^{(n)}\hspace*{-3pt}\restriction_{x_{i} = 0} &= t_{\tau, \bm{a}_{(i)}}^{(n - 1)}, \\
        t_{\tau, \bm{a}}^{(n)}\hspace*{-3pt}\restriction_{x_{i} = 1} &= t_{\tau - a_{i}, \bm{a}_{(i)}}^{(n - 1)}, \\
        t_{\tau, \bm{a}}^{(n)}\hspace*{-3pt}\restriction_{x_{i} = x_{j}} &= t_{\tau, \bm{a}[j \leftarrow\hspace*{-5pt}+ a_{i}]_{(i)}}^{(n - 1)}, \\
        t_{\tau, \bm{a}}^{(n)}\hspace*{-3pt}\restriction_{x_{i} = \neg x_{j}} &= t_{\tau - a_{i}, \bm{a}[j \leftarrow\hspace*{-5pt}+ - a_{i}]_{(i)}}^{(n - 1)}.
    \end{align*}
\end{proposition}

\begin{proof}
    Substituting $x_{i} = 0$ removes the term $a_{i} x_{i}$ from the left-hand side of the inequality, while substituting $x_{i} = 1$ turns it into the constant $a_{i}$, which is moved to the right-hand side.
    Substituting $x_{i} = x_{j}$ merges $a_{i} x_{i} + a_{j} x_{j}$ into $(a_{j} + a_{i}) x_{j}$.
    Lastly, substituting $x_{i} = \neg x_{j}$ turns $a_{i} x_{i} + a_{j} x_{j}$ into
    \begin{equation*}
        a_{i} (1 - x_{j}) + a_{j} x_{j} = a_{i} + (a_{j} - a_{i}) x_{j},
    \end{equation*}
    and the constant $a_{i}$ is again moved to the right-hand side.
    Re-indexing the remaining variables yields the four identities.
\end{proof}

\autoref{proposition:LinearInequalityReduction} also illustrates why statement iv) of \autoref{proposition:UnateReduction} fails to apply to the knapsack constraint:
If $a_{i} > a_{j} \geq 0$, the substitution $x_{i} = \neg x_{j}$ renders the coefficient $a_{j} - a_{i}$ of $x_{j}$ negative, so that the subproblem is no longer a knapsack problem and its feasibility predicate becomes positive rather than negative unate in $x_{j}$.
That is, the subproblem stays unate, just not in the same direction anymore, which again shows that the conditions in iii) and iv) of \autoref{proposition:UnateReduction} are sufficient but not necessary.
Accordingly, the unsafe values of the subproblem have to be re-determined from the signs of the updated coefficients $\bm{a}'$ via $b_{k} = 1$ if $a'_{k} > 0$ and $b_{k} = 0$ if $a'_{k} < 0$.
The control pattern $\Lambda_{t\restriction_{x_{k} = b_{k}}}$ of the $k$th partial mixer then checks whether $\bm{a}' \cdot \bm{x}[k \leftarrow b_{k}] \leq \tau'$ holds, i.e., for the knapsack constraint, whether item $k$ can be added without exceeding the capacity.
It may be evaluated coherently, e.g., by accumulating the weighted sum into an ancilla register and comparing it to the threshold.
Hence, compiling the subproblem mixers merely amounts to updating $(\tau, \bm{a})$ according to \autoref{proposition:LinearInequalityReduction} and the unsafe values $\bm{b}$ accordingly.
In particular, \autoref{algorithm:UnateRQAOA} can be used for single linear inequality constraints with the restriction to $\mathcal{A}$ dropped altogether, provided the unsafe values are re-determined from the updated coefficients after every substitution.
The algorithm's deletion step simply discards all items that do not fit into the residual capacity on their own.

For the ordinary 0--1 knapsack problem, this mixer layout has essentially appeared before, albeit not as a variational ansatz.
The quantum tree generator~(QTG)~\cite{Wilkening2025AQuantumAlgorithmForSolving01KnapsackProblems} prepares a superposition of all feasible solutions by processing the items one after another, applying a single-qubit rotation to the $m$th qubit, controlled on whether item $m$ still fits into the knapsack.
Applied to its initial state $\ket{\bm{0}} = \ket{\neg \bm{b}}$, the QTG thus coincides with a single application of the partial mixers $U_{\texttt{CX}, m}$, $m = 1, \ldots, n$, with an individual rotation angle per qubit, up to diagonal phases and the ancilla registers implementing the controls.
However, the QTG's angles are not optimised variationally, but fixed by a bias hyperparameter and a reference solution, since the QTG serves as a state preparation routine, both for amplitude amplification~\cite{Wilkening2025AQuantumAlgorithmForSolving01KnapsackProblems} and for Grover-mixer QAOA~\cite{Christiansen2025QuantumTreeGeneratorImprovesQaoaStateOfTheArtForTheKnapsackProblem}.
The unate mixer~\eqref{equation:UnateMixer} may be regarded as a variational and repeatable generalisation of the QTG, which, by \autoref{proposition:LinearInequalityReduction}, extends to all subproblems arising in the RQAOA.
We make this correspondence precise in \autoref{appendix:QuantumTreeGenerator}.
Conversely, upon exchanging variational parameters with fixed hyperparameters, the unate mixer design may serve as a blueprint for quantum tree generators for all COPs with unate feasibility predicate.

Lastly, note that the closure property of \autoref{proposition:LinearInequalityReduction} does not extend to conjunctions of several linear inequality constraints, such as the independence constraints $x_{i} + x_{j} \leq 1$, $\{i, j\} \in E$, of the MIS.
While each individual constraint stays a single linear inequality, the substitution $x_{i} = \neg x_{j}$ may render one of them positive unate in $x_{j}$ while another remains negative unate in $x_{j}$, as exemplified for the MIS in \autoref{subsection:UnateFeasibilityPredicates}.
Within the unate framework, such conjunctions thus have to be treated with the admissibility restriction of \autoref{algorithm:UnateRQAOA}.
Viewed as constraints on linear forms, in contrast, they are closed under all four substitutions, and we derive suitable mixers for them in \autoref{subsection:LinearFormConstraints}.

\subsection{\label{subsection:HammingWeightConstraints}Hamming weight constraints}

One of the most prominent examples of feasibility-preserving mixer constructions are the \texttt{XY} mixers introduced in~\cite{Hadfield2019FromTheQuantumApproximateOptimizationAlgorithmToAQuantumAlternatingOperatorAnsatz}.
They were designed for mixing through computational basis states of constant-Hamming weight, occurring during the one-hot encoding of $\kappa$-ary variables, where $\kappa > 2$.
Here, we will directly discuss COPs with constant-Hamming weight constraints without tracing them back to COPs with $\kappa$-ary variables, such as Max-$\kappa$-ColourableGraph.
In addition to being conceptually leaner, it isolates the basic building block of problems with several Hamming weight constraints and, more generally, of constraints on several linear forms, which we address in \autoref{subsection:LinearFormConstraints}.
Note, however, that this does not cover every such problem efficiently.
For instance, the prominent Travelling Salesperson Problem (TSP) is (in the quantum optimisation literature) commonly formulated using $2 n$ overlapping Hamming weight constraints on $n^{2}$ variables~\cite{Lucas2014IsingFormulationsOfManyNpProblems}; the number of constraints thus grows with the problem size, so that the efficiency guarantees derived below do not apply, and we leave the TSP to future work.
Furthermore, besides the \texttt{XY} mixers, there are also \texttt{SWAP} mixers addressing the very same constraint family~\cite{Binkowski2025SymmetryBasedQuantumAlgorithmsForOpenShopSchedulingWithHardConstraints}.
In the following, we will extend both mixers to valid RQAOA mixer families for COPs with a constant-Hamming weight constraint.

For given $n \in \N$ and $d \in \Z$, let $h_{d}^{(n)} : \bit^{n} \rightarrow \bit$ be the Boolean function defined by $h_{d}^{(n)}(\bm{x}) = 1$ if and only if $\abs{\bm{x}} = d$.
Note that it is $h_{d}^{(n)} \notequiv 0$ if and only if $0 \leq d \leq n$.
We will first study the effect that each of the four RQAOA substitutions has on the constraint structure.
If we assign $0$ to some variable $x_{i}$, $i \in [n]$, the remaining $n - 1$ bits still have to obey the exact same Hamming weight constraint, i.e.,
\begin{equation*}
    h_{d}^{(n)}\hspace*{-3pt}\restriction_{x_{i} = 0} = h_{d}^{(n - 1)}.
\end{equation*}
Instead, assigning $1$ to $x_{i}$ decrements the Hamming weight the remaining variables have to achieve, i.e.,
\begin{equation*}
    h_{d}^{(n)}\hspace*{-3pt}\restriction_{x_{i} = 1} = h_{d - 1}^{(n - 1)}.
\end{equation*}
If we only allowed these two substitutions, we would remain within the same problem class and could therefore employ the same mixer concepts for the root problem and all derived subproblems.
However, the binary substitutions $x_{i} = (\neg) x_{j}$, $j \in [n] \setminus \{i\}$, are not directly compatible with a plain Hamming weight constraint: for example, enforcing $x_{i} = \neg x_{j}$, the value of $x_{j}$ becomes irrelevant for the resulting weight constraint, since either $x_{i}$ (if $x_{j}$ is not set) or $x_{j}$ contributes $1$ to the total Hamming weight.
That is, the remaining $n - 2$ variables $x_{k}$, $k \in [n] \setminus \{i, j\}$, have to attain a Hamming weight of $d - 1$, while $x_{j}$ may be chosen arbitrarily.
If we enforce instead $x_{i} = x_{j}$, then $x_{j}$ contributes twice to the total Hamming weight after the variable reduction.
Via a further reduction involving $x_{j}$, these effects can spread to produce even more convoluted Hamming weight constraints.
For example, the chain of substitutions ``$x_{i} = \neg x_{j}$ and $x_{k} = \neg x_{j}$'' leads to $x_{j}$ contributing $1$ to the total Hamming weight when set, and---via $x_{i}$ and $x_{k}$---$2$ when unset.
The most elegant recourse is to view and embed constant-Hamming weight constraints as special cases of general 0--1 linear equality constraints.

For given $n \in \N$, $d \in \Z$, and weight vector $\bm{w} \in \Z^{n}$, let $h_{d, \bm{w}}^{(n)} : \bit^{n} \rightarrow \bit$ be the Boolean function defined by
\begin{equation*}
    h_{d, \bm{w}}^{(n)}(\bm{x}) = 1 \text{ if and only if } \sum_{i = 1}^{n} w_{i} x_{i} = d.
\end{equation*}
Then, $h_{d}^{(n)} = h_{d, \bm{1}}^{(n)}$ and the effect of the unary substitution $x_{i} = 0$ readily generalises to
\begin{equation*}
    h_{d, \bm{w}}^{(n)}\hspace*{-3pt}\restriction_{x_{i} = 0} = h_{d, \bm{w}_{(i)}}^{(n - 1)},
\end{equation*}
where $\bm{w}_{(i)} \in \Z^{n - 1}$ denotes $\bm{w}$ with its $i$th entry removed.
The substitution $x_{i} = 1$ requires us to subtract $w_{i}$ (rather than just $1$) from the right-hand side of the equality constraint, i.e.,
\begin{equation*}
    h_{d, \bm{w}}^{(n)}\hspace*{-3pt}\restriction_{x_{i} = 1} = h_{d - w_{i}, \bm{w}_{(i)}}^{(n - 1)}.
\end{equation*}
The generalised constraint structure now allows us to handle binary substitutions:
enforcing $x_{i} = x_{j}$ means to merge the weights $w_{i}$ and $w_{j}$ into a cumulative weight for $x_{j}$, since setting $x_{j}$ to $1$ now contributes also the weight associated with $x_{i}$.
Using the notation $\bm{w}[j \leftarrow\hspace*{-10pt}+ w_{i}]$ for adding $w_{i}$ to the $j$th entry, we obtain
\begin{equation*}
    h_{d, \bm{w}}^{(n)}\hspace*{-3pt}\restriction_{x_{i} = x_{j}} = h_{d, \bm{w}[j \leftarrow\hspace*{-5pt}+ w_{i}]_{(i)}}^{(n - 1)}.
\end{equation*}
Lastly, the substitution $x_{i} = \neg x_{j}$ entails that if $x_{j}$ is set to $0$, it contributes the weight $w_{i}$, while contributing its original weight $w_{j}$ if set to $1$.
Equivalently, we may subtract $w_{i}$ from both sides of the linear equality constraint, cancelling $x_{j}$'s contribution when it is set to $0$, reducing its contribution when set to $1$ to $w_{j} - w_{i}$, and leaving the right-hand side as $d - w_{i}$.
Therefore, we have just argued that
\begin{equation*}
    h_{d, \bm{w}}^{(n)}\hspace*{-3pt}\restriction_{x_{i} = \neg x_{j}} = h_{d - w_{i}, \bm{w}[j \leftarrow\hspace*{-5pt}+ - w_{i}]_{(i)}}^{(n - 1)}.
\end{equation*}

In summary, we have thus successfully identified a broader class of COP constraints---general 0--1 linear equality constraints---which is invariant under the four RQAOA-driven substitutions.
Note that these update rules coincide exactly with those for single linear inequality constraints in \autoref{proposition:LinearInequalityReduction}.
The next step is to formulate mixer families for this broader class together with efficient compilation rules to obtain the mixer for a derived subproblem from the root problem's mixer.
For inspiration, we start with Hadfield~\textit{et~al.}'s constructions for the ordinary constant-Hamming weight case.
Here, feasibility-preserving operations are precisely those which do not change the number of ones in the computational basis states, i.e., merely permute bit values.
The predestined operation for these bit swaps is the swap gate
\begin{align*}
    \texttt{S} &= \ket{00}\hspace*{-3pt}\bra{00} + \ket{01}\hspace*{-3pt}\bra{10} + \ket{10}\hspace*{-3pt}\bra{01} + \ket{11}\hspace*{-3pt}\bra{11} \\
    &= (\one \one + \texttt{X} \texttt{X} + \texttt{Y} \texttt{Y} + \texttt{Z} \texttt{Z}) / 2.
\end{align*}
The swap gate is not only unitary, but also Hermitian so that exponentiating it gives a valid swap mixer
\begin{equation*}
    U_{\texttt{S}}(\beta) \coloneqq e^{-i \beta \texttt{S}} = e^{-i \beta / 2} e^{-i \beta \texttt{X} \texttt{X} / 2} e^{-i \beta \texttt{Y} \texttt{Y} / 2} e^{-i \beta \texttt{Z} \texttt{Z} / 2},
\end{equation*}
where we have used that all the two-qubit Pauli strings $\one \one$, $\texttt{X} \texttt{X}$, $\texttt{Y} \texttt{Y}$, and $\texttt{Z} \texttt{Z}$ mutually commute.
The exponentiated identity only contributes a global phase, and omitting it, the diagonal \texttt{Z}\texttt{Z} term, and the normalisation factor leads to the \texttt{XY} mixer
\begin{equation*}
    U_{\texttt{s}}(\beta) \coloneqq e^{-i \beta \texttt{X} \texttt{X}} e^{-i \beta \texttt{Y} \texttt{Y}},
\end{equation*}
which simply acts as the identity on $\spn\{\ket{00}, \ket{11}\}$, and is the exponentiated Hamiltonian
\begin{equation*}
    \texttt{s} \coloneqq 2 (\ket{01}\hspace*{-3pt}\bra{10} + \ket{10}\hspace*{-3pt}\bra{01}) = \texttt{X} \texttt{X} + \texttt{Y} \texttt{Y}.
\end{equation*}
For the constant-Hamming weight base case, chaining together individual swap or \texttt{XY} mixers along the qubit indices produces a feasibility-preserving parameterised circuit which provides transitions between all pairs of feasible computational basis states~\cite{Hadfield2019FromTheQuantumApproximateOptimizationAlgorithmToAQuantumAlternatingOperatorAnsatz}.

For general linear equality constraints, swapping bit values is not always a feasibility-preserving operation or might fail to connect all feasible solutions.
For example, if $\bm{w} = (1, 1, 2, 1)$ and $d = 2$, the feasible solutions are $0010$, $0101$, $1001$, and $1100$.
The bit swap of, e.g., the two leftmost positions is feasibility-preserving, while exchanging $1 \leftrightarrow 3$ maps all feasible solutions except $0101$ to infeasible bitstrings.
In particular, there is no simple bit swap connecting $0010$ to the other feasible bitstrings; the operation connecting $0010$ to $1100$ is an unbalanced exchange of $1$s in the index subset $\{1, 2\}$ with the $1$ in the index singleton $\{3\}$.
However, such an exchange may only fire on bitstrings which are uniform among both targeted index subsets.
For example, the feasible bitstring $\bm{x} = 1001$ is not uniform among the index subset $\{1, 2\}$, so that there is no meaningful way of exchanging the $1$s between subsets $\{1, 2\}$ and $\{3\}$; hence this operation must act as the identity on $\bm{x}$.

Let us now formalise the general construct and argue its feasibility preservation.
The generalisation of the \texttt{XY} Hamiltonian for exchanging two uniform blocks of sizes $k$ and $\ell$ reads
\begin{align}\label{equation:GeneralisedXYHamiltonian}
    \begin{split}
        \texttt{s}^{\{k, \ell\}} &\coloneqq 2^{k + \ell - 1} \big(\ket{\bm{0}^{k} \bm{1}^{\ell}}\hspace*{-3pt}\bra{\bm{1}^{k} \bm{0}^{\ell}} + \ket{\bm{1}^{k} \bm{0}^{\ell}}\hspace*{-3pt}\bra{\bm{0}^{k} \bm{1}^{\ell}}\big) \\
        &= \sum_{\substack{A \subseteq [k + \ell] \\ \abs{A} \text{ even}}} (-1)^{\abs{A} / 2 + \abs{A \cap [k]}} \bm{P}_{A},
    \end{split}
\end{align}
where
\begin{equation*}
    \bm{P}_{A} \coloneqq \bigotimes_{j = 1}^{k + \ell} P_{j}, \qquad P_{j} = \begin{cases} \texttt{Y}, & j \in A \\
        \texttt{X}, & j \notin A \\
    \end{cases}.
\end{equation*}
For $k = \ell = 1$,~\eqref{equation:GeneralisedXYHamiltonian} reproduces the \texttt{XY} Hamiltonian, $\texttt{s}^{\{1, 1\}} = \texttt{s}$, since the only even-sized subsets of $[2]$ are $\emptyset$ and $[2]$, contributing $\texttt{X} \texttt{X}$ and $(-1)^{1 + 1} \texttt{Y} \texttt{Y}$, respectively.
For $(k, \ell) = (1, 0)$, it reduces to a plain bit flip, $\texttt{s}^{\{1, 0\}} = \ket{0}\hspace*{-3pt}\bra{1} + \ket{1}\hspace*{-3pt}\bra{0} = \texttt{X}$.
In general, the Pauli expansion follows from $\ket{0}\hspace*{-3pt}\bra{1} = (\texttt{X} + i \texttt{Y}) / 2$ and $\ket{1}\hspace*{-3pt}\bra{0} = (\texttt{X} - i \texttt{Y}) / 2$.
Writing $a_{A} \coloneqq \abs{A \cap [k]}$ and $b_{A} \coloneqq \abs{A \setminus [k]}$ for $A \subseteq [k + \ell]$ and expanding the tensor products yields
\begin{equation*}
    \ket{\bm{0}^{k} \bm{1}^{\ell}}\hspace*{-3pt}\bra{\bm{1}^{k} \bm{0}^{\ell}} = 2^{-(k + \ell)} \sum_{A \subseteq [k + \ell]} i^{a_{A} - b_{A}} \bm{P}_{A},
\end{equation*}
while its Hermitian conjugate $\ket{\bm{1}^{k} \bm{0}^{\ell}}\hspace*{-3pt}\bra{\bm{0}^{k} \bm{1}^{\ell}}$ contributes the complex conjugate coefficients, since all $\bm{P}_{A}$ are Hermitian.
Hence, $\bm{P}_{A}$ appears in $\texttt{s}^{\{k, \ell\}}$ with coefficient $2^{k + \ell - 1} \cdot 2^{-(k + \ell)} \cdot 2 \operatorname{Re}(i^{a_{A} - b_{A}})$.
It vanishes whenever $\abs{A} = a_{A} + b_{A}$ is odd, and otherwise equals $(-1)^{(a_{A} - b_{A}) / 2} = (-1)^{\abs{A} / 2 + a_{A}}$, where we have used that $(a_{A} - b_{A}) / 2 = \abs{A} / 2 - b_{A}$ and $a_{A} \equiv b_{A} \pmod{2}$.
In particular, the prefactor $2^{k + \ell - 1}$ ensures that all non-vanishing Pauli coefficients are $\pm 1$, as for the \texttt{XY} Hamiltonian.

Furthermore, $\texttt{s}^{\{k, \ell\}}$ is symmetric in $k$ and $\ell$:
Moving the first $k$ qubits behind the last $\ell$ ones maps the pair $\{\ket{\bm{0}^{k} \bm{1}^{\ell}}, \ket{\bm{1}^{k} \bm{0}^{\ell}}\}$ onto $\{\ket{\bm{1}^{\ell} \bm{0}^{k}}, \ket{\bm{0}^{\ell} \bm{1}^{k}}\}$, i.e., onto the pair defining $\texttt{s}^{\{\ell, k\}}$, and~\eqref{equation:GeneralisedXYHamiltonian} is symmetric in the two basis states.
On the level of the Pauli expansion, this is reflected by $(-1)^{a_{A}} = (-1)^{b_{A}}$ for all even-sized $A$.
This justifies the set notation $\{k, \ell\}$.

\begin{figure*}[t]
    \centering
    \begin{quantikz}[row sep={0.75cm,between origins}, column sep=0.3cm, align equals at=3]
        \lstick{$k_{1}$} & \gate[5]{e^{-i \beta \texttt{s}_{K}^{L}}} & \\
        \lstick{$k_{2}$} & & \\
        \lstick{$\ell_{1}$} & & \\
        \lstick{$\ell_{2}$} & & \\
        \lstick{$\ell_{3}$} & &
    \end{quantikz}\hspace*{0.1cm}$=$\hspace*{0.1cm}
    \begin{quantikz}[row sep={0.75cm,between origins}, column sep=0.3cm, align equals at=3]
        & \ctrl{4}\gategroup[5,steps=1,style={dashed,rounded corners,inner xsep=2pt},label style={label position=below,anchor=north,yshift=-0.2cm}]{$C$} & & \gate{e^{-i 16 \beta \texttt{X}}}\gategroup[5,steps=1,style={dashed,rounded corners,inner xsep=2pt},label style={label position=below,anchor=north,yshift=-0.2cm}]{$\Lambda$} & & \ctrl{4}\gategroup[5,steps=1,style={dashed,rounded corners,inner xsep=2pt},label style={label position=below,anchor=north,yshift=-0.2cm}]{$C$} & \\
        & \targ{} & & \ocontrol{} & & \targ{} & \\
        & \targ{} & & \control{} & & \targ{} & \\
        & \targ{} & & \control{} & & \targ{} & \\
        & \targ{} & & \ctrl{-4} & & \targ{} &
    \end{quantikz}
    \hspace*{0.8cm}
    \begin{quantikz}[row sep={0.75cm,between origins}, column sep=0.3cm, align equals at=3]
        \lstick{$k$} & \gate[5]{e^{-i \beta \texttt{s}_{K'}^{L'}}} & \\
        \lstick{$\ell_{1}$} & & \\
        \lstick{$\ell_{2}$} & & \\
        \lstick{$\ell_{3}$} & & \\
        \lstick{$\ell_{4}$} & &
    \end{quantikz}\hspace*{0.1cm}$=$\hspace*{0.1cm}
    \begin{quantikz}[row sep={0.75cm,between origins}, column sep=0.3cm, align equals at=3]
        & \ctrl{4}\gategroup[5,steps=1,style={dashed,rounded corners,inner xsep=2pt},label style={label position=below,anchor=north,yshift=-0.2cm}]{$C$} & & \gate{e^{-i 16 \beta \texttt{X}}}\gategroup[5,steps=1,style={dashed,rounded corners,inner xsep=2pt},label style={label position=below,anchor=north,yshift=-0.2cm}]{$\Lambda'$} & & \ctrl{4}\gategroup[5,steps=1,style={dashed,rounded corners,inner xsep=2pt},label style={label position=below,anchor=north,yshift=-0.2cm}]{$C$} & \\
        & \targ{} & & \control{} & & \targ{} & \\
        & \targ{} & & \control{} & & \targ{} & \\
        & \targ{} & & \control{} & & \targ{} & \\
        & \targ{} & & \ctrl{-4} & & \targ{} &
    \end{quantikz}
    \caption{\label{figure:GeneralisedXYDecomposition}
        Decomposition of exponentiated generalised \texttt{XY} Hamiltonians via \texttt{CNOT}-cascades and a single multi-controlled rotation.
        Left: $\texttt{s}_{K}^{L}$ for $K = \{k_{1}, k_{2}\}$ and $L = \{\ell_{1}, \ell_{2}, \ell_{3}\}$, i.e., $(k, \ell) = (2, 3)$.
        The \texttt{CNOT}-cascade $C$, controlled on qubit $k_{1}$, maps the two exchanged basis states $\ket{\bm{0}\bm{0}\bm{1}\bm{1}\bm{1}}$ and $\ket{\bm{1}\bm{1}\bm{0}\bm{0}\bm{0}}$ to $\ket{b}\ket{\bm{0}\bm{1}\bm{1}\bm{1}}$ with $b = 0$ and $b = 1$, respectively, and every other basis state to one whose last four qubits differ from $\ket{\bm{0}\bm{1}\bm{1}\bm{1}}$.
        Hence, the rotation $\Lambda$ applying the $\texttt{X}$-rotation by an angle $2^{k + \ell - 1} \beta = 16 \beta$ to qubit $k_{1}$, controlled negatively on $k_{2}$ (open circle) and positively on $\ell_{1}, \ell_{2}, \ell_{3}$ (filled circles), only acts on these two states, and conjugation with $C$ restores the original basis, so that the circuit acts non-trivially only on $\spn\{\ket{\bm{0}\bm{0}\bm{1}\bm{1}\bm{1}}, \ket{\bm{1}\bm{1}\bm{0}\bm{0}\bm{0}}\}$.
        In contrast, the Pauli factorisation of the left-hand side consists of $2^{k + \ell - 1} = 16$ commuting exponentiated Pauli strings of weight $5$.
        Right: $\texttt{s}_{K'}^{L'}$ for $K' = \{k\}$ and $L' = \{\ell_{1}, \ell_{2}, \ell_{3}, \ell_{4}\}$, i.e., $(k, \ell) = (1, 4)$.
        Here, the cascade $C$ maps the exchanged states $\ket{\bm{1}\bm{0}\bm{0}\bm{0}\bm{0}}$ and $\ket{\bm{0}\bm{1}\bm{1}\bm{1}\bm{1}}$ to $\ket{b}\ket{\bm{1}\bm{1}\bm{1}\bm{1}}$ with $b = 1$ and $b = 0$, respectively, so that $\Lambda'$ applies $e^{-i 16 \beta \texttt{X}}$ to $k$, controlled positively on all of $\ell_{1}, \ldots, \ell_{4}$.
    }
\end{figure*}
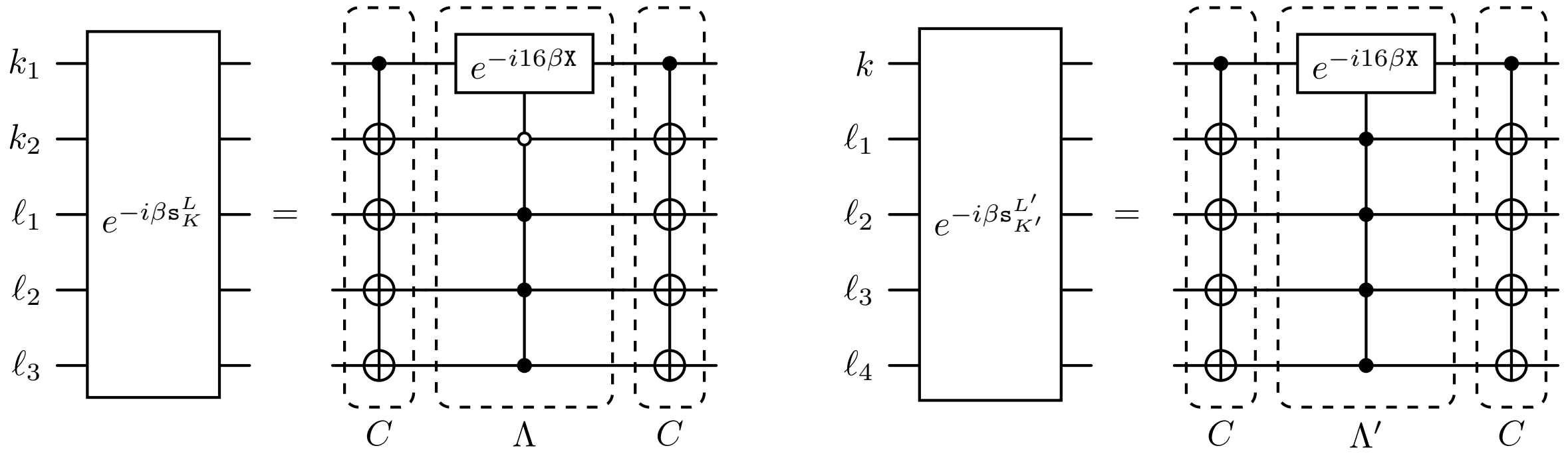

Lastly, any two Pauli strings $\bm{P}_{A}$ and $\bm{P}_{B}$ with $\abs{A}$ and $\abs{B}$ even differ in exactly
\begin{equation*}
    \abs{A \triangle B} = \abs{A} + \abs{B} - 2 \abs{A \cap B}
\end{equation*}
positions, which is an even number.
Since each of these positions contributes an anti-commuting pair of \texttt{X} and \texttt{Y}, the two Pauli strings commute.
Hence, $e^{-i \beta \texttt{s}^{\{k, \ell\}}}$ factorises into $2^{k + \ell - 1}$ exponentiated Pauli strings (for $k + \ell \geq 1$), generalising the factorisation of the \texttt{XY} mixer $U_{\texttt{s}}(\beta)$.
Since the number of factors grows exponentially in the block size, it is often preferable to exploit that $\texttt{s}^{\{k, \ell\}}$ acts non-trivially only on the two-dimensional subspace $\spn\{\ket{\bm{0}^{k} \bm{1}^{\ell}}, \ket{\bm{1}^{k} \bm{0}^{\ell}}\}$, on which it equals $2^{k + \ell - 1} \texttt{X}$:
For $k \geq 1$, a cascade $C$ of $k + \ell - 1$ \texttt{CNOT}-gates, controlled on the first qubit and targeting each of the other qubits, maps these two basis states to $\ket{0} \ket{\bm{0}^{k - 1} \bm{1}^{\ell}}$ and $\ket{1} \ket{\bm{0}^{k - 1} \bm{1}^{\ell}}$, respectively.
Therefore,
\begin{equation*}
    e^{-i \beta \texttt{s}^{\{k, \ell\}}} = C\, \Lambda\big(e^{-i \beta 2^{k + \ell - 1} \texttt{X}_{1}}\big)\, C,
\end{equation*}
where $\Lambda$ controls on the remaining $k + \ell - 1$ qubits being in state $\ket{\bm{0}^{k - 1} \bm{1}^{\ell}}$.
Such a multi-controlled single-qubit rotation can be decomposed into $O(k + \ell)$ elementary gates using ancilla qubits~\cite[Section~4.3]{Nielsen2012QuantumComputationAndQuantumInformation}.
\autoref{figure:GeneralisedXYDecomposition} illustrates this decomposition for $k = 2$ and $\ell = 3$, as well as for $k = 1$ and $\ell = 4$.

For two disjoint index subsets $K, L \subseteq [n]$, we denote by $\texttt{s}_{K}^{L}$ the embedding of $\texttt{s}^{\{\abs{K}, \abs{L}\}}$ into $\lo(\qubit^{\otimes n})$, implementing the exchange of the uniform blocks formed by the index subsets $K$ and $L$.
By the symmetry argued above, it holds that $\texttt{s}_{K}^{L} = \texttt{s}_{L}^{K}$.
As a special case, we also allow one of the index subsets, say $L$, to be empty, and introduce the uniform bit flip $\texttt{s}_{K} \coloneqq \texttt{s}_{K}^{\emptyset}$, where the embedded $\texttt{s}^{\{\abs{K}, 0\}}$ is also defined via~\eqref{equation:GeneralisedXYHamiltonian}.
As the next result shows, the block structure of the just introduced operations is directly compatible with the feasibility structure of linear equality constraints.
Its derivation compares the restrictions $\bm{x}_{K}$ and $\bm{x}_{L}$ of feasible bitstrings $\bm{x} \in \bit^{n}$ to the two exchanged blocks.

\begin{lemma}\label{lemma:FeasibilityPreservingGeneralisedXY}
    Let $\bm{w} \in \Z^{n}$, $d \in \Z$, and $(n, f, h_{d, \bm{w}}^{(n)})$ be a COP.
    Then, for every $K, L \subseteq [n]$ with $K \cap L = \emptyset$, $\texttt{{\upshape s}}_{K}^{L}$ is feasibility-preserving if $\sum_{k \scriptin K} w_{k} = \sum_{\ell \scriptin L} w_{\ell}$, where we define $\sum_{\ell \scriptin \emptyset} w_{\ell} \coloneqq 0$.
\end{lemma}

\begin{proof}
    Let $\bm{x} \in \bit^{n}$ be any feasible solution and let $K, L \subseteq [n]$, $K \cap L = \emptyset$, so that $\sum_{k \scriptin K} w_{k} = \sum_{\ell \scriptin L} w_{\ell}$.
    If $(\bm{x}_{K}, \bm{x}_{L}) \notin \{(\bm{1}, \bm{0}), (\bm{0}, \bm{1})\}$, then $\texttt{s}_{K}^{L} \ket{\bm{x}} = 0$, which trivially lies in the feasible subspace.
    Otherwise, consider the case that $\bm{x}_{K} = \bm{1}$ and $\bm{x}_{L} = \bm{0}$ and write $\tilde{\bm{x}}$ so that $2^{\abs{K} + \abs{L} - 1} \ket{\tilde{\bm{x}}} = \texttt{s}_{K}^{L} \ket{\bm{x}}$, i.e., it is $\tilde{\bm{x}}_{K} = \bm{0}$, $\tilde{\bm{x}}_{L} = \bm{1}$, and $\tilde{x}_{j} = x_{j}$ for all $j \in (K \cup L)^{c}$.
    Then:
    \begin{align*}
        d &= \bm{x} \cdot \bm{w} = \sum_{j = 1}^{n} x_{j} w_{j} = \sum_{k \scriptin K} w_{k} + \sum_{j \scriptin (K \cup L)^{c}} x_{j} w_{j} \\
        &= \sum_{\ell \scriptin L} w_{\ell} + \sum_{j \scriptin (K \cup L)^{c}} x_{j} w_{j} = \tilde{\bm{x}} \cdot \bm{w}.
    \end{align*}
    Hence, the resulting bitstring $\tilde{\bm{x}}$ is again feasible, so that $\texttt{s}_{K}^{L}$ maps to the feasible subspace.
    Proving preservation of feasibility for the case that $\bm{x}_{K} = \bm{0}$ and $\bm{x}_{L} = \bm{1}$ is completely analogous.
    In summary, $\texttt{s}_{K}^{L}$ maps each feasible computational basis state again to a feasible computational basis state, and general feasibility preservation follows by linearity.
\end{proof}

Next, we show that for every linear equality constraint there is always a collection of generalised \texttt{XY} mixers which provides transitions between all pairs of feasible computational basis states.
This fact is grounded in a deeper framework, combining aspects of graph theory and linear algebra, which will also prove indispensable for further considerations.
Accordingly, we first establish the abstract background and derive important results, such as the existence of a transition-providing mixer family and criteria for an efficient fabrication, alongside the deeper theory.

\begin{definition}\label{definition:FibreGraph}
    Let $\bm{w} \in \Z^{n}$, $d \in \Z$, $(n, f, h_{d, \bm{w}}^{(n)})$ be a COP, and let
    \begin{equation*}
        \mathcal{E} \coloneqq\hspace*{-1pt} \left\{\hspace*{-2pt}(K, L) : K, L \subseteq [n], K \cap L = \emptyset,\hspace*{-1pt} \sum_{k \scriptin K}\hspace*{-2pt} w_{k}\hspace*{-1pt} =\hspace*{-1pt} \sum_{\ell \scriptin L}\hspace*{-2pt} w_{\ell}\hspace*{-2pt}\right\}
    \end{equation*}
    be the collection of all pairs of index subsets, constituting feasibility-preserving operations.
    The COP's \emph{fibre graph} $F_{\mathcal{E}'} = (V, E)$, given some $\mathcal{E}' \subseteq \mathcal{E}$, is defined via
    \begin{equation*}
        V \coloneqq \{\bm{x} \in \bit^{n} \defcolon h_{d, \bm{w}}^{(n)}(\bm{x}) = 1\}
    \end{equation*}
    and
    \begin{equation*}
        E \coloneqq \bigcup_{(K, L) \scriptin \mathcal{E}'} \{\{\bm{x}, \bm{y}\} \defcolon \texttt{s}_{K}^{L} \ket{\bm{x}} = 2^{\abs{K} + \abs{L} - 1} \ket{\bm{y}}\}.
    \end{equation*}
\end{definition}

The fibre graph's vertex set consists of feasible solutions to the underlying COP.
A feasibility-preserving operation $\texttt{s}_{K}^{L}$ can constitute one or more edges, each connecting two feasible solutions the operation maps to each other (modulo prefactor).
Since the underlying bit exchange is involutive, the edges (and thus the graph) are undirected.
Note that, for every $\bm{x} \in \bit^{n}$, $\texttt{s}_{K}^{L} \ket{\bm{x}}$ will have non-zero overlap with at most one computational basis state, i.e., $\texttt{s}_{K}^{L}$ constitutes at most one edge per vertex.

The most important connection between the fibre graph's properties and QAOA mixer properties is captured in the following result.

\begin{proposition}\label{proposition:FibreGraphConnectivity}
    Let $\bm{w} \in \Z^{n}$, $d \in \Z$, $(n, f, h_{d, \bm{w}}^{(n)})$ be a COP, and $\mathcal{E}' \subseteq \mathcal{E}$.
    The associated fibre graph $F_{\mathcal{E}'}$ is connected if and only if the parameterised mixer
    \begin{equation}\label{equation:GeneralisedXYParameterisedMixer}
        U_{\mathcal{E}'}(\beta) \coloneqq \prod_{(K, L) \scriptin \mathcal{E}'} e^{-i \beta \texttt{{\upshape s}}_{K}^{L}}
    \end{equation}
    provides transitions between all pairs of feasible computational basis states for any product ordering in~\eqref{equation:GeneralisedXYParameterisedMixer}.
\end{proposition}

\begin{proof}
    First, assume $F_{\mathcal{E}'}$ connected and let $\bm{x}, \bm{y} \in S$.
    Consider the primitive uniform block exchange $\texttt{s}^{\{k, \ell\}}$, as defined in~\eqref{equation:GeneralisedXYHamiltonian} and write $\ket{\psi} \coloneqq \ket{\bm{0}^{k} \bm{1}^{\ell}}$, as well as $\ket{\phi} \coloneqq \ket{\bm{1}^{k} \bm{0}^{\ell}}$:
    Since $\braket{\phi \vert \psi} = 0$, it is
    \begin{align*}
        &\big(\ket{\psi}\hspace*{-3pt}\bra{\phi} + \ket{\phi}\hspace*{-3pt}\bra{\psi}\big)^{2} = \ket{\psi}\hspace*{-3pt}\bra{\psi} + \ket{\phi}\hspace*{-3pt}\bra{\phi} \text{ and} \\
        &\big(\ket{\psi}\hspace*{-3pt}\bra{\phi} + \ket{\phi}\hspace*{-3pt}\bra{\psi}\big)^{3} = \ket{\psi}\hspace*{-3pt}\bra{\phi} + \ket{\phi}\hspace*{-3pt}\bra{\psi}
    \end{align*}
    so that
    \begin{align*}
        e^{-i \beta \texttt{s}^{\{k, \ell\}}} &= \one + \big(\cos(2^{k + \ell - 1} \beta) - 1\big) \big(\ket{\psi}\hspace*{-3pt}\bra{\psi} + \ket{\phi}\hspace*{-3pt}\bra{\phi}\big) \\
        &\quad - i \sin(2^{k + \ell - 1} \beta) \texttt{s}^{\{k, \ell\}} / 2^{k + \ell - 1}.
    \end{align*}
    Since $\texttt{s}_{K}^{L}$ is merely the embedding of $\texttt{s}^{\{\abs{K}, \abs{L}\}}$ into $\lo(\qubit^{\otimes n})$, the same identity holds for every factor $e^{-i \beta \texttt{s}_{K}^{L}}$ of $U_{\mathcal{E}'}(\beta)$.
    Hence, every such factor is a trigonometric polynomial in $\beta$, and so is every overlap $\braket{\bm{y} \vert U_{\mathcal{E}'}^{t}(\beta) \vert \bm{x}}$, $t \in \N$.
    Moreover, every $\texttt{s}_{K}^{L}$ has non-negative entries with respect to the computational basis.
    If $\bm{x} = \bm{y}$, let $t \coloneqq 1$.
    Otherwise, let $t$ denote the distance of $\bm{x}$ and $\bm{y}$ in $F_{\mathcal{E}'}$, so that there are $(K_{1}, L_{1}), \ldots, (K_{t}, L_{t}) \in \mathcal{E}'$ and $\alpha > 0$ with
    \begin{equation*}
        \ket{\bm{y}} = \alpha\, \texttt{s}_{K_{t}}^{L_{t}} \cdots \texttt{s}_{K_{1}}^{L_{1}} \ket{\bm{x}}.
    \end{equation*}
    Since every application of the mixer $U_{\mathcal{E}'}(\beta)$ contains every move, the $r$th application may execute $(K_{r}, L_{r})$, so that some term in the power series expansion of $\braket{\bm{y} \vert U_{\mathcal{E}'}^{t}(\beta) \vert \bm{x}}$ does not vanish.
    Exactly as in the proof of \autoref{proposition:BitFlipMixersForUnateFeasibilityPredicates}, all terms of lowest order $D$ carry the same phase $(-i)^{D}$ and non-negative coefficients, so that the overlap does not vanish identically.
    Hence, it is non-zero for all but finitely many $\beta \in [0, 2 \pi)$, for any product ordering in~\eqref{equation:GeneralisedXYParameterisedMixer}.
    Conversely, if $F_{\mathcal{E}'}$ is not connected, there exists a pair $\bm{x}, \bm{y} \in S$, $\bm{x} \neq \bm{y}$, so that, for every sequence $(K_{1}, L_{1}), \ldots, (K_{t}, L_{t}) \in \mathcal{E}'$, it is $\braket{\bm{y} \vert \texttt{s}_{K_{t}}^{L_{t}} \cdots \texttt{s}_{K_{1}}^{L_{1} } \vert \bm{x}} = 0$.
    However, by the above closed form, the only term of any $\exp(-i \beta \texttt{s}_{K}^{L})$, $\beta \in \R$ arbitrary, that drives transitions between computational basis states is proportional to $\texttt{s}_{K}^{L}$, while the remaining terms are diagonal.
    Hence, since $\braket{\bm{y} \vert \bm{x}} = 0$, repeated applications of $U_{\mathcal{E}'}(\beta)$ with $\beta \in \R$ arbitrary remain in the orthogonal complement of $\ket{\bm{y}}$, i.e., $U_{\mathcal{E}'}(\beta)$ fails to provide transitions between $\ket{\bm{x}}$ and $\ket{\bm{y}}$.
\end{proof}

That is, a connected fibre graph always contains a description for a parameterised mixer, adhering to Hadfield~\textit{et~al.}'s two mixer requirements.
Therefore, showing that every COP with linear equality constraint has a connected fibre graph settles the existence of suitable QAOA mixers for the entire problem class.
The following lemma establishes exactly this fact in the graph-theoretical framework.

\begin{lemma}\label{lemma:ExistenceOfConnectedFibreGraph}
    Let $\bm{w} \in \Z^{n}$, $d \in \Z$, and $(n, f, h_{d, \bm{w}}^{(n)})$ be a COP.
    Then, the full fibre graph $F_{\mathcal{E}}$ is connected.
\end{lemma}

\begin{proof}
    If $\abs{S} \leq 1$, there is nothing to show.
    Otherwise, let $\bm{x}, \bm{y} \in S$, $\bm{x} \neq \bm{y}$, be arbitrary.
    Define
    \begin{equation*}
        K \coloneqq \{i \in [n] \defcolon x_{i} = 1\, \land\, y_{i} = 0\}
    \end{equation*}
    and
    \begin{equation*}
        L \coloneqq \{j \in [n] \defcolon x_{j} = 0\, \land\, y_{j} = 1\}.
    \end{equation*}
    Since $\bm{x} \neq \bm{y}$, at least one of these sets is non-empty.
    In any case, it is $K \cap L = \emptyset$.
    Most importantly,
    \begin{align*}
        \sum_{i \scriptin K} w_{i} - \sum_{j \scriptin L} w_{j} &= \sum_{i = 1}^{n} (x_{i} - y_{i}) w_{i} \\
        &= \sum_{i = 1}^{n} x_{i} w_{i} - \sum_{j = 1}^{n} y_{j} w_{j} = d - d = 0.
    \end{align*}
    Therefore, $(K, L) \in \mathcal{E}$.
    Additionally, by construction, $\texttt{s}_{K}^{L} \ket{\bm{x}} = 2^{\abs{K} + \abs{L} - 1} \ket{\bm{y}}$, so that $\{\bm{x}, \bm{y}\}$ is an edge in $F_{\mathcal{E}}$; in particular, $\bm{x}$ and $\bm{y}$ are connected in $F_{\mathcal{E}}$.
    Since $\bm{x}$ and $\bm{y}$ were chosen arbitrarily, this establishes that $F_{\mathcal{E}}$ is connected.
\end{proof}

Combining \autoref{lemma:FeasibilityPreservingGeneralisedXY}, \autoref{proposition:FibreGraphConnectivity}, and \autoref{lemma:ExistenceOfConnectedFibreGraph}, we infer the existence of admissible, i.e., feasibility-preserving and transition-providing, QAOA mixers for COPs with linear equality constraint, built from uniform block exchange operations.

\begin{corollary}\label{corollary:ExistenceOfQAOAMixer}
    Let $\bm{w} \in \Z^{n}$, $d \in \Z$, and $(n, f, h_{d, \bm{w}}^{(n)})$ be a COP.
    Then, there exists a feasibility-preserving parameterised mixer of the form~\eqref{equation:GeneralisedXYParameterisedMixer}, providing transitions between all pairs of feasible computational basis states.
\end{corollary}

We now turn to our main objective of formulating efficient mechanical rules for deriving (sub)problem mixers from the (sub)problem's feasibility structure.
While the mixer design~\eqref{equation:GeneralisedXYParameterisedMixer} is valid for all COPs with linear equality constraint, two things are not guaranteed in the general case: the existence of a polynomial-sized $\mathcal{E}'$ for which $F_{\mathcal{E}'}$ is connected, and---even where such an $\mathcal{E}'$ exists---an efficient routine for constructing it.
In \autoref{appendix:ProblematicInstances}, we discuss such problematic examples in detail.
Both guarantees are restored once the weights are controlled in magnitude.
For a linear equality constraint $h_{d, \bm{w}}^{(n)}$, we call $\sum_{i = 1}^{n} \abs{w_{i}}$ its \emph{weight extent}, and we restrict to the subclass of problems arising via repeated RQAOA-driven substitution from a root problem $h_{d, \bm{w}}^{(n)}$ of \emph{polynomially bounded} weight extent
\begin{equation*}
    W \coloneqq \sum_{i = 1}^{n} \abs{w_{i}} = O(\mathrm{poly}(n)).
\end{equation*}
This is perfectly equivalent to demanding polynomially bounded weights $\max_{i} \abs{w_{i}} = O(\mathrm{poly}(n))$, since evidently $\max_{i} \abs{w_{i}} \leq W \leq n \max_{i} \abs{w_{i}}$.
This especially covers the Hamming-weight root $\bm{w} = \bm{1} \in \Z^{n}$, where $W = n$.
The point of the extent is that it never grows under substitution: any of the four reductions only redistributes or eliminates portions of the cumulative weight.
More precisely, the following result holds true.

\begin{lemma}\label{lemma:SubstitutionWeightRestriction}
    Let $n \in \N$, $d \in \Z$, and $\bm{w} \in \Z^{n}$, and let $d' \in \Z$ and $\bm{w}' \in \Z^{n - 1}$ so that $h_{d, \bm{w}}^{(n)}\hspace*{-3pt}\restriction_{x_{i} = y} = h_{d', \bm{w}'}^{(n - 1)}$ for some $y \in \{0, 1, x_{j}, \neg x_{j}\}$, where $i, j \in [n]$, $i \neq j$.
    Then,
    \begin{equation}\label{equation:WeightInequality}
        \sum_{k = 1}^{n - 1} \abs{w'_{k}} \leq \sum_{\ell = 1}^{n} \abs{w_{\ell}}.
    \end{equation}
\end{lemma}

\begin{proof}
    For $y = 0 / 1$, we simply have $\bm{w}' = \bm{w}_{(i)}$ and $d' = d$ or $d' = d - w_{i}$, respectively, hence
    \begin{equation*}
        \sum_{k = 1}^{n - 1} \abs{w'_{k}} = \sum_{\substack{\ell = 1 \\ \ell \neq i}}^{n} \abs{w_{\ell}} \leq \sum_{\ell = 1}^{n} \abs{w_{\ell}}.
    \end{equation*}
    For $y = x_{j} / \neg x_{j}$, again $d' = d$ or $d' = d - w_{i}$, respectively, whereas the weight vector update reads $\bm{w}' = \bm{w}[j \raisebox{0.8pt}{$\leftarrow$}\hspace*{-8pt}\pm w_{i}]_{(i)}$, implying
    \begin{align*}
        \sum_{k = 1}^{n - 1} \abs{w'_{k}} &= \sum_{\substack{\ell = 1 \\ \ell \neq i, j}}^{n} \abs{w_{\ell}} + \abs{w_{j} \pm w_{i}} \\
        &\leq \sum_{\substack{\ell = 1 \\ \ell \neq i, j}}^{n} \abs{w_{\ell}} + \abs{w_{j}} + \abs{w_{i}} = \sum_{\ell = 1}^{n} \abs{w_{\ell}}.
    \end{align*}
    Therefore, in summary, the inequality~\eqref{equation:WeightInequality} holds for all four possible substitutions.
\end{proof}

Iterating \autoref{lemma:SubstitutionWeightRestriction} along any substitution chain shows that the weight extent is non-increasing, so every problem $h_{d', \bm{w}'}^{(n')}$ descending from the root stays within the root's extent, i.e., satisfies
\begin{equation}\label{equation:WeightRestriction}
    \sum_{i = 1}^{n'} \abs{w'_{i}} \leq W.
\end{equation}
Having tied every subproblem's weight extent to the root's weight extent, we now investigate its impact on the size and producibility of connecting families $\mathcal{E}'$ for the reduced problems.
Note that the full family $\mathcal{E}$ in \autoref{definition:FibreGraph} is generally exponentially large, so \autoref{corollary:ExistenceOfQAOAMixer} only settles existence, not efficiency.
Under the weight restriction~\eqref{equation:WeightRestriction}, however, we can do markedly better.
The major insight will be that constructing a connecting $\mathcal{E}'$ separates cleanly into building ordinary singleton exchanges for variables of the same weight, and proper multi-bit uniform exchanges between same-weight classes of variables.
The former are ``for free'', while the latter require additional structural analysis of the number of attainable, distinct subset sums over the multiset $\bm{w}$.
As it turns out, the weight extent $W$ provides a sufficiently tight bound for this important quantity.

For the first step of dividing the variables into equal-weight classes, we require additional notation:
For each $v \in \Z$, let $K_{v} \coloneqq \{i \in [n] \defcolon w_{i} = v\}$ and $n_{v} \coloneqq \abs{K_{v}}$.
The classes partition $[n]$, so the weight restriction~\eqref{equation:WeightRestriction} distributes over them as
\begin{equation*}
    \sum_{v \neq 0} \abs{v}\, n_{v} = \sum_{i = 1}^{n} \abs{w_{i}} \leq W.
\end{equation*}
A bitstring's feasibility depends only on the number of ones it places in each class.
Accordingly, we write $m_{v}(\bm{x}) \coloneqq \abs{\{i \in K_{v} \defcolon x_{i} = 1\}}$ and define the \emph{grading} of $\bm{x}$ as $\bm{g}(\bm{x}) \coloneqq (m_{v}(\bm{x}))_{v \neq 0}$.
Then, it indeed holds that
\begin{equation*}
    \bm{w} \cdot \bm{x} = \sum_{v \neq 0} v\, m_{v}(\bm{x}),
\end{equation*}
so that $\bm{x}$ is feasible if and only if $\sum_{v \neq 0} v\, m_{v}(\bm{x}) = d$; the weight-$0$ coordinates in $K_{0}$ never affect feasibility.

The decomposition splits the block exchanges of~\eqref{equation:GeneralisedXYHamiltonian} into two distinct roles.
Intra-class transpositions $\texttt{s}_{\{i\}}^{\{j\}}$ with $i, j \in K_{v}$ and zero-flips $\texttt{s}_{\{i\}}$ with $i \in K_{0}$ are feasibility-preserving for free and leave the grading fixed.
In contrast, cross-class exchanges $\texttt{s}_{K}^{L}$, where $K, L \subseteq K_{v}$ for no $v \in \Z$, redistribute $1$s between distinct classes.
As in the constant-Hamming weight case, adjacency swaps within each class, together with zero-flips, already connect all feasible solutions of the same grading.

\begin{corollary}\label{corollary:AdjacencyIntraClassMoves}
    Let $\bm{w} \in \Z^{n}$, $d \in \Z$, and $(n, f, h_{d, \bm{w}}^{(n)})$ be a COP.
    Define
    \begin{equation}\label{equation:BitFlip}
        \mathcal{T}_{0} \coloneqq \{(\{i\}, \emptyset) \defcolon i \in K_{0}\}
    \end{equation}
    and, for every $v \neq 0$,
    \begin{equation}\label{equation:AdjacencySwap}
        \mathcal{T}_{v} \coloneqq \big\{\hspace*{-1pt}(\{i\}, \{{\textstyle\successor_{K_{v}}}(i)\}) : i \in K_{v}\hspace*{-2pt} \setminus\hspace*{-2pt} \{\max(K_{v})\}\big\},
    \end{equation}
    where $\successor_{K_{v}}(i)$ is the successor of $i$ in $K_{v} \subseteq [n]$.
    Then, any $\bm{x}, \bm{y} \in S$, $\bm{x} \neq \bm{y}$, with $\bm{g}(\bm{x}) = \bm{g}(\bm{y})$ are connected in $F_{\cup_{v \scriptin \Z} \mathcal{T}_{v}}$.
\end{corollary}

Next, we isolate the action of cross-class exchanges.

\begin{definition}\label{definition:GradingQuotient}
    Let $\bm{w} \in \Z^{n}$, $d \in \Z$, $(n, f, h_{d, \bm{w}}^{(n)})$ be a COP, and $\mathcal{E}' \subseteq \mathcal{E}$.
    The COP's \emph{grading quotient} $\Gamma_{\mathcal{E}'} = (V, E)$ is defined via
    \begin{equation*}
        V \coloneqq \Big\{ (m_{v})_{v \neq 0} \defcolon \sum_{v \neq 0} v\, m_{v} = d,\ 0 \leq m_{v} \leq n_{v} \Big\}
    \end{equation*}
    and
    \begin{equation*}
        E \coloneqq \hspace*{-2pt}\bigcup_{(K, L) \scriptin \mathcal{E}'}\hspace*{-2pt} \{\{\bm{g}(\bm{x}), \bm{g}(\bm{y})\} \defcolon \texttt{s}_{K}^{L} \ket{\bm{x}} = 2^{\abs{K} + \abs{L} - 1} \ket{\bm{y}}\}.
    \end{equation*}
\end{definition}

The grading quotient collapses each grading to a single vertex, turning intra-class moves from~\eqref{equation:BitFlip} or~\eqref{equation:AdjacencySwap} into self-loops.
By the next lemma, this collapse loses no connectivity information.

\begin{lemma}\label{lemma:GradingFaithfulness}
    Let $\bm{w} \in \Z^{n}$, $d \in \Z$, $(n, f, h_{d, \bm{w}}^{(n)})$ be a COP, and $\mathcal{E}' \subseteq \mathcal{E}$ such that
    \begin{equation*}
        \mathcal{T} \coloneqq \bigcup_{v \scriptin \Z} \mathcal{T}_{v} \subseteq \mathcal{E}'.
    \end{equation*}
    $F_{\mathcal{E}'}$ is connected if and only if $\Gamma_{\mathcal{E}' \setminus \mathcal{T}}$ is connected.
\end{lemma}

\begin{proof}
    Let $\bm{x}, \bm{x}' \in V(F_{\mathcal{E}'})$ with gradings $m \coloneqq \bm{g}(\bm{x})$, and $m' \coloneqq \bm{g}(\bm{x}')$; every vertex of $\Gamma_{\mathcal{E}' \setminus \mathcal{T}}$ arises this way.
    First, assume $F_{\mathcal{E}'}$ to be connected, i.e., there exists a sequence $s_{1}, \ldots, s_{p} \in \mathcal{E}'$, constituting a path between $\bm{x}$ and $\bm{x}'$ in $F_{\mathcal{E}'}$.
    W.l.o.g., all vertices of the same grading are visited consecutively, since, by \autoref{corollary:AdjacencyIntraClassMoves}, any subpath between $\bm{y}$ and $\bm{y}'$ with $\bm{g}(\bm{y}) = \bm{g}(\bm{y}')$ can be replaced by one consisting of moves in $\mathcal{T}$.
    Along such a path, any two neighbouring vertices $\bm{y}, \bm{z}$ with connecting move $s_{i}$ either share their grading, i.e., their vertex in $\Gamma_{\mathcal{E}' \setminus \mathcal{T}}$, or $s_{i} \notin \mathcal{T}$, so that $\bm{g}(\bm{y})$ and $\bm{g}(\bm{z})$ remain connected in $\Gamma_{\mathcal{E}' \setminus \mathcal{T}}$.
    Hence, $m$ and $m'$ are connected in $\Gamma_{\mathcal{E}' \setminus \mathcal{T}}$.

    Second, assume $\Gamma_{\mathcal{E}' \setminus \mathcal{T}}$ to be connected, i.e., there exists a sequence $s_{1}, \ldots, s_{q} \in \mathcal{E}' \setminus \mathcal{T}$, inducing a path between $m$ and $m'$ in $\Gamma_{\mathcal{E}' \setminus \mathcal{T}}$.
    Let $\nu, \mu, \nu'$ be consecutive vertices on this path.
    By definition of $\Gamma_{\mathcal{E}' \setminus \mathcal{T}}$, we can pick vertices $\bm{y}, \bm{y}', \bm{z}, \bm{z}' \in V(F_{\mathcal{E}'})$ with $\bm{g}(\bm{y}) = \bm{g}(\bm{y}') = \mu$, $\bm{g}(\bm{z}) = \nu$, and $\bm{g}(\bm{z}') = \nu'$, so that $\{\bm{y}, \bm{z}\}, \{\bm{y}', \bm{z}'\} \in E(F_{\mathcal{E}'})$.
    By \autoref{corollary:AdjacencyIntraClassMoves}, $\bm{y}$ and $\bm{y}'$ are connected via moves in $\mathcal{T} \subseteq \mathcal{E}'$, and hence so are $\bm{z}$ and $\bm{z}'$.
    Inductively, this also connects $\bm{x}$ and $\bm{x}'$ in $F_{\mathcal{E}'}$.
\end{proof}

The set $\mathcal{T}$ of intra-class transpositions and zero-flips, defined in \autoref{lemma:GradingFaithfulness}, consists of
\begin{equation*}
    \abs{\mathcal{T}} = \abs{K_{0}} + \sum_{v \neq 0} \max\{0, \abs{K_{v}} - 1\} \leq n
\end{equation*}
elements, with equality only if $K_{0} = [n]$.
Especially, $\abs{\mathcal{E}'}$ is polynomial in $n$ if and only if $\abs{\mathcal{E}' \setminus \mathcal{T}}$ is.
By \autoref{lemma:GradingFaithfulness}, fabricating a connecting family of polynomial size therefore reduces to fabricating a polynomial-size collection $\mathcal{C}$ of cross-class exchanges with connected grading quotient $\Gamma_{\mathcal{C}}$.
The following theorem fabricates such exchanges via dynamic programming.

\begin{theorem}\label{theorem:ConnectingMoves}
    Let $\bm{w} \in \Z^{n}$, $d \in \Z$, and $(n, f, h_{d, \bm{w}}^{(n)})$ be a COP with weight extent $W = \sum_{i = 1}^{n} \abs{w_{i}}$ and $c$ distinct non-zero elements within $\bm{w}$.
    Then, a family $\mathcal{E}' \subseteq \mathcal{E}$ with
    \begin{equation}\label{equation:CountConnectingMoves}
        \abs{\mathcal{E}'} \leq n + (W + 1) (n - \abs{K_{0}} + c),
    \end{equation}
    so that $F_{\mathcal{E}'}$ is connected, can be constructed in time $O(n + (W + 1)(n - \abs{K_{0}})^{2})$, counting arithmetic operations and memory accesses at unit cost.\footnote{
        If each witness $C_{i}(\sigma)$ constructed in the proof is instead stored implicitly through its back pointer, and each move $(K_{i}(\sigma, q), L_{i}(\sigma, q))$ is stored through the triple $(i, \sigma, q)$, the construction time reduces to $O(n + (W + 1)(n - \abs{K_{0}} + c))$.
        In particular, a move is trivial if and only if $(\sigma, q)$ is the back pointer of $\sigma + q u_{i}$, which is checked in constant time.
        The explicit index sets of any single move can then be materialised on demand in $O(n - \abs{K_{0}})$ time, e.g., when compiling the corresponding mixer gate.}
    Moreover, any two feasible solutions are joined in $F_{\mathcal{E}'}$ by a path containing at most $2c$ moves from $\mathcal{E}' \setminus \mathcal{T}$.
\end{theorem}

\begin{figure*}[t]
    \centering
    \footnotesize
    \setlength{\tabcolsep}{3pt}
    \renewcommand{\arraystretch}{1.1}
    \begin{tabular}[t]{cc>{$}l<{$}c>{$}l<{$}}
        \multicolumn{5}{l}{Stage $i = 1$: $K_{u_{1}} = \{1,2\}$, $u_{1} = 1$} \\
        \toprule
        $\sigma$ & $q$ & C_{0}(\sigma) \cup V_{1}(q) & $\sigma'$ & (K_{1}, L_{1}) \\
        \midrule
        $0$ & $0$ & \emptyset & $0$ & \text{---} \\
        $0$ & $1$ & \{1\} & $1$ & \text{---} \\
        $0$ & $2$ & \{1,2\} & $2$ & \text{---} \\
        \bottomrule
    \end{tabular}
    \hfill
    \begin{tabular}[t]{cc>{$}l<{$}c>{$}l<{$}}
        \multicolumn{5}{l}{Stage $i = 2$: $K_{u_{2}} = \{3,4\}$, $u_{2} = 2$} \\
        \toprule
        $\sigma$ & $q$ & C_{1}(\sigma) \cup V_{2}(q) & $\sigma'$ & (K_{2}, L_{2}) \\
        \midrule
        $0$ & $0$ & \emptyset & $0$ & \text{---} \\
        $0$ & $1$ & \{3\} & $2$ & \text{---} \\
        $0$ & $2$ & \{3,4\} & $4$ & \text{---} \\
        $1$ & $0$ & \{1\} & $1$ & \text{---} \\
        $1$ & $1$ & \{1,3\} & $3$ & \text{---} \\
        $1$ & $2$ & \{1,3,4\} & $5$ & \text{---} \\
        \rowcolor{Orange!15}[\tabcolsep][\tabcolsep] $2$ & $0$ & \{1,2\} & $2$ & (\{1,2\}, \{3\}) \\
        \rowcolor{Orange!15}[\tabcolsep][\tabcolsep] $2$ & $1$ & \{1,2,3\} & $4$ & (\{1,2\}, \{4\}) \\
        $2$ & $2$ & \{1,2,3,4\} & $6$ & \text{---} \\
        \bottomrule
    \end{tabular}
    \hfill
    \begin{tabular}[t]{cc>{$}l<{$}c>{$}l<{$}}
        \multicolumn{5}{l}{Stage $i = 3$: $K_{u_{3}} = \{5\}$, $u_{3} = 3$} \\
        \toprule
        $\sigma$ & $q$ & C_{2}(\sigma) \cup V_{3}(q) & $\sigma'$ & (K_{3}, L_{3}) \\
        \midrule
        $0$ & $0$ & \emptyset & $0$ & \text{---} \\
        $0$ & $1$ & \{5\} & $3$ & \text{---} \\
        $1$ & $0$ & \{1\} & $1$ & \text{---} \\
        $1$ & $1$ & \{1,5\} & $4$ & \text{---} \\
        $2$ & $0$ & \{3\} & $2$ & \text{---} \\
        $2$ & $1$ & \{3,5\} & $5$ & \text{---} \\
        \rowcolor{Orange!15}[\tabcolsep][\tabcolsep] $3$ & $0$ & \{1,3\} & $3$ & (\{1,3\}, \{5\}) \\
        $3$ & $1$ & \{1,3,5\} & $6$ & \text{---} \\
        \rowcolor{Orange!15}[\tabcolsep][\tabcolsep] $4$ & $0$ & \{3,4\} & $4$ & (\{3,4\}, \{1,5\}) \\
        $4$ & $1$ & \{3,4,5\} & $7$ & \text{---} \\
        \rowcolor{Orange!15}[\tabcolsep][\tabcolsep] $5$ & $0$ & \{1,3,4\} & $5$ & (\{1,4\}, \{5\}) \\
        $5$ & $1$ & \{1,3,4,5\} & $8$ & \text{---} \\
        \rowcolor{Orange!15}[\tabcolsep][\tabcolsep] $6$ & $0$ & \{1,2,3,4\} & $6$ & (\{2,4\}, \{5\}) \\
        $6$ & $1$ & \{1,2,3,4,5\} & $9$ & \text{---} \\
        \bottomrule
    \end{tabular}
    \caption{\label{figure:DynamicProgram}
        The dynamic program from the proof of \autoref{theorem:ConnectingMoves} for $\bm{w} = (1, 1, 2, 2, 3)$, i.e., $c = 3$ classes, listing every step $(\sigma, q)$, $\sigma \in \Sigma_{i - 1}$, $q \in [\abs{K_{u_{i}}}] \cup \{0\}$, of stage $i$ in processing order ($\sigma$ ascending, then $q$ ascending).
        The candidate $C_{i - 1}(\sigma) \cup V_{i}(q)$ realises the sum $\sigma' \coloneqq \sigma + q u_{i}$.
        If this sum is attained for the first time, $(\sigma, q)$ becomes its back pointer and the candidate becomes the witness $C_{i}(\sigma')$; the resulting move is trivial~(---).
        Otherwise, the candidate is compared with the stored witness, and removing their common indices yields the balanced move $(K_{i}, L_{i}) \coloneqq (K_{i}(\sigma, q), L_{i}(\sigma, q))$ (shaded).
        The $26$ steps fix $20$ witnesses and produce the $6$ moves of $\mathcal{C}$.
    }
\end{figure*}

\begin{proof}
    Impose an arbitrary order on the distinct non-zero elements $u_{1}, \ldots, u_{c}$ of $\bm{w}$ and write $N \coloneqq n - \abs{K_{0}}$ for the number of non-zero weights, so that $c \leq N$.

    Since $\abs{w_{k}} \leq W$ for all $k \in [n]$, bucket sort over $\{-W, \ldots, W\}$ distributes the indices into the classes $K_{v}$, each in ascending order, in $O(n + W)$ time.
    From these, $\mathcal{T}$ is read off in $O(n)$ time.
    For every $i \in [c]$ and $q \in [\abs{K_{u_{i}}}] \cup \{0\}$, let $V_{i}(q) \subseteq K_{u_{i}}$ denote the set of the first $q$ elements of $K_{u_{i}}$.
    For every $i \in [c] \cup \{0\}$, define
    \begin{equation*}
        \Sigma_{i} \coloneqq \bigg\{\sum_{j = 1}^{i} q_{j} u_{j} \defcolon q_{j} \in \big[\abs{K_{u_{j}}}\big] \cup \{0\}\bigg\}
    \end{equation*}
    so that $\Sigma_{0} = \{0\}$ and
    \begin{equation*}
        \Sigma_{i} = \{\sigma + q u_{i} \defcolon \sigma \in \Sigma_{i - 1}, q \in [\abs{K_{u_{i}}}] \cup \{0\}\}
    \end{equation*}
    for $i \in [c]$.
    For each $i \in [c]$, we have that
    \begin{align*}
        \sigma \in \Sigma_{i} &\implies \sum_{v < 0} \abs{K_{v}} v \leq \sigma \leq \sum_{v > 0} \abs{K_{v}} v \\
        &\implies \abs{\Sigma_{i}} \leq W + 1,
    \end{align*}
    so that every $\Sigma_{i}$ can simply be stored as a Boolean array over this window of $W + 1$ integers.
    We now fix, for every $i \in [c] \cup \{0\}$ and $\sigma \in \Sigma_{i}$, a witness $C_{i}(\sigma) \subseteq K_{u_{1}} \cup \cdots \cup K_{u_{i}}$ with
    \begin{equation*}
        \sum_{k \scriptin C_{i}(\sigma)} w_{k} = \sigma
    \end{equation*}
    via the standard dynamic program for bounded-multiplicity subset sum:
    Set $C_{0}(0) \coloneqq \emptyset$ and compute $\Sigma_{i}$ from $\Sigma_{i - 1}$ according to the above recursion.
    Whenever some $\sigma' \in \Sigma_{i}$ is attained for the first time as $\sigma' = \sigma + q u_{i}$, store the back pointer $(\sigma, q)$ and set
    \begin{equation*}
        C_{i}(\sigma') \coloneqq C_{i - 1}(\sigma) \cup V_{i}(q).
    \end{equation*}
    Inductively, $C_{i}(\sigma')$ has the required properties.
    Computing $\Sigma_{i}$ from $\Sigma_{i - 1}$ takes $O(\abs{\Sigma_{i - 1}} (\abs{K_{u_{i}}} + 1))$ time, so all back pointers are determined in
    \begin{equation}\label{equation:CountDynamicProgram}
        \sum_{i = 1}^{c} \abs{\Sigma_{i - 1}} (\abs{K_{u_{i}}} + 1) \leq (W + 1) (N + c)
    \end{equation}
    steps.
    We now define, for every index $i \in [c]$, attainable sum $\sigma \in \Sigma_{i - 1}$, and multiplicity $q \in [\abs{K_{u_{i}}}] \cup \{0\}$, the two sets
    \begin{align*}
        K_{i}(\sigma, q) &\coloneqq (C_{i - 1}(\sigma) \cup V_{i}(q)) \setminus C_{i}(\sigma + q u_{i}) \text{ and} \\
        L_{i}(\sigma, q) &\coloneqq C_{i}(\sigma + q u_{i}) \setminus (C_{i - 1}(\sigma) \cup V_{i}(q)).
    \end{align*}
    Both sets are, by construction, disjoint.
    Furthermore, summing weights over all indices in $C_{i - 1}(\sigma) \cup V_{i}(q)$ readily yields $\sigma + q u_{i}$ and thus the same value as when summing over the indices of the witness set $C_{i}(\sigma + q u_{i})$.
    The sets $K_{i}(\sigma, q)$ and $L_{i}(\sigma, q)$ arise from these two sets by removing their common elements, hence their weight sums coincide as well, establishing that $(K_{i}(\sigma, q), L_{i}(\sigma, q)) \in \mathcal{E}$.
    Moves with $K_{i}(\sigma, q) = L_{i}(\sigma, q) = \emptyset$ act trivially and are discarded.
    \autoref{figure:DynamicProgram} lists all steps of this dynamic program, together with the resulting moves, for a small example.
    Let $\mathcal{C} \subseteq \mathcal{E}$ denote the totality of all remaining moves and $\mathcal{E}' \coloneqq \mathcal{T} \cup \mathcal{C}$.
    Since $\abs{\mathcal{T}} \leq n$ and, by~\eqref{equation:CountDynamicProgram}, $\abs{\mathcal{C}} \leq (W + 1)(N + c)$, the bound~\eqref{equation:CountConnectingMoves} follows.
    To write out a move explicitly, we materialise both witnesses by unrolling at most $c$ back pointers and concatenating the pairwise disjoint sets $V_{j}(q_{j})$, and then form both set differences using a marker array over $[n]$, initialised once in $O(n)$ time.
    Since $\abs{C_{i}(\sigma)} \leq N$ and $c \leq N$, this takes $O(N)$ time per move and $O((W + 1)(N + c) N) = O((W + 1) N^{2})$ time for all of $\mathcal{C}$.
    Together with the previous steps, the claimed running time follows.

    What remains to show is that $\Gamma_{\mathcal{C}}$ is connected.
    Here, it is useful to overload the grading definition to index subsets via $\bm{g}(A) \coloneqq (\abs{A \cap K_{v}})_{v \neq 0}$, $A \subseteq [n]$.
    We construct paths from all vertices of $V(\Gamma_{\mathcal{C}})$ to the hub $\bm{g}(C_{c}(d))$.
    Let $(m_{v})_{v \neq 0} \in V(\Gamma_{\mathcal{C}})$ and define, for every $i \in [c] \cup \{0\}$,
    \begin{equation*}
        \sigma_{i} \coloneqq \sum_{j = 1}^{i} m_{u_{j}} u_{j} \text{ and } A_{i} \coloneqq C_{i}(\sigma_{i}) \cup \bigcup_{j > i} V_{j}(m_{u_{j}})
    \end{equation*}
    so that indeed $\bm{g}(A_{i}) \in V(\Gamma_{\mathcal{C}})$ for all $i \in [c] \cup \{0\}$ with $\bm{g}(A_{0}) = (m_{v})_{v \neq 0}$ and $A_{c} = C_{c}(d)$.
    Then, by construction, for every $i \in [c]$, it holds that $K_{i}(\sigma_{i - 1}, m_{u_{i}}) \subseteq A_{i - 1}$, $L_{i}(\sigma_{i - 1}, m_{u_{i}}) \cap A_{i - 1} = \emptyset$, and
    \begin{equation*}
        (A_{i - 1} \setminus K_{i}(\sigma_{i - 1}, m_{u_{i}})) \cup L_{i}(\sigma_{i - 1}, m_{u_{i}}) = A_{i}.
    \end{equation*}
    If the move $(K_{i}(\sigma_{i - 1}, m_{u_{i}}), L_{i}(\sigma_{i - 1}, m_{u_{i}}))$ was discarded as trivial, then $A_{i - 1} = A_{i}$ and the step is void.
    Hence, $(m_{v})_{v \neq 0}$ and $\bm{g}(C_{c}(d))$ are connected in $\Gamma_{\mathcal{C}}$ as endpoints of a path along $\bm{g}(A_{i})$, $i \in [c] \cup \{0\}$, of at most $c$ edges.
    Here, $(m_{v})_{v \neq 0} \in V(\Gamma_{\mathcal{C}})$ was chosen arbitrarily.
    Now every two $(m_{v})_{v \neq 0}, (m'_{v})_{v \neq 0} \in V(\Gamma_{\mathcal{C}})$ are connected via the concatenation of their respective paths to $\bm{g}(C_{c}(d))$, proving that $\Gamma_{\mathcal{C}}$ is connected.
    By \autoref{lemma:GradingFaithfulness}, $F_{\mathcal{E}'}$ is thus connected as well.
    Furthermore, any two feasible gradings are joined in $\Gamma_{\mathcal{C}}$ by a path of at most $2c$ edges.
    Lifting this path to $F_{\mathcal{E}'}$ as in the proof of \autoref{lemma:GradingFaithfulness} only inserts moves from $\mathcal{T}$, so the lifted path contains at most $2c$ moves from $\mathcal{C} = \mathcal{E}' \setminus \mathcal{T}$.
\end{proof}

Note that $c \leq n - \abs{K_{0}} \leq n$ in~\eqref{equation:CountConnectingMoves} and in the runtime analysis, so that a polynomially bounded weight extent already implies a polynomial-sized, connecting $\mathcal{E}'$, which can be fabricated in polynomial time.

\begin{corollary}\label{corollary:PolyConnectingMoves}
    Let $\bm{w} \in \Z^{n}$, $d \in \Z$, and $(n, f, h_{d, \bm{w}}^{(n)})$ be a COP with weight extent $W \in O(\mathrm{poly}(n))$.
    Then, a family $\mathcal{E}' \subseteq \mathcal{E}$ with $\abs{\mathcal{E}'} \in O(\mathrm{poly}(n))$, so that $F_{\mathcal{E}'}$ is connected, can be constructed in $O(\mathrm{poly}(n))$ time.
    Moreover, any two feasible solutions are joined in $F_{\mathcal{E}'}$ by a path containing at most $2 n$ moves from $\mathcal{E}' \setminus \mathcal{T}$.
\end{corollary}

Let us also consider the deletion of forced variables, i.e., variables $x_{k}$ which attain the same value $b \in \bit$ in every feasible solution $\bm{x}$, i.e., $x_{k} = b$.
For instance, for $\bm{w} = (1, 1, 2, 2, 3)$ and $d = 8$, every feasible solution has to omit exactly one of the two variables of weight $1$, so that $x_{3}$, $x_{4}$, and $x_{5}$ are forced to $1$, while neither $x_{1}$ nor $x_{2}$ is forced.
As for unate feasibility predicates, every feasible state yields $\abs{M_{k}} = 1$ for a forced variable $x_{k}$, so that the selection rule would spend subsequent RQAOA iterations on eliminating forced variables, whose qubits would be carried along until then.
Since feasibility only depends on the grading, variables in $K_{0}$ are never forced, while all variables of a class $K_{v}$, $v \neq 0$, are forced simultaneously or not at all.
Which classes are forced can be read off from the dynamic program of \autoref{theorem:ConnectingMoves}, complemented by a second pass over the classes in reverse order.

\begin{proposition}\label{proposition:ForcedVariablesEquality}
    Let $\bm{w} \in \Z^{n}$, $d \in \Z$, and $(n, f, h_{d, \bm{w}}^{(n)})$ be a COP with $h_{d, \bm{w}}^{(n)} \notequiv 0$ and weight extent $W$, and let $u_{1}, \ldots, u_{c}$ and $\Sigma_{0}, \ldots, \Sigma_{c}$ be as in the proof of \autoref{theorem:ConnectingMoves}.
    Define $\bar{\Sigma}_{c + 1} \coloneqq \{0\}$ and, for $i = c, \ldots, 1$,
    \begin{equation*}
        \bar{\Sigma}_{i} \coloneqq \{\sigma + q u_{i} \defcolon \sigma \in \bar{\Sigma}_{i + 1}, q \in [\abs{K_{u_{i}}}] \cup \{0\}\},
    \end{equation*}
    as well as, for $i \in [c]$,
    \begin{equation*}
        Q_{i} \coloneqq \big\{q \in [\abs{K_{u_{i}}}] \cup \{0\} \defcolon d - q u_{i} \in \Sigma_{i - 1} + \bar{\Sigma}_{i + 1}\big\},
    \end{equation*}
    where $\Sigma + \bar{\Sigma} \coloneqq \{\sigma + \bar{\sigma} \defcolon \sigma \in \Sigma, \bar{\sigma} \in \bar{\Sigma}\}$.
    Then, a variable $x_{k}$ is forced to $0$ (resp.\ to $1$) if and only if $k \in K_{u_{i}}$ for some $i \in [c]$ with $Q_{i} = \{0\}$ (resp.\ with $Q_{i} = \{\abs{K_{u_{i}}}\}$).
    All forced variables can be determined in $O(n + (W + 1)(n - \abs{K_{0}}))$ time.
\end{proposition}

\begin{proof}
    By construction, $\bar{\Sigma}_{i + 1}$ consists of all sums $\sum_{j > i} q_{j} u_{j}$ with $q_{j} \in [\abs{K_{u_{j}}}] \cup \{0\}$, i.e., of all sums attainable by the classes $K_{u_{i + 1}}, \ldots, K_{u_{c}}$, just as $\Sigma_{i - 1}$ consists of all sums attainable by $K_{u_{1}}, \ldots, K_{u_{i - 1}}$.
    First, we show that $q \in Q_{i}$ if and only if some feasible solution $\bm{x}$ satisfies $m_{u_{i}}(\bm{x}) = q$.
    For such an $\bm{x}$, the sums $\sigma \coloneqq \sum_{j < i} m_{u_{j}}(\bm{x}) u_{j}$ and $\bar{\sigma} \coloneqq \sum_{j > i} m_{u_{j}}(\bm{x}) u_{j}$ satisfy $\sigma \in \Sigma_{i - 1}$, $\bar{\sigma} \in \bar{\Sigma}_{i + 1}$, and $\sigma + \bar{\sigma} = d - q u_{i}$, hence $q \in Q_{i}$.
    Conversely, if $q \in Q_{i}$, there are multiplicities $q_{j} \in [\abs{K_{u_{j}}}] \cup \{0\}$, $j \in [c] \setminus \{i\}$, with $\sum_{j \neq i} q_{j} u_{j} = d - q u_{i}$, so that the indicator vector of $V_{i}(q) \cup \bigcup_{j \neq i} V_{j}(q_{j})$ is a feasible solution with $q$ ones in $K_{u_{i}}$.
    Second, since feasibility only depends on the grading, every permutation of the bits within $K_{u_{i}}$ maps feasible solutions to feasible solutions.
    Hence, for $k \in K_{u_{i}}$, some feasible solution satisfies $x_{k} = 1$ if and only if $Q_{i}$ contains some $q \geq 1$, and some feasible solution satisfies $x_{k} = 0$ if and only if $Q_{i}$ contains some $q \leq \abs{K_{u_{i}}} - 1$.
    As $h_{d, \bm{w}}^{(n)} \notequiv 0$, it is $Q_{i} \neq \emptyset$, so that $x_{k}$ is forced to $0$ if and only if $Q_{i} = \{0\}$, and forced to $1$ if and only if $Q_{i} = \{\abs{K_{u_{i}}}\}$.
    For $k \in K_{0}$, flipping $x_{k}$ preserves feasibility, so that $x_{k}$ attains both values among the feasible solutions and is not forced.

    For the running time, let $N \coloneqq n - \abs{K_{0}}$, and recall from the proof of \autoref{theorem:ConnectingMoves} that the classes are formed in $O(n + W)$ time.
    All $\Sigma_{i}$ and $\bar{\Sigma}_{i}$ are stored as Boolean arrays over the same window of $W + 1$ integers, so that, by the same count as in~\eqref{equation:CountDynamicProgram}, each of the two passes takes at most $(W + 1)(N + c)$ steps.
    For every $i \in [c]$ and every candidate $q$, testing $q \in Q_{i}$ amounts to scanning all $\sigma \in \Sigma_{i - 1}$ and looking up $d - q u_{i} - \sigma$ in the array of $\bar{\Sigma}_{i + 1}$, which takes $O(W + 1)$ time.
    Hence, determining all sets $Q_{i}$ takes \mbox{$O((W + 1)(N + c))$} time.
    Reading off the forced variables takes $O(n)$ time.
    Since $c \leq N$, and $W = 0$ whenever $N = 0$, the claimed running time follows.
\end{proof}

\autoref{algorithm:ConnectingFamily} summarises the constructions from the proofs of \autoref{theorem:ConnectingMoves} and \autoref{proposition:ForcedVariablesEquality} in pseudo-code, where $V_{i}(q)$ again denotes the set of the first $q$ elements of $K_{u_{i}}$ and the sets $\Sigma_{i}$ and $\bar{\Sigma}_{i}$ are stored as Boolean arrays over the window of at most $W + 1$ attainable sums.
The forward sums are thus computed twice, first without witnesses to detect the forced variables, and then with witnesses for the reduced COP, which does not affect the asymptotic running time.
The problem-specific mixer is then given by $U_{\mathcal{E}'}(\beta)$ as in~\eqref{equation:GeneralisedXYParameterisedMixer}, where each factor $e^{-i \beta \texttt{s}_{K}^{L}}$ is compiled, e.g., via the \texttt{CNOT}-cascade construction discussed after~\eqref{equation:GeneralisedXYHamiltonian}.
Note that, apart from the selection of forced classes, $\mathcal{C}$ is computed without reference to $d$, which otherwise only enters the very last step.

\begin{algorithm}[t]
    \caption{\label{algorithm:ConnectingFamily}Dynamic program for $h_{d, \bm{w}}^{(n)}$}
    \begin{algorithmic}[1]
        \Require weights $\bm{w} \in \Z^{n}$, $d \in \Z$ with $h_{d, \bm{w}}^{(n)} \notequiv 0$
        \Ensure forced values $\bm{b} \in \bit^{J}$, and $\mathcal{E}' \subseteq \mathcal{E}$ with $F_{\mathcal{E}'}$ connected, feasible $\bm{x}_{\iota}$ on $[n] \setminus J$
        \State bucket sort $[n]$ into the classes $K_{v}$, each in ascending order
        \State order the distinct non-zero weights as $u_{1}, \ldots, u_{c}$
        \State compute all $\Sigma_{i}$, $\bar{\Sigma}_{i}$, and $Q_{i}$ as in \autoref{proposition:ForcedVariablesEquality}
        \State $J \gets \emptyset$
        \For{$i \in [c]$ with $Q_{i} = \{q\} \subseteq \{0, \abs{K_{u_{i}}}\}$}
        \State $J \gets J \cup K_{u_{i}}$,\ $d \gets d - q u_{i}$
        \State $b_{k} \gets q / \abs{K_{u_{i}}}$ for all $k \in K_{u_{i}}$
        \EndFor
        \State discard the classes contained in $J$
        \State renumber the remaining weights as $u_{1}, \ldots, u_{c}$
        \State read off $\mathcal{T}$ as in \autoref{lemma:GradingFaithfulness}
        \State $\Sigma_{0} \gets \{0\}$,\ $C_{0}(0) \gets \emptyset$,\ $\mathcal{C} \gets \emptyset$
        \For{$i = 1, \ldots, c$}
        \State $\Sigma_{i} \gets \emptyset$
        \For{$\sigma \in \Sigma_{i - 1}$ and $q = 0, \ldots, \abs{K_{u_{i}}}$}
        \State $\sigma' \gets \sigma + q u_{i}$,\ $B \gets C_{i - 1}(\sigma) \cup V_{i}(q)$
        \If{$\sigma' \notin \Sigma_{i}$}
        \State $\Sigma_{i} \gets \Sigma_{i} \cup \{\sigma'\}$,\ $C_{i}(\sigma') \gets B$
        \EndIf
        \State $(K, L) \gets (B \setminus C_{i}(\sigma'), C_{i}(\sigma') \setminus B)$
        \If{$K \cup L \neq \emptyset$}
        \State $\mathcal{C} \gets \mathcal{C} \cup \{(K, L)\}$
        \EndIf
        \EndFor
        \EndFor
        \State $\bm{x}_{\iota} \gets$ indicator vector of $C_{c}(d)$ on $[n] \setminus J$
        \State \Return $J$, $\bm{b}$, $\mathcal{E}' \gets \mathcal{T} \cup \mathcal{C}$, and $\bm{x}_{\iota}$
    \end{algorithmic}
\end{algorithm}

This last step also settles the initial state preparation.
By construction, $d \in \Sigma_{c}$ holds if and only if the COP has a feasible solution, and in this case the witness $C_{c}(d)$---which already served as the hub in the proof of \autoref{theorem:ConnectingMoves}---satisfies $\sum_{k \scriptin C_{c}(d)} w_{k} = d$.
Hence, its indicator vector $\bm{x}_{\iota}$ is feasible, and the computational basis state $\ket{\bm{x}_{\iota}}$ is prepared by at most $n - \abs{K_{0}}$ \texttt{X}-gates.
Materialising $C_{c}(d)$ merely requires unrolling at most $c$ back pointers, i.e., $O(n)$ additional time.
Thus, for polynomially bounded weight extent, the root problem and every subproblem admit an efficiently computable feasible initial state.

For general weights, in contrast, no efficient initial state preparation can be expected:
Already for non-negative weights, deciding whether the COP has a feasible solution is the NP-complete subset sum problem~\cite{Karp1972ReducibilityAmongCombinatorialProblems}.
Any polynomial-time routine preparing a state $\ket{\iota} \in \mathcal{S}$ whenever $\mathcal{S} \neq \{0\}$ would decide this problem, since measuring $\ket{\iota}$ in the computational basis and checking the outcome for feasibility accepts if and only if a feasible solution exists.
This remains true under the promise that a feasible solution exists, as guaranteed for RQAOA subproblems by \autoref{lemma:LocalConsistency}, since the routine's output can be verified.
Therefore, unless $\mathrm{NP} \subseteq \mathrm{BQP}$, there is no efficient feasible initial state preparation for general linear equality constraints.
Of course, the reuse of a current problem's initial state, as discussed in \autoref{subsection:InitialStatePreparation}, remains possible whenever it obeys the enforced substitution.

Together with \autoref{proposition:FibreGraphConnectivity}, \autoref{corollary:PolyConnectingMoves} settles the existence of a polynomial number of generalised \texttt{XY} mixers for a problem with polynomially bounded weight extent.
By \autoref{lemma:SubstitutionWeightRestriction}, this polynomial existence guarantee extends to all potential subproblems derived via RQAOA-powered substitutions.

Three remarks on the use of this result within RQAOA are in order.
First, the connecting family of a subproblem is not inferred from the root problem's mixer, but fabricated anew from the reduced weights via \autoref{theorem:ConnectingMoves}.
The naive inference---restricting the current problem's family to those moves not involving the eliminated variable $x_{i}$---preserves feasibility, but in general not connectivity, and repairing it by combining parent moves can require exponentially many applications of parent moves, even for the Hamming-weight root (see \autoref{appendix:InferenceFromParent}).
Fabrication, instead, is guaranteed to be efficient: by \autoref{lemma:SubstitutionWeightRestriction}, every subproblem inherits the root's weight extent bound $W$, so that its connecting family is constructed in $O(n + (W + 1) n^{2})$ time.
For the Hamming-weight root with $W = n$, this yields, for every subproblem, a connecting family of $O(n^{2})$ moves in $O(n^{3})$ time.

Second, forced variables are eliminated by unary substitutions $x_{k} = b_{k}$ at the root and after every RQAOA-driven substitution, exactly as in the deletion step of \autoref{algorithm:UnateRQAOA}.
These deletions are classical preprocessing and do not discard any information the quantum circuit could provide, since feasibility-preserving circuits never alter a forced variable, i.e., $M_{k} = \pm 1$ holds irrespective of the parameters.
A single pass suffices:
Fixing all forced variables at their forced values removes no feasible solution, so that the restriction to the remaining variables is a bijection between the feasible sets, and no remaining variable becomes forced.
The resulting problem is again a linear equality constraint on $n - \abs{J}$ variables with weight extent at most $W$ by \autoref{lemma:SubstitutionWeightRestriction}, so that the output of \autoref{algorithm:ConnectingFamily} indeed constitutes a valid subproblem with connecting family and feasible initial state.
The detection adds two passes of a Boolean dynamic program and the tests of \autoref{proposition:ForcedVariablesEquality}, whose cost $O(n + (W + 1) n)$ is dominated by the $O(n + (W + 1) n^{2})$ time for fabricating the connecting family; for the Hamming-weight root, it amounts to $O(n^{2})$ time per subproblem.
Note that a plain Hamming weight constraint $h_{d}^{(n)}$ only has forced variables in the degenerate cases $d \in \{0, n\}$, where all variables are forced; non-trivial forced variables, as in the above example, only emerge from non-uniform weights, e.g., after binary substitutions.

Third, fabrication does not have to start entirely from scratch.
Up to the deletion of forced classes, $\mathcal{E}'$ depends on $\bm{w}$ but not on $d$, so that the unary substitutions $x_{i} = 0$ and $x_{i} = 1$ produce the very same family whenever they leave the same classes unforced; the right-hand side otherwise merely determines the hub $\bm{g}(C_{c}(d))$ in the proof of \autoref{theorem:ConnectingMoves}.
Moreover, a unary substitution only removes $i$ from its class, while a binary substitution $x_{i} = (\neg) x_{j}$ additionally moves $j$ from $K_{w_{j}}$ to $K_{w_{j} \pm w_{i}}$, so that at most three classes change.
Accordingly, all but $O(1)$ moves of $\mathcal{T}$ carry over.
If the order $u_{1}, \ldots, u_{c}$ of the distinct weights is retained, the same holds for the sums $\Sigma_{i}$, the witnesses $C_{i}(\sigma)$, and the moves $(K_{i}(\sigma, q), L_{i}(\sigma, q))$ of all stages $i$ preceding the first changed or newly forced class, and, symmetrically, for the backward sums $\bar{\Sigma}_{i}$ of all stages succeeding the last changed class, so that only the remaining stages of the dynamic programs need to be recomputed.
In the worst case, these comprise the entire tables, i.e., reuse saves compilation effort without improving the asymptotic cost.

\subsection{\label{subsection:LinearFormConstraints}Constraints on linear forms}

Many COPs impose several constraints at once.
A prominent example is the one-hot encoding of $\kappa$-ary variables, e.g., in Max-$\kappa$-ColourableSubgraph~\cite{Hadfield2019FromTheQuantumApproximateOptimizationAlgorithmToAQuantumAlternatingOperatorAnsatz}, where every vertex carries $\kappa$ binary variables, exactly one of which has to be set, i.e., every vertex contributes one Hamming weight constraint.
The natural first step beyond \autoref{subsection:HammingWeightConstraints} is therefore to impose several linear equality constraints simultaneously.
Formally, for given $n, m \in \N$ and $A \in \Z^{m \times n}$, let $A_{i \ast} \in \Z^{n}$ denote the $i$th row of $A$ and let $A_{\ast j} \in \Z^{m}$ denote its $j$th column.
For given $\bm{d} \in \Z^{m}$, we denote the \emph{system of linear equality constraints} by
\begin{equation*}
    h_{\bm{d}, A}^{(n)}(\bm{x}) = 1 \text{ if and only if } A \bm{x} = \bm{d},
\end{equation*}
i.e., $h_{\bm{d}, A}^{(n)} = \bigwedge_{r = 1}^{m} h_{d_{r}, A_{r \ast}}^{(n)}$, and call $W_{r} \coloneqq \sum_{i = 1}^{n} \abs{A_{r i}}$, $r \in [m]$, the \emph{row extents} of $A$.
We call the vectors $\bm{y} \in \Z^{m}$ \emph{values}, and the sets $\{\bm{x} \in \bit^{n} \defcolon A \bm{x} = \bm{y}\}$ the \emph{fibres} of $A$, so that the feasible set of $h_{\bm{d}, A}^{(n)}$ is the fibre of $\bm{d}$.
For such systems, most of the mixer theory of \autoref{subsection:HammingWeightConstraints} carries over with little effort.
\autoref{lemma:FeasibilityPreservingGeneralisedXY}, \autoref{definition:FibreGraph}, \autoref{proposition:FibreGraphConnectivity}, \autoref{lemma:ExistenceOfConnectedFibreGraph}, and \autoref{corollary:ExistenceOfQAOAMixer} remain valid verbatim upon replacing the scalar balance condition $\sum_{k \scriptin K} w_{k} = \sum_{\ell \scriptin L} w_{\ell}$ by the vector condition
\begin{equation*}
    \sum_{k \scriptin K} A_{\ast k} = \sum_{\ell \scriptin L} A_{\ast \ell},
\end{equation*}
since their proofs only use that exchanging two uniform blocks of equal total weight leaves the value of the linear map $\bm{x} \mapsto A \bm{x}$ invariant.
We call such pairs $(K, L)$ \emph{balanced} (with respect to $A$).
Likewise, the grading construction carries over with classes indexed by column vectors, $K_{\bm{v}} \coloneqq \{i \in [n] \defcolon A_{\ast i} = \bm{v}\}$ for $\bm{v} \in \Z^{m}$, where $K_{\bm{0}}$ collects the variables not appearing in any linear form.
In particular, \autoref{corollary:AdjacencyIntraClassMoves} and \autoref{lemma:GradingFaithfulness} remain valid, since intra-class transpositions and flips of variables in $K_{\bm{0}}$ leave $A \bm{x}$ invariant.
What changes is the number of attainable sums in the dynamic program, which are now vectors in $\Z^{m}$.

\begin{theorem}\label{theorem:ConnectingMovesSystems}
    Let $\bm{d} \in \Z^{m}$, $A \in \Z^{m \times n}$, $(n, f, h_{\bm{d}, A}^{(n)})$ be a COP with row extents $W_{1}, \ldots, W_{m}$, and $c$ distinct non-zero columns, and let $\Pi \coloneqq \prod_{r = 1}^{m} (W_{r} + 1)$.
    Then, a family $\mathcal{E}' \subseteq \mathcal{E}$ with
    \begin{equation*}
        \abs{\mathcal{E}'} \leq n + \Pi (n - \abs{K_{\bm{0}}} + c),
    \end{equation*}
    so that $F_{\mathcal{E}'}$ is connected, can be constructed in time $O(m n \log(n) + \Pi (n - \abs{K_{\bm{0}}}) (n - \abs{K_{\bm{0}}} + m))$.
    Moreover, any two feasible solutions are joined in $F_{\mathcal{E}'}$ by a path containing at most $2 c$ moves from $\mathcal{E}' \setminus \mathcal{T}$, and a feasible solution can be computed within the same time bound whenever one exists.
\end{theorem}

\begin{figure*}[t]
    \centering
    \footnotesize
    \setlength{\tabcolsep}{2.5pt}
    \renewcommand{\arraystretch}{1.1}
    \begin{tabular}[t]{@{}cc>{$}l<{$}c>{$}l<{$}@{}}
        \multicolumn{5}{@{}l}{Stage $i = 1$: $\bm{v}_{1} = (1, 1)$, $K_{\bm{v}_{1}} = \{1,2\}$} \\
        \toprule
        $\bm{\sigma}$ & $q$ & C_{0}(\bm{\sigma}) \cup V_{1}(q) & $\bm{\sigma}'$ & (K_{1}, L_{1}) \\
        \midrule
        $(0,0)$ & $0$ & \emptyset & $(0,0)$ & \text{---} \\
        $(0,0)$ & $1$ & \{1\} & $(1,1)$ & \text{---} \\
        $(0,0)$ & $2$ & \{1,2\} & $(2,2)$ & \text{---} \\
        \bottomrule
    \end{tabular}
    \hfill
    \begin{tabular}[t]{@{}cc>{$}l<{$}c>{$}l<{$}@{}}
        \multicolumn{5}{@{}l}{Stage $i = 2$: $\bm{v}_{2} = (1, 2)$, $K_{\bm{v}_{2}} = \{3,4\}$} \\
        \toprule
        $\bm{\sigma}$ & $q$ & C_{1}(\bm{\sigma}) \cup V_{2}(q) & $\bm{\sigma}'$ & (K_{2}, L_{2}) \\
        \midrule
        $(0,0)$ & $0$ & \emptyset & $(0,0)$ & \text{---} \\
        $(0,0)$ & $1$ & \{3\} & $(1,2)$ & \text{---} \\
        $(0,0)$ & $2$ & \{3,4\} & $(2,4)$ & \text{---} \\
        $(1,1)$ & $0$ & \{1\} & $(1,1)$ & \text{---} \\
        $(1,1)$ & $1$ & \{1,3\} & $(2,3)$ & \text{---} \\
        $(1,1)$ & $2$ & \{1,3,4\} & $(3,5)$ & \text{---} \\
        $(2,2)$ & $0$ & \{1,2\} & $(2,2)$ & \text{---} \\
        $(2,2)$ & $1$ & \{1,2,3\} & $(3,4)$ & \text{---} \\
        $(2,2)$ & $2$ & \{1,2,3,4\} & $(4,6)$ & \text{---} \\
        \bottomrule
    \end{tabular}
    \hfill
    \begin{tabular}[t]{@{}cc>{$}l<{$}c>{$}l<{$}@{}}
        \multicolumn{5}{@{}l}{Stage $i = 3$: $\bm{v}_{3} = (1, 3)$, $K_{\bm{v}_{3}} = \{5\}$} \\
        \toprule
        $\bm{\sigma}$ & $q$ & C_{2}(\bm{\sigma}) \cup V_{3}(q) & $\bm{\sigma}'$ & (K_{3}, L_{3}) \\
        \midrule
        $(0,0)$ & $0$ & \emptyset & $(0,0)$ & \text{---} \\
        $(0,0)$ & $1$ & \{5\} & $(1,3)$ & \text{---} \\
        $(1,1)$ & $0$ & \{1\} & $(1,1)$ & \text{---} \\
        $(1,1)$ & $1$ & \{1,5\} & $(2,4)$ & \text{---} \\
        $(1,2)$ & $0$ & \{3\} & $(1,2)$ & \text{---} \\
        $(1,2)$ & $1$ & \{3,5\} & $(2,5)$ & \text{---} \\
        $(2,2)$ & $0$ & \{1,2\} & $(2,2)$ & \text{---} \\
        $(2,2)$ & $1$ & \{1,2,5\} & $(3,5)$ & \text{---} \\
        $(2,3)$ & $0$ & \{1,3\} & $(2,3)$ & \text{---} \\
        $(2,3)$ & $1$ & \{1,3,5\} & $(3,6)$ & \text{---} \\
        \rowcolor{Orange!15}[\tabcolsep][\tabcolsep] $(2,4)$ & $0$ & \{3,4\} & $(2,4)$ & (\{3,4\}, \{1,5\}) \\
        $(2,4)$ & $1$ & \{3,4,5\} & $(3,7)$ & \text{---} \\
        $(3,4)$ & $0$ & \{1,2,3\} & $(3,4)$ & \text{---} \\
        $(3,4)$ & $1$ & \{1,2,3,5\} & $(4,7)$ & \text{---} \\
        \rowcolor{Orange!15}[\tabcolsep][\tabcolsep] $(3,5)$ & $0$ & \{1,3,4\} & $(3,5)$ & (\{3,4\}, \{2,5\}) \\
        $(3,5)$ & $1$ & \{1,3,4,5\} & $(4,8)$ & \text{---} \\
        $(4,6)$ & $0$ & \{1,2,3,4\} & $(4,6)$ & \text{---} \\
        $(4,6)$ & $1$ & \{1,2,3,4,5\} & $(5,9)$ & \text{---} \\
        \bottomrule
    \end{tabular}
    \caption{\label{figure:DynamicProgramSystems}
        The dynamic program from the proof of \autoref{theorem:ConnectingMovesSystems} for the two rows $(1, 1, 1, 1, 1)$ and $(1, 1, 2, 2, 3)$ of $A$, i.e., the example of \autoref{figure:DynamicProgram} with an additional cardinality constraint, presented in the same way with the $i$th distinct non-zero column $\bm{v}_{i}$ of $A$ in place of $u_{i}$ and $\bm{\sigma}' \coloneqq \bm{\sigma} + q \bm{v}_{i}$.
        The additional row separates sums that coincide in \autoref{figure:DynamicProgram}, e.g., $\{3,4\}$ and $\{1,2,3\}$ now attain $(2,4)$ and $(3,4)$, respectively.
        Hence, only $2$ of the $30$ steps produce a move, and both moves are balanced with respect to both rows, whereas $\Sigma_{3}$ grows from $10$ to $16$ attainable sums.
    }
\end{figure*}

\begin{proof}
    We follow the proof of \autoref{theorem:ConnectingMoves} with two minor modifications.
    First, the classes $K_{\bm{v}}$ are formed by sorting the columns lexicographically in time $O(m n \log n)$, since every comparison costs $O(m)$.
    Second, every attainable sum $\bm{\sigma} \in \Sigma_{i}$ now lies in the box
    \begin{equation*}
        \bigtimes_{r = 1}^{m} \left[\sum_{i \scriptin [n] \defcolon A_{r i} < 0} A_{r i}, \sum_{i \scriptin [n] \defcolon A_{r i} > 0} A_{r i}\right] \cap \Z^{m},
    \end{equation*}
    which contains at most $\Pi$ integer points.
    Storing $\Sigma_{i}$ as a Boolean array over this box, indexed via a mixed-radix encoding, every one of the at most $\Pi (N + c)$ steps of the dynamic program costs $O(m)$ time for the vector addition and the index computation, where $N \coloneqq n - \abs{K_{\bm{0}}}$.
    Materialising the moves takes $O(N)$ time per move, i.e., $O(\Pi (N + c) N)$ time in total.
    Since $c \leq N$, the claimed running time follows, and the bound on $\abs{\mathcal{E}'}$ follows as well.
    The connectivity argument via the hub $\bm{g}(C_{c}(\bm{d}))$ and the bound on the path length carry over verbatim, since they only rely on the sums of the witnesses and on \autoref{lemma:GradingFaithfulness}.
    \autoref{figure:DynamicProgramSystems} illustrates the dynamic program for a system of two linear equality constraints.
    Lastly, a feasible solution exists if and only if $\bm{d} \in \Sigma_{c}$, given by the indicator vector of $C_{c}(\bm{d})$.
\end{proof}

Inspecting the proof, the right-hand side $\bm{d}$ enters at a single point only:
The dynamic program, and hence the family $\mathcal{E}'$, only depends on $A$, while $\bm{d}$ merely selects the hub $C_{c}(\bm{d})$.
Therefore, a single run of the dynamic program yields one family $\mathcal{E}'$ which connects every non-empty fibre of $A$ simultaneously, together with the set $\Sigma_{c}$ of attainable values and a hub $C_{c}(\bm{y})$ for every $\bm{y} \in \Sigma_{c}$.
This suggests replacing the right-hand side $\bm{d}$ by a set $D \subseteq \Z^{m}$ of admissible values, i.e., considering the Boolean function defined by
\begin{equation*}
    g_{D, A}^{(n)}(\bm{x}) = 1 \text{ if and only if } A \bm{x} \in D,
\end{equation*}
whose feasible set is the union of all fibres with values in $D$, so that $h_{\bm{d}, A}^{(n)} = g_{\{\bm{d}\}, A}^{(n)}$.
This covers virtually all constraints whose feasibility only depends on the values of a few linear forms of $\bm{x}$:
A single linear inequality constraint $t_{\tau, \bm{a}}^{(n)}$ is obtained for $m = 1$, $A = \bm{a}^{\top}$, and $D = \{y \in \Z \defcolon y \leq \tau\}$, interval constraints $\ell \leq \bm{a} \cdot \bm{x} \leq \tau$ for $D = [\ell, \tau] \cap \Z$, parity conditions on $\bm{a} \cdot \bm{x}$ for $D = 2 \Z$ or $D = 2 \Z + 1$, and conjunctions of such row-wise conditions for Cartesian products of the respective sets.
We do not impose any structure on $D$, but merely assume that membership in $D$ can be decided in time $T_{D}$.
For $\bm{v} \in \Z^{m}$, let $D - \bm{v} \coloneqq \{\bm{y} - \bm{v} \defcolon \bm{y} \in D\}$ denote the shifted set.
Then, the update rules of \autoref{subsection:HammingWeightConstraints} and \autoref{proposition:LinearInequalityReduction} carry over column-wise.

\begin{proposition}\label{proposition:LinearFormReduction}
    Let $n, m \in \N$, $A \in \Z^{m \times n}$, and $D \subseteq \Z^{m}$.
    For every $i \in [n]$ and every $j \in [n] \setminus \{i\}$,
    \begin{align*}
        g_{D, A}^{(n)}\hspace*{-3pt}\restriction_{x_{i} = 0} &= g_{D, A_{(i)}}^{(n - 1)}, \\
        g_{D, A}^{(n)}\hspace*{-3pt}\restriction_{x_{i} = 1} &= g_{D - A_{\ast i}, A_{(i)}}^{(n - 1)}, \\
        g_{D, A}^{(n)}\hspace*{-3pt}\restriction_{x_{i} = x_{j}} &= g_{D, A[j \leftarrow\hspace*{-5pt}+ A_{\ast i}]_{(i)}}^{(n - 1)}, \\
        g_{D, A}^{(n)}\hspace*{-3pt}\restriction_{x_{i} = \neg x_{j}} &= g_{D - A_{\ast i}, A[j \leftarrow\hspace*{-5pt}+ - A_{\ast i}]_{(i)}}^{(n - 1)},
    \end{align*}
    where $A_{(i)}$ denotes $A$ with its $i$th column removed and $A[j \leftarrow\hspace*{-10pt}+ A_{\ast i}]$ is $A$ with $A_{\ast i}$ added to its $j$th column.
\end{proposition}

\begin{proof}
    Write $A \bm{x} = \sum_{k = 1}^{n} A_{\ast k} x_{k}$.
    Substituting $x_{i} = 0$ removes the term $A_{\ast i} x_{i}$, while substituting $x_{i} = 1$ turns it into the constant $A_{\ast i}$, which is absorbed into the shifted set $D - A_{\ast i}$.
    Substituting $x_{i} = x_{j}$ merges $A_{\ast i} x_{i} + A_{\ast j} x_{j}$ into $(A_{\ast j} + A_{\ast i}) x_{j}$, and substituting $x_{i} = \neg x_{j}$ turns it into $A_{\ast i} + (A_{\ast j} - A_{\ast i}) x_{j}$, whose constant part is again absorbed into $D - A_{\ast i}$.
    Re-indexing the remaining variables yields the four identities.
\end{proof}

In particular, constraints on linear forms are closed under all four substitutions without any admissibility restriction.
Furthermore, by \autoref{lemma:SubstitutionWeightRestriction} applied to every row individually, each row extent $W_{r}$ is non-increasing along any substitution chain.

Moves within a fibre are taken care of by, e.g., the family $\mathcal{E}'$ constructed in \autoref{theorem:ConnectingMovesSystems}, so that it only remains to connect the feasible fibres to each other.
To this end, let
\begin{equation*}
    \mathcal{B} \coloneqq \{(K, L) \defcolon K, L \subseteq [n], K \cap L = \emptyset\} \supseteq \mathcal{E}
\end{equation*}
denote the set of all implementable (not necessarily feasibility-preserving) block exchanges.
Moves between distinct fibres change the value $A \bm{x}$ and are therefore not unconditionally feasibility-preserving, i.e., they are elements of the unbalanced set $\mathcal{B} \setminus \mathcal{E}$.
To still employ them, we condition these additional block exchanges on the feasibility of both bitstrings they exchange.
For $\bm{x} \in \bit^{n}$ and $(K, L) \in \mathcal{B}$, i.e., not necessarily balanced, let $\bm{x}^{K \vert L}$ denote the bitstring obtained from $\bm{x}$ by setting all bits indexed by $K$ to $1$ and all bits indexed by $L$ to $0$, and, for an arbitrary predicate $g : \bit^{n} \rightarrow \bit$, define
\begin{equation*}
    \eta_{K, L}(\bm{x}) \coloneqq g(\bm{x}^{K \vert L}) \land g(\bm{x}^{L \vert K}).
\end{equation*}
As $\eta_{K, L}$ only depends on the bits of $\bm{x}$ outside $K \cup L$, the controlled operation $U_{K, L}(\beta) \coloneqq \Lambda_{\eta_{K, L}}(e^{-i \beta \texttt{s}_{K}^{L}})$ is well-defined, with control qubits disjoint from the target qubits $K \cup L$, and feasibility-preserving, since it only acts if both exchanged bitstrings are feasible.
If $(K, L) \in \mathcal{E}$, both exchanged bitstrings lie in the same fibre, so that, within the feasible subspace, $U_{K, L}(\beta)$ reduces to its uncontrolled version.
Enlarging the set of admissible block exchanges from $\mathcal{E}$ to $\mathcal{B}$ in \autoref{definition:FibreGraph}, we obtain a suitable generalisation of the fibre graph for COPs with more general feasibility predicates $g_{D, A}^{(n)}$.
Given this notion of a fibre graph, \autoref{proposition:FibreGraphConnectivity} carries over to products of controlled block exchanges $U_{K, L}(\beta)$, $(K, L) \in \mathcal{B}'$, for any $\mathcal{B}' \subseteq \mathcal{B}$, while \autoref{lemma:ExistenceOfConnectedFibreGraph} carries over to the full set $\mathcal{B}$ (instead of $\mathcal{E}$).
Finally, we construct a connecting family for the generalised fibre graph from the balanced family of \autoref{theorem:ConnectingMovesSystems} by complementing it with a star of controlled unbalanced moves in $\mathcal{B} \setminus \mathcal{E}$ between the hubs of all feasible fibres.

\begin{theorem}\label{theorem:ConnectingMovesLinearForms}
    Let $A \in \Z^{m \times n}$, $D \subseteq \Z^{m}$, and $(n, f, g_{D, A}^{(n)})$ be a COP, and let $c$ and $\Pi$ be as in \autoref{theorem:ConnectingMovesSystems}.
    Then, a family $\mathcal{B}' \subseteq \mathcal{B}$ with
    \begin{equation*}
        \abs{\mathcal{B}'} \leq n + \Pi (n - \abs{K_{\bm{0}}} + c + 1),
    \end{equation*}
    and $\abs{\mathcal{B}' \setminus \mathcal{E}} \leq \Pi - 1$, so that $F_{\mathcal{B}'}$ is connected, can be constructed in time
    \begin{equation*}
        O\Big(m n \log(n) + \Pi \big((n - \abs{K_{\bm{0}}}) (n - \abs{K_{\bm{0}}} + m) + T_{D}\big)\Big).
    \end{equation*}
    Moreover, any two feasible solutions are joined in $F_{\mathcal{B}'}$ by a path containing at most $2$ moves from $\mathcal{B}' \setminus \mathcal{E}$ and $2 c$ moves from $(\mathcal{B}' \cap \mathcal{E}) \setminus \mathcal{T}$, and a feasible solution can be computed within the same time bound whenever one exists.
\end{theorem}

\begin{proof}
    Let $\mathcal{E}'$ be the family constructed in the proof of \autoref{theorem:ConnectingMovesSystems}, which only depends on $A$, and let $\Sigma_{c}$ and $H_{\bm{y}} \coloneqq C_{c}(\bm{y})$, $\bm{y} \in \Sigma_{c}$, be the attainable values and hubs computed alongside.
    Deciding membership in $D$ for all of $\Sigma_{c}$ determines the set $Y \coloneqq \Sigma_{c} \cap D$ of feasible values in $O(\Pi T_{D})$ time.
    \autoref{figure:HubsLinearForms} shows the hubs, the feasible values, and the resulting star for an example.
    If $Y = \emptyset$, there is no feasible solution and $\mathcal{B}' \coloneqq \emptyset$ trivially suffices.
    Otherwise, fix some $\bm{y}_{0} \in Y$ and define the star
    \begin{equation*}
        \mathcal{R} \coloneqq \{(H_{\bm{y}_{0}} \setminus H_{\bm{y}}, H_{\bm{y}} \setminus H_{\bm{y}_{0}}) \defcolon \bm{y} \in Y \setminus \{\bm{y}_{0}\}\},
    \end{equation*}
    each of whose at most $\Pi - 1$ moves is materialised in $O(N)$ time by unrolling back pointers, as in the proof of \autoref{theorem:ConnectingMoves}.
    Set $\mathcal{B}' \coloneqq \mathcal{E}' \cup \mathcal{R}$; the bounds on $\abs{\mathcal{B}'}$ and on the running time follow from \autoref{theorem:ConnectingMovesSystems}.

    Now let $\bm{x}$ be feasible and $\bm{y} \coloneqq A \bm{x} \in Y$.
    All moves in $\mathcal{E}'$ are balanced, so that both bitstrings exchanged by any of them lie in the same fibre.
    Hence, on the fibre of $\bm{y}$, which consists of feasible solutions only, every control of a move in $\mathcal{E}'$ evaluates to $1$, and \autoref{theorem:ConnectingMovesSystems}, applied to $h_{\bm{y}, A}^{(n)}$, provides a path from $\bm{x}$ to the indicator vector of $H_{\bm{y}}$ with at most $c$ moves from $\mathcal{E}' \setminus \mathcal{T}$.
    If $\bm{y} \neq \bm{y}_{0}$, the move $(K, L) \coloneqq (H_{\bm{y}_{0}} \setminus H_{\bm{y}}, H_{\bm{y}} \setminus H_{\bm{y}_{0}}) \in \mathcal{R}$ exchanges the indicator vectors of $H_{\bm{y}}$ and $H_{\bm{y}_{0}}$, both of which are feasible, so that its control evaluates to $1$ on them, as $\bm{y}, \bm{y}_{0} \in D$.
    Therefore, every feasible solution is connected to the indicator vector of $H_{\bm{y}_{0}}$ by a path containing at most $c$ moves from $\mathcal{E}' \setminus \mathcal{T}$ and one move from $\mathcal{R}$, which establishes connectivity and the bound on the path length.
    Lastly, a feasible solution exists if and only if $Y \neq \emptyset$, and is then given by the indicator vector of $H_{\bm{y}_{0}}$.
\end{proof}

\begin{figure}[t]
    \centering
    \footnotesize
    \setlength{\tabcolsep}{3pt}
    \renewcommand{\arraystretch}{1.1}
    \begin{tabular}{c>{$}l<{$}>{$}c<{$}>{$}l<{$}}
        \toprule
        $\bm{y} \in \Sigma_{3}$ & H_{\bm{y}} = C_{3}(\bm{y}) & \bm{y} \in D & (H_{\bm{y}_{0}} \setminus H_{\bm{y}}, H_{\bm{y}} \setminus H_{\bm{y}_{0}}) \\
        \midrule
        $(0, 0)$ & \emptyset & \text{no} & \text{---} \\
        $(1, 1)$ & \{1\} & \text{no} & \text{---} \\
        $(1, 2)$ & \{3\} & \text{no} & \text{---} \\
        $(1, 3)$ & \{5\} & \text{no} & \text{---} \\
        $(2, 2)$ & \{1,2\} & \text{no} & \text{---} \\
        \rowcolor{NavyBlue!10}[\tabcolsep][\tabcolsep] $(2, 3)$ & \{1,3\} & \text{yes} & \text{centre } \bm{y}_{0} \\
        \rowcolor{NavyBlue!10}[\tabcolsep][\tabcolsep] $(2, 4)$ & \{1,5\} & \text{yes} & (\{3\}, \{5\}) \\
        \rowcolor{NavyBlue!10}[\tabcolsep][\tabcolsep] $(2, 5)$ & \{3,5\} & \text{yes} & (\{1\}, \{5\}) \\
        $(3, 4)$ & \{1,2,3\} & \text{no} & \text{---} \\
        $(3, 5)$ & \{1,2,5\} & \text{no} & \text{---} \\
        $(3, 6)$ & \{1,3,5\} & \text{no} & \text{---} \\
        $(3, 7)$ & \{3,4,5\} & \text{no} & \text{---} \\
        $(4, 6)$ & \{1,2,3,4\} & \text{no} & \text{---} \\
        $(4, 7)$ & \{1,2,3,5\} & \text{no} & \text{---} \\
        $(4, 8)$ & \{1,3,4,5\} & \text{no} & \text{---} \\
        $(5, 9)$ & \{1,2,3,4,5\} & \text{no} & \text{---} \\
        \bottomrule
    \end{tabular}
    \caption{\label{figure:HubsLinearForms}
        The final stage of the construction in the proof of \autoref{theorem:ConnectingMovesLinearForms} for the matrix $A$ of \autoref{figure:DynamicProgramSystems} and right-hand side $D = \{2\} \times \{3, 4, \ldots\}$, i.e., $\bm{x}$ has Hamming weight $2$ and satisfies $(1, 1, 2, 2, 3) \cdot \bm{x} \geq 3$.
        The dynamic program of \autoref{figure:DynamicProgramSystems} is reused unchanged and yields a hub $H_{\bm{y}}$ for every attainable value $\bm{y} \in \Sigma_{3}$, of which $D$ merely selects the feasible values $Y$ (shaded).
        With centre $\bm{y}_{0} = (2, 3)$, the star $\mathcal{R}$ consists of the 2 controlled moves in the last column, which connect the hubs of all feasible fibres.
    }
\end{figure}

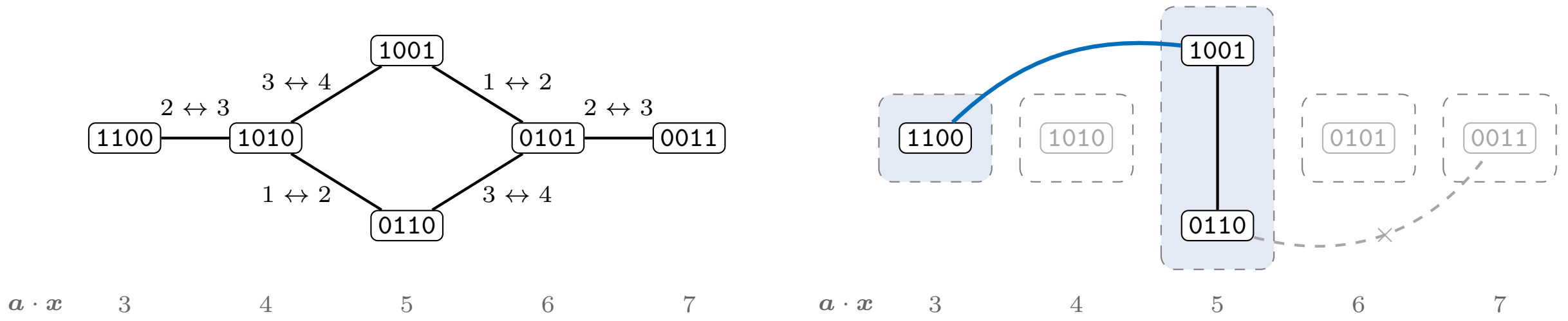
\begin{figure*}[t]
    \centering
    \tikzset{
        bs/.style={draw, rounded corners=2pt, inner sep=2pt, fill=white, font=\footnotesize\ttfamily},
        infeasible/.style={bs, draw=black!30, text=black!35},
        fibre/.style={draw=black!50, dashed, rounded corners=4pt},
        feasiblefibre/.style={fibre, fill=NavyBlue!10},
        xy/.style={thick},
        controlled/.style={very thick, NavyBlue},
        blocked/.style={thick, dashed, black!35},
    }
    \begin{tikzpicture}[x=1.45cm, y=0.9cm]
        \node[bs] (a) at (0, 0) {1100};
        \node[bs] (b) at (1, 0) {1010};
        \node[bs] (c) at (2, 1) {1001};
        \node[bs] (d) at (2, -1) {0110};
        \node[bs] (e) at (3, 0) {0101};
        \node[bs] (f) at (4, 0) {0011};
        \draw[xy] (a) -- node[above=3pt, font=\scriptsize] {$2 \leftrightarrow 3$} (b);
        \draw[xy] (b) -- node[above left, font=\scriptsize, inner sep=1pt] {$3 \leftrightarrow 4$} (c);
        \draw[xy] (b) -- node[below left, font=\scriptsize, inner sep=1pt] {$1 \leftrightarrow 2$} (d);
        \draw[xy] (c) -- node[above right, font=\scriptsize, inner sep=1pt] {$1 \leftrightarrow 2$} (e);
        \draw[xy] (d) -- node[below right, font=\scriptsize, inner sep=1pt] {$3 \leftrightarrow 4$} (e);
        \draw[xy] (e) -- node[above=3pt, font=\scriptsize] {$2 \leftrightarrow 3$} (f);
        \foreach \v in {3, ..., 7} {\node[font=\scriptsize, black!60] at (\v - 3, -1.9) {$\v$};}
        \node[font=\scriptsize, black!60, anchor=east] at (-0.35, -1.9) {$\bm{a} \cdot \bm{x}$};
    \end{tikzpicture}
    \hspace*{0.6cm}
    \begin{tikzpicture}[x=1.45cm, y=0.9cm]
        \draw[feasiblefibre] (-0.4, -0.5) rectangle (0.4, 0.5);
        \draw[fibre] (0.6, -0.5) rectangle (1.4, 0.5);
        \draw[feasiblefibre] (1.6, -1.5) rectangle (2.4, 1.5);
        \draw[fibre] (2.6, -0.5) rectangle (3.4, 0.5);
        \draw[fibre] (3.6, -0.5) rectangle (4.4, 0.5);
        \node[bs] (a) at (0, 0) {1100};
        \node[infeasible] (b) at (1, 0) {1010};
        \node[bs] (c) at (2, 1) {1001};
        \node[bs] (d) at (2, -1) {0110};
        \node[infeasible] (e) at (3, 0) {0101};
        \node[infeasible] (f) at (4, 0) {0011};
        \draw[xy] (c) -- (d);
        \draw[controlled] (a) to[bend left=25] (c);
        \draw[blocked] (d) to[bend right=35] node[pos=0.5, solid, font=\footnotesize] {$\times$} (f);
        \foreach \v in {3, ..., 7} {\node[font=\scriptsize, black!60] at (\v - 3, -1.9) {$\v$};}
        \node[font=\scriptsize, black!60, anchor=east] at (-0.35, -1.9) {$\bm{a} \cdot \bm{x}$};
    \end{tikzpicture}
    \caption{\label{figure:FibreGraphLinearForms}
        Left: fibre graph $F_{\mathcal{E}'}$ of the Hamming weight constraint $h_{2}^{(4)}$, where $\mathcal{E}' = \mathcal{T}$ consists of the adjacency swaps $i \leftrightarrow i + 1$, i.e., the chain of \texttt{XY} mixers.
        The bitstrings are arranged according to the value of the linear form $\bm{a} \cdot \bm{x}$ with $\bm{a} = (1, 2, 3, 4)$.
        Right: controlled fibre graph $F_{\mathcal{B}'}$ of $g_{D, A}^{(4)}$ with rows $A_{1 \ast} = (1, 1, 1, 1)$ and $A_{2 \ast} = \bm{a}$, and $D = \{2\} \times \{3, 5\}$, so that only two of the five fibres within the Hamming weight layer are feasible (shaded); infeasible bitstrings are greyed out.
        As all columns of $A$ are distinct, $\mathcal{T} = \emptyset$, and the dynamic program of \autoref{theorem:ConnectingMovesSystems} yields the single balanced move $(\{2, 3\}, \{1, 4\})$ (black), which connects the fibre $\bm{y} = (2, 5)$ without any control.
        The star of \autoref{theorem:ConnectingMovesLinearForms}, centred at $\bm{y}_{0} = (2, 5)$ with hub $H_{\bm{y}_{0}} = \{1, 4\}$, consists of the controlled transposition $(\{4\}, \{2\})$ to the hub $H_{(2, 3)} = \{1, 2\}$ (blue).
        Its control blocks the exchange of the feasible $0110$ with the infeasible $0011$ (dashed).
    }
\end{figure*}

\autoref{figure:FibreGraphLinearForms} illustrates the construction for a Hamming weight constraint with an additional, non-convex condition on a second linear form.
For $D = \{\bm{d}\}$, the star is empty and \autoref{theorem:ConnectingMovesLinearForms} reduces to \autoref{theorem:ConnectingMovesSystems}.
Only the moves of the star are controlled, and their controls, $\eta_{K, L}(\bm{x}) = 1$ if and only if $A \bm{x}^{K \vert L} \in D$ and $A \bm{x}^{L \vert K} \in D$, only depend on the linear forms, so that they can be evaluated coherently by accumulating $A \bm{x}$ into ancilla registers, as discussed for the unate mixer in \autoref{appendix:QuantumTreeGenerator}.
For a constant number of linear forms with polynomially bounded row extents and polynomial-time membership testing for $D$, the root problem and every subproblem admit polynomial-sized connecting families and feasible initial states, both constructible in polynomial time.
Forced variables are detected as in \autoref{proposition:ForcedVariablesEquality}, where the condition defining $Q_{i}$ becomes $\bm{\sigma} + q \bm{v}_{i} + \bar{\bm{\sigma}} \in D$ for some $\bm{\sigma} \in \Sigma_{i - 1}$ and $\bar{\bm{\sigma}} \in \bar{\Sigma}_{i + 1}$.
Since $\bm{\sigma}$ and $\bar{\bm{\sigma}}$ now range independently over up to $\Pi$ values each, these tests become quadratic rather than linear in $\Pi$, which still keeps them polynomial in this regime.
In general, however, $\Pi \geq 2^{m'}$, where $m'$ denotes the number of non-zero rows, so that the bound grows exponentially in the number of linear forms.
This cannot be avoided in full generality:
The exact cover problem~\cite{Karp1972ReducibilityAmongCombinatorialProblems} is expressed by $h_{\bm{1}, A}^{(n)}$, where $A \in \bit^{m \times n}$ has one row per ground element and one column per subset, so that all row extents are at most $n$.
Deciding feasibility is thus NP-complete, and, by the same argument as in \autoref{subsection:HammingWeightConstraints}, there is no efficient feasible initial state preparation for general systems of linear equality constraints unless $\mathrm{NP} \subseteq \mathrm{BQP}$, even for polynomially bounded row extents.
Similarly, the polynomial bound on the row extents cannot be dropped, since already a single linear form with unbounded coefficients expresses the subset sum problem, e.g., via an equality constraint or, equivalently, via the two inequality constraints $\bm{a} \cdot \bm{x} \leq \tau$ and $- \bm{a} \cdot \bm{x} \leq - \tau$.

\subsection{\label{subsection:EqualityAndInequalityConstraints}Equality and inequality constraints}

\begin{figure*}[t]
    \centering
    \footnotesize
    \setlength{\tabcolsep}{2.5pt}
    \renewcommand{\arraystretch}{1.1}
    \begin{tabular}[t]{@{}cc>{$}l<{$}cc>{$}l<{$}@{}}
        \multicolumn{6}{@{}l}{Stage $i = 1$: $K_{u_{1}} = (2, 1)$, $u_{1} = 1$} \\
        \toprule
        $\sigma$ & $q$ & C_{0}(\sigma) \cup V_{1}(q) & $\bm{a}(\cdot)$ & $\sigma'$ & (K_{1}, L_{1}) \\
        \midrule
        $0$ & $0$ & \emptyset & $0$ & $0$ & \text{---} \\
        $0$ & $1$ & \{2\} & $1$ & $1$ & \text{---} \\
        $0$ & $2$ & \{1,2\} & $5$ & $2$ & \text{---} \\
        \bottomrule
    \end{tabular}
    \hfill
    \begin{tabular}[t]{@{}cc>{$}l<{$}cc>{$}l<{$}@{}}
        \multicolumn{6}{@{}l}{Stage $i = 2$: $K_{u_{2}} = (4, 3)$, $u_{2} = 2$} \\
        \toprule
        $\sigma$ & $q$ & C_{1}(\sigma) \cup V_{2}(q) & $\bm{a}(\cdot)$ & $\sigma'$ & (K_{2}, L_{2}) \\
        \midrule
        $0$ & $0$ & \emptyset & $0$ & $0$ & \text{---} \\
        $0$ & $1$ & \{4\} & $2$ & $2$ & \text{---} \\
        $0$ & $2$ & \{3,4\} & $5$ & $4$ & \text{---} \\
        $1$ & $0$ & \{2\} & $1$ & $1$ & \text{---} \\
        $1$ & $1$ & \{2,4\} & $3$ & $3$ & \text{---} \\
        $1$ & $2$ & \{2,3,4\} & $6$ & $5$ & \text{---} \\
        \rowcolor{Orange!15}[\tabcolsep][\tabcolsep] $2$ & $0$ & \{1,2\} & $5$ & $2$ & (\{1,2\}, \{4\}) \\
        \rowcolor{Orange!15}[\tabcolsep][\tabcolsep] $2$ & $1$ & \{1,2,4\} & $7$ & $4$ & (\{1,2\}, \{3\}) \\
        $2$ & $2$ & \{1,2,3,4\} & $10$ & $6$ & \text{---} \\
        \bottomrule
    \end{tabular}
    \hfill
    \begin{tabular}[t]{@{}cc>{$}l<{$}cc>{$}l<{$}@{}}
        \multicolumn{6}{@{}l}{Stage $i = 3$: $K_{u_{3}} = (5)$, $u_{3} = 3$} \\
        \toprule
        $\sigma$ & $q$ & C_{2}(\sigma) \cup V_{3}(q) & $\bm{a}(\cdot)$ & $\sigma'$ & (K_{3}, L_{3}) \\
        \midrule
        $0$ & $0$ & \emptyset & $0$ & $0$ & \text{---} \\
        \rowcolor{Orange!15}[\tabcolsep][\tabcolsep] $0$ & $1$ & \{5\} & $5$ & $3$ & (\{5\}, \{2,4\}) \\
        $1$ & $0$ & \{2\} & $1$ & $1$ & \text{---} \\
        \rowcolor{Orange!15}[\tabcolsep][\tabcolsep] $1$ & $1$ & \{2,5\} & $6$ & $4$ & (\{2,5\}, \{3,4\}) \\
        $2$ & $0$ & \{4\} & $2$ & $2$ & \text{---} \\
        \rowcolor{Orange!15}[\tabcolsep][\tabcolsep] $2$ & $1$ & \{4,5\} & $7$ & $5$ & (\{5\}, \{2,3\}) \\
        $3$ & $0$ & \{2,4\} & $3$ & $3$ & \text{---} \\
        $3$ & $1$ & \{2,4,5\} & $8$ & $6$ & \text{---} \\
        $4$ & $0$ & \{3,4\} & $5$ & $4$ & \text{---} \\
        $4$ & $1$ & \{3,4,5\} & $10$ & $7$ & \text{---} \\
        $5$ & $0$ & \{2,3,4\} & $6$ & $5$ & \text{---} \\
        $5$ & $1$ & \{2,3,4,5\} & $11$ & $8$ & \text{---} \\
        \rowcolor{Orange!15}[\tabcolsep][\tabcolsep] $6$ & $0$ & \{1,2,3,4\} & $10$ & $6$ & (\{1,3\}, \{5\}) \\
        $6$ & $1$ & \{1,2,3,4,5\} & $15$ & $9$ & \text{---} \\
        \bottomrule
    \end{tabular}
    \caption{\label{figure:DynamicProgramEqualityInequality}
        The dynamic program from the proof of \autoref{theorem:ConnectingMovesEqualityInequality} for the weights $\bm{w} = (1, 1, 2, 2, 3)$ of \autoref{figure:DynamicProgram} and the inequality coefficients $\bm{a} = (4, 1, 3, 2, 5)$, presented in the same way with the additional column $\bm{a}(C_{i - 1}(\sigma) \cup V_{i}(q))$.
        Every class is now listed in ascending order of $\bm{a}$, so that $V_{i}(q)$ consists of its first $q$ elements in this order.
        The witness $C_{i}(\sigma')$ is the candidate of minimal $\bm{a}$-weight among all steps reaching $\sigma'$ (---), which need not be the first one, e.g., $C_{3}(3) = \{2, 4\}$ is attained only after $\{5\}$.
        Hence, every move $(K_{i}, L_{i})$ (shaded) satisfies $\bm{a}(L_{i}) \leq \bm{a}(K_{i})$, so that replacing $K_{i}$ by $L_{i}$ never increases $\bm{a} \cdot \bm{x}$.
        The dynamic program depends neither on $d$ nor on $\tau$; for $d = 4$, the initial state is the indicator vector of $C_{3}(4) = \{3, 4\}$, which minimises $\bm{a} \cdot \bm{x}$ subject to $\bm{w} \cdot \bm{x} = 4$.
    }
\end{figure*}

Lastly, we consider COPs whose feasibility predicate
\begin{equation*}
    g = h_{\bm{d}, A}^{(n)} \land t_{\tau, \bm{a}}^{(n)}
\end{equation*}
combines a system of linear equality constraints with a single linear inequality constraint.
By \autoref{proposition:LinearFormReduction} and \autoref{proposition:LinearInequalityReduction}, this class is closed under all four substitutions, and, by \autoref{lemma:SubstitutionWeightRestriction}, the row extents of $A$ are non-increasing along any substitution chain.
Evidently, $g = g_{D, A'}^{(n)}$ is a constraint on the $m + 1$ linear forms given by the rows of $A' \coloneqq \big(\begin{smallmatrix} A \\ \bm{a}^{\top} \end{smallmatrix}\big)$, with $D = \{\bm{d}\} \times \{y \in \Z \defcolon y \leq \tau\}$, so that \autoref{theorem:ConnectingMovesLinearForms} applies.
However, the additional row multiplies $\Pi$ by $W_{\bm{a}} + 1$, where $W_{\bm{a}} \coloneqq \sum_{i = 1}^{n} \abs{a_{i}}$, and contributes up to $W_{\bm{a}}$ moves between distinct fibres, so that efficiency requires polynomially bounded coefficients $\bm{a}$.
In contrast, a single linear inequality does not suffer from the subset sum obstruction, since its feasible set always contains the solutions minimising its left-hand side.
We now show that this allows us to dispense with the additional row entirely, so that the coefficients $\bm{a}$ only enter the construction through comparisons.

Unsurprisingly, neither of our two mixer constructions suffices on its own:
Balanced block exchanges may violate $t_{\tau, \bm{a}}^{(n)}$, and the controlled bit flips of the unate mixer may violate $h_{\bm{d}, A}^{(n)}$.
Hence, we employ the controlled block exchanges $U_{K, L}(\beta)$ of \autoref{subsection:LinearFormConstraints} with respect to $g$, noting that neither their definition nor that of the generalised fibre graph $F_{\mathcal{B}'}$, $\mathcal{B}' \subseteq \mathcal{B}$, relies on the specific form of the feasibility predicate.
For moves $(K, L) \in \mathcal{B}$ balanced with respect to $A$, both exchanged bitstrings satisfy $h_{\bm{d}, A}^{(n)}$ whenever one of them does, so that, on the feasible subspace, the control only needs to check $t_{\tau, \bm{a}}^{(n)}$ for both of them.
The family of \autoref{theorem:ConnectingMovesSystems} nevertheless does not suffice, since its connecting paths may pass through intermediate bitstrings satisfying $h_{\bm{d}, A}^{(n)}$ but not $t_{\tau, \bm{a}}^{(n)}$.
Instead, we steer these paths such that the inequality's left-hand side never increases.
For $\bm{a} \in \Z^{n}$ and $K \subseteq [n]$, we write $\bm{a}(K) \coloneqq \sum_{k \scriptin K} a_{k}$ for the sum of the inequality constraint's coefficients over a given index set.

\begin{theorem}\label{theorem:ConnectingMovesEqualityInequality}
    Let $\bm{d}\hspace*{-1pt} \in\hspace*{-1pt} \Z^{m}, A\hspace*{-1pt} \in\hspace*{-1pt} \Z^{m \times n}, \tau \in \Z, \bm{a} \in \Z^{n}$, $(n, f, h_{\bm{d}, A}^{(n)} \land t_{\tau, \bm{a}}^{(n)})$ be a COP, and let $c$ and $\Pi$ be as in \autoref{theorem:ConnectingMovesSystems}.
    Then, a family $\mathcal{B}' \subseteq \mathcal{B}$ of moves balanced with respect to $A$ with
    \begin{equation*}
        \abs{\mathcal{B}'} \leq n + \Pi (n - \abs{K_{\bm{0}}} + c),
    \end{equation*}
    so that $F_{\mathcal{B}'}$ is connected, can be constructed in time $O(m n \log(n) + \Pi (n - \abs{K_{\bm{0}}}) (n - \abs{K_{\bm{0}}} + m))$.
    On the feasible subspace, the controls reduce to the threshold conditions
    \begin{equation*}
        \eta_{K, L}(\bm{x}) = 1 \iff \hspace*{-8pt} \sum_{k \notin K \cup L} \hspace*{-6pt} a_{k} x_{k} + \max\{\bm{a}(K), \bm{a}(L)\} \leq \tau.
    \end{equation*}
    Moreover, any two feasible solutions are joined in $F_{\mathcal{B}'}$ by a path containing at most $2 c$ moves from $\mathcal{B}' \setminus \mathcal{T}$, and a feasible solution can be computed within the same time bound whenever one exists.
\end{theorem}

\begin{proof}
    First of all, the two bitstrings $\bm{x}^{K \vert L}$ and $\bm{x}^{L \vert K}$ agree outside $K \cup L$ and attain the left-hand sides $\sum_{k \notin K \cup L} a_{k} x_{k} + \bm{a}(K)$ and $\sum_{k \notin K \cup L} a_{k} x_{k} + \bm{a}(L)$, respectively, which shows the form of the controls.

    We follow the proof of \autoref{theorem:ConnectingMovesSystems} with two modifications, both ensuring that the value of $\bm{a} \cdot \bm{x}$ is non-increasing along the connecting paths.
    Let $\bm{v}_{1}, \ldots, \bm{v}_{c}$ denote the distinct non-zero columns of $A$ in the order processed by the dynamic program.
    First, we order the indices within every class $K_{\bm{v}}$ by ascending $\bm{a}$-values (ties broken arbitrarily) instead of by index, define $\mathcal{T}_{\bm{v}}$, $\bm{v} \neq \bm{0}$, as the adjacency swaps with respect to this order, and let $V_{i}(q)$ denote the $q$ indices of $K_{\bm{v}_{i}}$ with the smallest $\bm{a}$-values.
    Second, when computing $\Sigma_{i}$ from its predecessor $\Sigma_{i - 1}$, we store, for every $\bm{\sigma}' \in \Sigma_{i}$, a back pointer $(\bm{\sigma}, q)$ with $\bm{\sigma}' = \bm{\sigma} + q \bm{v}_{i}$ minimising the expression $\bm{a}(C_{i - 1}(\bm{\sigma})) + \bm{a}(V_{i}(q))$, and set $C_{i}(\bm{\sigma}') \coloneqq C_{i - 1}(\bm{\sigma}) \cup V_{i}(q)$.
    Inductively, $C_{i}(\bm{\sigma})$ minimises $\bm{a}(P)$ over all $P \subseteq K_{\bm{v}_{1}} \cup \cdots \cup K_{\bm{v}_{i}}$ with $\sum_{k \scriptin P} A_{\ast k} = \bm{\sigma}$:
    Every such set $P$ decomposes into $P \cap K_{\bm{v}_{i}}$ of some size $q$, whose $\bm{a}$-weight is at least $\bm{a}(V_{i}(q))$, and a remainder with column sum $\bm{\sigma} - q \bm{v}_{i}$, whose $\bm{a}$-weight is at least $\bm{a}(C_{i - 1}(\bm{\sigma} - q \bm{v}_{i}))$ by the induction hypothesis.
    The moves $(K_{i}(\bm{\sigma}, q), L_{i}(\bm{\sigma}, q))$ are defined as before for all sums $\bm{\sigma} \in \Sigma_{i - 1}$ and multiplicities $q \in [\abs{K_{\bm{v}_{i}}}] \cup \{0\}$, so that they are balanced with respect to $A$ and the bound on $\abs{\mathcal{B}'}$ is unchanged.
    \autoref{figure:DynamicProgramEqualityInequality} illustrates the modified dynamic program for the example of \autoref{figure:DynamicProgram}.
    Sorting the classes takes $O(n \log n)$ time, and storing $\bm{a}(C_{i}(\bm{\sigma}))$ alongside every witness and prefix sums of the sorted $\bm{a}$-values provides the $\bm{a}$-weight of every candidate in constant time, so that the dynamic program and the materialisation of the moves obey the bounds of \autoref{theorem:ConnectingMovesSystems}.

    Let $Z \coloneqq \{k \in K_{\bm{0}} \defcolon a_{k} < 0\}$ and, if $\bm{d} \in \Sigma_{c}$, let $\bm{x}^{*}$ be the indicator vector of $C_{c}(\bm{d}) \cup Z$.
    By construction, $\bm{x}^{*}$ minimises $\bm{a} \cdot \bm{x}$ over all $\bm{x}$ satisfying $h_{\bm{d}, A}^{(n)}$.
    Now let $\bm{x}$ be feasible.
    Using only balanced moves, we construct a path from $\bm{x}$ to $\bm{x}^{*}$ along which $\bm{a} \cdot \bm{x}$ is non-increasing.
    Then, every bitstring on this path also satisfies $t_{\tau, \bm{a}}^{(n)}$, since its left-hand side is bounded by $\bm{a} \cdot \bm{x} \leq \tau$.
    Hence, both endpoints of every traversed edge are feasible, so that the respective control evaluates to $1$ and the path lies in $F_{\mathcal{B}'}$.
    First, flip every bit $x_{k}$, $k \in K_{\bm{0}}$, which disagrees with the indicator vector of $Z$, each flip changing $\bm{a} \cdot \bm{x}$ by $-\abs{a_{k}}$.
    Second, as long as the ones within some class $K_{\bm{v}}$, $\bm{v} \neq \bm{0}$, do not occupy its first $m_{\bm{v}}(\bm{x})$ positions, there are two consecutive positions $p$, $p'$ with $x_{p} = 0$ and $x_{p'} = 1$, and the adjacency swap $p' \leftrightarrow p$ changes $\bm{a} \cdot \bm{x}$ by $a_{p} - a_{p'} \leq 0$.
    Since each such swap decreases the sum of the occupied positions, this process terminates in the indicator vector of $P_{0} \cup Z$, where $\bm{\sigma}_{i}$ and $P_{i}$ are defined as $\sigma_{i}$ and $A_{i}$ in the proof of \autoref{theorem:ConnectingMoves}, now with $a$-minimal witnesses.
    Third, as in the proof of \autoref{theorem:ConnectingMoves}, the move $(K_{i}(\bm{\sigma}_{i - 1}, m_{\bm{v}_{i}}), L_{i}(\bm{\sigma}_{i - 1}, m_{\bm{v}_{i}}))$ maps $P_{i - 1} \cup Z$ to $P_{i} \cup Z$ for every $i \in [c]$, and
    \begin{align*}
        \bm{a}(P_{i}) - \bm{a}(P_{i - 1}) &= \bm{a}(C_{i}(\bm{\sigma}_{i})) - \bm{a}(C_{i - 1}(\bm{\sigma}_{i - 1})) \\
        &\quad - \bm{a}(V_{i}(m_{\bm{v}_{i}})) \leq 0
    \end{align*}
    by minimality of $C_{i}(\bm{\sigma}_{i})$, since $C_{i - 1}(\bm{\sigma}_{i - 1}) \cup V_{i}(m_{\bm{v}_{i}})$ is a candidate for the sum $\bm{\sigma}_{i}$.
    As $P_{c} = C_{c}(\bm{d})$, the path ends in $\bm{x}^{*}$, and it contains at most $c$ moves from $\mathcal{B}' \setminus \mathcal{T}$.
    Hence, any two feasible solutions are joined via $\bm{x}^{*}$ in $F_{\mathcal{B}'}$ by a path containing at most $2 c$ such moves.
    Lastly, since $\bm{x}^{*}$ minimises $\bm{a} \cdot \bm{x}$ subject to $h_{\bm{d}, A}^{(n)}$, it is feasible whenever any feasible solution exists, and it is computed by unrolling at most $c$ back pointers.
\end{proof}

Together with the closure under all four substitutions and the non-increasing row extents, \autoref{theorem:ConnectingMovesEqualityInequality} yields RQAOA circuit families without any admissibility restriction:
The initial state is $\ket{\bm{x}^{*}}$, and the mixer is the product of controlled block exchanges $U_{K, L}(\beta)$, $(K, L) \in \mathcal{B}'$, fabricated anew for every subproblem.
In contrast to \autoref{theorem:ConnectingMovesLinearForms} applied to the $m + 1$ linear forms, the coefficients $\bm{a}$ are not required to be polynomially bounded, and all moves are balanced with respect to $A$, at the expense of a threshold control on every move.
These controls can be evaluated coherently by accumulating the weighted sum into an ancilla register, as discussed for the unate mixer in \autoref{appendix:QuantumTreeGenerator}.
For $i \in K_{\bm{0}}$, in particular, unateness of $t_{\tau, \bm{a}}^{(n)}$ simplifies the control of the flip $(\{i\}, \emptyset) \in \mathcal{T}_{\bm{0}}$ to
\begin{align*}
    \eta_{\{i\}, \emptyset}(\bm{x}) &= t_{\tau, \bm{a}}^{(n)}(\bm{x}[i \leftarrow 1]) \land t_{\tau, \bm{a}}^{(n)}(\bm{x}[i \leftarrow 0]) \\
    &= t_{\tau, \bm{a}}^{(n)}(\bm{x}[i \leftarrow b_{i}]),
\end{align*}
where $b_{i}$ denotes the unsafe value of $x_{i}$.
Recalling that $\texttt{s}_{\{i\}} = \texttt{X}_{i}$, we obtain $U_{\{i\}, \emptyset}(\beta) = \Lambda_{t\restriction_{x_{i} = b_{i}}}(e^{-i \beta \texttt{{\upshape X}}_{i}})$, i.e., the partial mixer of~\eqref{equation:UnateMixer}.
Hence, the construction subsumes both the generalised \texttt{XY} mixers of \autoref{subsection:LinearFormConstraints}, recovered for $\bm{a} = \bm{0}$, and the unate mixer of \autoref{subsection:SingleLinearInequalityConstraints}, recovered for the variables not appearing in any equality constraint.
For the cardinality-constrained knapsack problem, all variables form a single class at the root, so that the construction reduces to a chain of controlled \texttt{XY} mixers along the items sorted by ascending $a_{i}$; only subsequent binary substitutions split the variables into several classes.

Beyond a single inequality, no comparable guarantee can be expected in general.
Further inequalities can still be treated as additional linear forms via \autoref{theorem:ConnectingMovesLinearForms}, but the resulting exponential dependence on their number cannot be avoided, as discussed in \autoref{subsection:LinearFormConstraints}.
For instance, the independence constraints $x_{i} + x_{j} \leq 1$, $\{i, j\} \in E$, of a graph $G$ together with $\abs{\bm{x}} = d$ are feasible if and only if $G$ has an independent set of size $d$, which is NP-complete to decide~\cite{Karp1972ReducibilityAmongCombinatorialProblems}; hence, there is no efficient feasible initial state preparation unless $\mathrm{NP} \subseteq \mathrm{BQP}$, by the same argument as in \autoref{subsection:HammingWeightConstraints}.

\section{\label{section:NumericalExperiments}Numerical experiments}

We complement the circuit constructions of \autoref{section:CatalogueOfRQAOACircuitFamilies} by numerical proof-of-principle experiments on two constrained problem classes.
We demonstrate how feasible initial states and feasibility-preserving mixers are compiled for every subproblem along a chain of RQAOA-driven substitutions, and draw performance comparisons to the non-recursive setting and to the established, penalty-based treatment of constraints.
We simulate all circuits exactly and compare four methods on instances with up to $n = 24$ variables.

\subsection{\label{subsection:BenchmarkProblems}Benchmark problems}

First, as the prototypical representative for single linear inequalities, we consider the 0--1 knapsack problem, which asks for a subset of $n$ items with profits $\bm{p} \in \N^{n}$ and weights $\bm{a} \in \N^{n}$, which maximises the total profit subject to the capacity $\tau \in \N$.
As a minimisation problem, it is the COP $(n, f_{\text{KP}}, t_{\tau, \bm{a}}^{(n)})$ with objective function $f_{\text{KP}}(\bm{x}) \coloneqq -\bm{p} \cdot \bm{x}$ and the knapsack constraint of \autoref{subsection:SingleLinearInequalityConstraints}.
By \autoref{proposition:LinearInequalityReduction}, every subproblem is again subject to a single linear inequality constraint, so that the controlled bit flip mixer of \autoref{proposition:BitFlipMixersForUnateFeasibilityPredicates} applies in every recursion step.
We use strongly correlated instances from the KBDS generator~\cite{Williams202601KnapsackBenchmarkDataset}, i.e., $p_{i} = a_{i} + 100$ with weights $a_{i} \leq 1000$.
The capacities $\tau$ range between $40\,\%$ and $58\,\%$ of the total weight.
For larger numbers of items than we can consider with our state-vector-based simulation, these instances are known to be hard for classical branch-and-bound solvers~\cite{Pisinger2005WhereAreTheHardKnapsackProblems}.

Second, as a candidate exercising our constructions for a single, non-uniform equality constraint, we consider the vertex-weighted max-bisection problem, which asks, for an undirected graph $G = ([n], E)$ with vertex weights $\bm{v} \in \N^{n}$, for a cut of maximum size whose two shores carry the same weight, up to rounding.
As a minimisation problem, it is the COP $(n, f_{\text{MC}}, h_{d, \bm{v}}^{(n)})$, where $f_{\text{MC}}$ is the MaxCut objective function of \autoref{subsection:RecursiveQAOAStateOfTheArt} with unit edge weights and $d \coloneqq \lfloor \sum_{i = 1}^{n} v_{i} / 2 \rfloor$.
For $\bm{v} = \bm{1}$, it reduces to the ordinary max-bisection problem with a Hamming weight constraint; non-uniform vertex weights, in contrast, yield a linear equality constraint with weight extent $W = \sum_{i = 1}^{n} v_{i}$.
By \autoref{lemma:SubstitutionWeightRestriction}, every subproblem is again subject to a single linear equality constraint with weight extent at most $W$, so that \autoref{algorithm:ConnectingFamily} supplies a connecting family and a feasible initial state in every recursion step.
We use random 3-regular graphs, the standard test case for QAOA on MaxCut since its original proposal~\cite{Farhi2014AQuantumApproximateOptimizationAlgorithm}, with vertex weights $v_{i}$ drawn independently and uniformly at random from $[10]$, so that $W \leq 10 n$.

\subsection{\label{subsection:NumericalSetup}Setup}

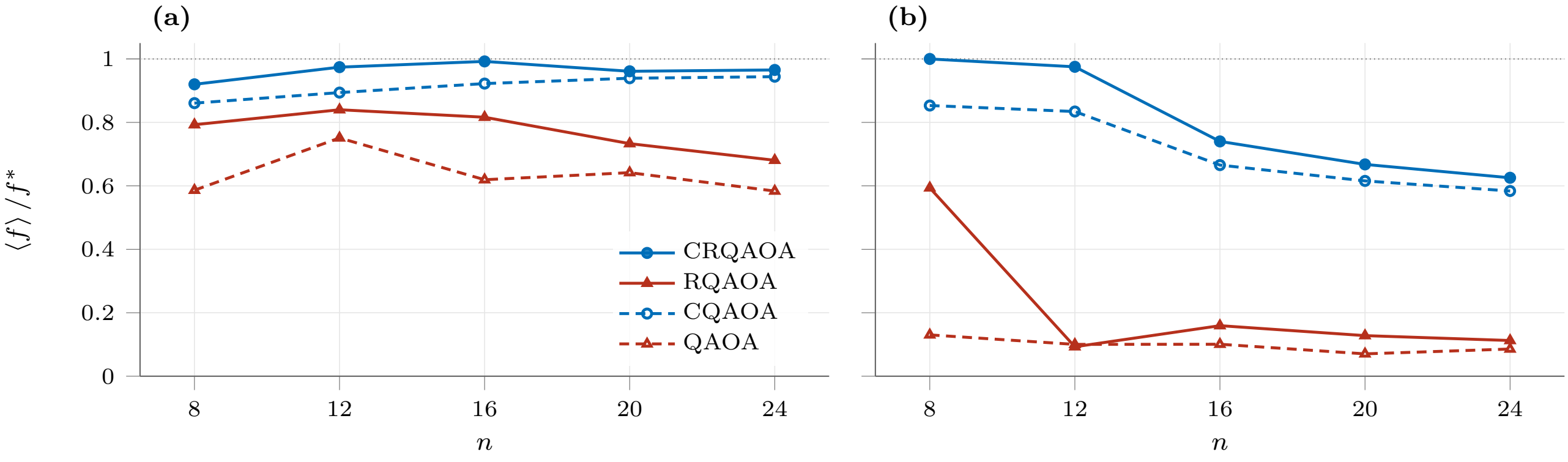
\begin{figure*}[t]
    \centering
    \begin{tikzpicture}
        \begin{groupplot}[
            numerics,
            group style={group size=2 by 1, horizontal sep=0.5cm, y descriptions at=edge left},
            scale only axis, width={(\textwidth - 1.6cm - 0.5cm) / 2}, height=3.6cm,
            xmin=6.5, xmax=25.5, xtick={8, 12, 16, 20, 24},
            xlabel={$n$}, ylabel={$\braket{f} / f^{*}$},
            ]
            \nextgroupplot[legend style={at={(0.97, 0.03)}, anchor=south east}]
            \addplot[crqaoa] table[x=n, y=CRQAOA, col sep=comma] {data/final_knapsack.csv};
            \addplot[rqaoa] table[x=n, y=RQAOA, col sep=comma] {data/final_knapsack.csv};
            \addplot[cqaoa] table[x=n, y=CQAOA, col sep=comma] {data/final_knapsack.csv};
            \addplot[qaoa] table[x=n, y=QAOA, col sep=comma] {data/final_knapsack.csv};
            \legend{CRQAOA, RQAOA, CQAOA, QAOA}
            \nextgroupplot
            \addplot[crqaoa] table[x=n, y=CRQAOA, col sep=comma] {data/final_bisection.csv};
            \addplot[rqaoa] table[x=n, y=RQAOA, col sep=comma] {data/final_bisection.csv};
            \addplot[cqaoa] table[x=n, y=CQAOA, col sep=comma] {data/final_bisection.csv};
            \addplot[qaoa] table[x=n, y=QAOA, col sep=comma] {data/final_bisection.csv};
        \end{groupplot}
        \node[font=\footnotesize\bfseries, anchor=south west] at (group c1r1.north west) {(a)};
        \node[font=\footnotesize\bfseries, anchor=south west] at (group c2r1.north west) {(b)};
    \end{tikzpicture}
    \caption{\label{figure:NumericsFinal}
        Final approximation ratio $\braket{f} / f^{*}$ of the four methods at depth $p = 2$ versus the number of variables $n$ for \textbf{(a)} knapsack and \textbf{(b)} vertex-weighted max-bisection instances, where infeasible outcomes of QAOA and RQAOA are scored as $0$.
        CRQAOA and RQAOA perform three substitutions; CQAOA and QAOA are optimised from four starts.
        The penalty-based methods suffer from small feasible-state probabilities, in particular for max-bisection.
    }
\end{figure*}

We compare four methods.
CRQAOA denotes the generalised RQAOA of \autoref{section:GeneralisedRecursiveQAOA}, which compiles a feasible initial state and a feasibility-preserving mixer for every subproblem.
The initial state is the all-zero string for knapsack and the indicator vector of the hub $C_{c}(d)$ from \autoref{algorithm:ConnectingFamily} for max-bisection.
CQAOA applies the same circuit family to the root problem only, i.e., without recursion.
The baselines RQAOA and QAOA are their unconstrained counterparts.
Irrespective of the problem, they start in $\ket{+}^{\otimes n}$ and use the standard mixer~\eqref{equation:StandardQAOAMixer}.
They account for the constraint by a penalty term in the phase separator, i.e., via
\begin{equation*}
    U_{\mathrm{P}}(\gamma) = e^{-i \gamma (H_{f} + \lambda H_{\text{pen}})}.
\end{equation*}
For max-bisection, the quadratic penalty $(\bm{v} \cdot \bm{x} - d)^{2}$ is weighted by $\lambda = \abs{E} + 1$, which exceeds every possible cut value.
For knapsack, we use the indicator penalty $[\bm{a} \cdot \bm{x} > \tau]$ with $\lambda$ given by $1.1 (\sum_{i = 1}^{n} p_{i} + 1)$ rounded to the nearest integer, which exceeds every total profit.
Its Pauli expansion is dense in general, but it avoids the slack variables of a quadratic penalty and can be applied directly in exact simulation.

For the knapsack problem in particular, several more sophisticated QAOA variants have been proposed, such as alternative penalty formulations~\cite{GrandRive2019KnapsackProblemVariantsOfQaoaForBatteryRevenueOptimisation}, resource-efficient QUBO formulations with fewer or no slack variables~\cite{Heuvel2026ResourceEfficientFormulationsOfTheKnapsackProblemForQuantumComputing}, warm-started variants with tailored mixers~\cite{Dam2021QuantumOptimizationHeuristicsWithAnApplicationToKnapsackProblems}, and Grover-mixer QAOA~\cite{Bartschi2020GroverMixersForQaoaShiftingComplexityFromMixerDesignToStatePreparation} with the QTG as state preparation~\cite{Christiansen2025QuantumTreeGeneratorImprovesQaoaStateOfTheArtForTheKnapsackProblem}.
For a mere proof of principle, we do not compare against these knapsack-specific methods, since our framework applies to entire constraint classes.
The natural references are the general QAOA and RQAOA, which share the circuit depth, the optimisation budget, and, for RQAOA, the recursion scheme with our methods.

All four methods use circuits of depth $p = 2$.
Since the initial state of CRQAOA and CQAOA is a computational basis state, the first phase separator only contributes a global phase, so that their circuits effectively comprise $p$ mixers and $p - 1$ phase separators.
The recursive methods perform $R = 3$ substitutions and thus fit four circuits, while the non-recursive methods fit a single circuit from four starts, so that all methods spend the same number of circuit fits.
The angles are optimised by L-BFGS with exact gradients.
A root circuit is initialised by fitting a circuit of depth $p - 1$ and extending its angles via the INTERP heuristic~\cite{Zhou2020QuantumApproximateOptimizationAlgorithmPerformanceMechanismAndImplementationOnNearTermDevices}; further starts are drawn uniformly at random.
Every subsequent recursion step is warm-started from the slightly perturbed optimal angles of the previous step.
Substitutions are selected by the rule of \autoref{algorithm:UnateRQAOA}, but over all pairs $i \neq j$ rather than the admissible pairs $\mathcal{A}$, since both constraint classes are closed under all four substitutions.
Before every recursion step, CRQAOA deletes forced variables as described in \autoref{subsection:UnateFeasibilityPredicates} and \autoref{subsection:HammingWeightConstraints}, respectively.

We measure the quality of a method by the approximation ratio $\braket{f} / f^{*}$, where $f^{*} < 0$ is the optimal objective value, computed by exhaustive enumeration, and where infeasible outcomes are scored as $0$.
For the recursive methods, we report $\braket{f}$ for the final circuit after $R$ substitutions, evaluated with respect to the root problem's objective function, i.e., including the contributions of all eliminated variables.
That is, we do not solve the final subproblem exhaustively as in \autoref{algorithm:UnateRQAOA}, which would trivialise the comparison for the considered problem sizes.
All circuits are simulated exactly as state vectors, where the constrained circuits only act on the feasible subspace $\mathcal{S}$.

\begin{figure*}[t]
    \centering
    \begin{tikzpicture}
        \begin{groupplot}[
            numerics,
            group style={group size=5 by 2, horizontal sep=0.25cm, vertical sep=0.35cm, y descriptions at=edge left, x descriptions at=edge bottom},
            scale only axis, width={(\textwidth - 1.6cm - 4 * 0.25cm) / 5}, height=2.6cm,
            xmin=-0.3, xmax=3.3, xtick={0, 1, 2, 3},
            ylabel={$\braket{f} / f^{*}$},
            ]
            \nextgroupplot
            \addplot[crqaoa] table[x=step, y=CRQAOA8, col sep=comma] {data/recursion_knapsack.csv};
            \addplot[rqaoa] table[x=step, y=RQAOA8, col sep=comma] {data/recursion_knapsack.csv};
            \addplot[cqaoa, mark=none] table[x=step, y=CQAOA8, col sep=comma] {data/recursion_knapsack.csv};
            \addplot[qaoa, mark=none] table[x=step, y=QAOA8, col sep=comma] {data/recursion_knapsack.csv};
            \nextgroupplot[legend style={at={(0.97, 0.03)}, anchor=south east}]
            \addplot[crqaoa] table[x=step, y=CRQAOA12, col sep=comma] {data/recursion_knapsack.csv};
            \addplot[rqaoa] table[x=step, y=RQAOA12, col sep=comma] {data/recursion_knapsack.csv};
            \addplot[cqaoa, mark=none] table[x=step, y=CQAOA12, col sep=comma] {data/recursion_knapsack.csv};
            \addplot[qaoa, mark=none] table[x=step, y=QAOA12, col sep=comma] {data/recursion_knapsack.csv};
            \legend{CRQAOA, RQAOA, CQAOA, QAOA}
            \nextgroupplot
            \addplot[crqaoa] table[x=step, y=CRQAOA16, col sep=comma] {data/recursion_knapsack.csv};
            \addplot[rqaoa] table[x=step, y=RQAOA16, col sep=comma] {data/recursion_knapsack.csv};
            \addplot[cqaoa, mark=none] table[x=step, y=CQAOA16, col sep=comma] {data/recursion_knapsack.csv};
            \addplot[qaoa, mark=none] table[x=step, y=QAOA16, col sep=comma] {data/recursion_knapsack.csv};
            \nextgroupplot
            \addplot[crqaoa] table[x=step, y=CRQAOA20, col sep=comma] {data/recursion_knapsack.csv};
            \addplot[rqaoa] table[x=step, y=RQAOA20, col sep=comma] {data/recursion_knapsack.csv};
            \addplot[cqaoa, mark=none] table[x=step, y=CQAOA20, col sep=comma] {data/recursion_knapsack.csv};
            \addplot[qaoa, mark=none] table[x=step, y=QAOA20, col sep=comma] {data/recursion_knapsack.csv};
            \nextgroupplot
            \addplot[crqaoa] table[x=step, y=CRQAOA24, col sep=comma] {data/recursion_knapsack.csv};
            \addplot[rqaoa] table[x=step, y=RQAOA24, col sep=comma] {data/recursion_knapsack.csv};
            \addplot[cqaoa, mark=none] table[x=step, y=CQAOA24, col sep=comma] {data/recursion_knapsack.csv};
            \addplot[qaoa, mark=none] table[x=step, y=QAOA24, col sep=comma] {data/recursion_knapsack.csv};
            \nextgroupplot
            \addplot[crqaoa] table[x=step, y=CRQAOA8, col sep=comma] {data/recursion_bisection.csv};
            \addplot[rqaoa] table[x=step, y=RQAOA8, col sep=comma] {data/recursion_bisection.csv};
            \addplot[cqaoa, mark=none] table[x=step, y=CQAOA8, col sep=comma] {data/recursion_bisection.csv};
            \addplot[qaoa, mark=none] table[x=step, y=QAOA8, col sep=comma] {data/recursion_bisection.csv};
            \nextgroupplot
            \addplot[crqaoa] table[x=step, y=CRQAOA12, col sep=comma] {data/recursion_bisection.csv};
            \addplot[rqaoa] table[x=step, y=RQAOA12, col sep=comma] {data/recursion_bisection.csv};
            \addplot[cqaoa, mark=none] table[x=step, y=CQAOA12, col sep=comma] {data/recursion_bisection.csv};
            \addplot[qaoa, mark=none] table[x=step, y=QAOA12, col sep=comma] {data/recursion_bisection.csv};
            \nextgroupplot
            \addplot[crqaoa] table[x=step, y=CRQAOA16, col sep=comma] {data/recursion_bisection.csv};
            \addplot[rqaoa] table[x=step, y=RQAOA16, col sep=comma] {data/recursion_bisection.csv};
            \addplot[cqaoa, mark=none] table[x=step, y=CQAOA16, col sep=comma] {data/recursion_bisection.csv};
            \addplot[qaoa, mark=none] table[x=step, y=QAOA16, col sep=comma] {data/recursion_bisection.csv};
            \nextgroupplot
            \addplot[crqaoa] table[x=step, y=CRQAOA20, col sep=comma] {data/recursion_bisection.csv};
            \addplot[rqaoa] table[x=step, y=RQAOA20, col sep=comma] {data/recursion_bisection.csv};
            \addplot[cqaoa, mark=none] table[x=step, y=CQAOA20, col sep=comma] {data/recursion_bisection.csv};
            \addplot[qaoa, mark=none] table[x=step, y=QAOA20, col sep=comma] {data/recursion_bisection.csv};
            \nextgroupplot
            \addplot[crqaoa] table[x=step, y=CRQAOA24, col sep=comma] {data/recursion_bisection.csv};
            \addplot[rqaoa] table[x=step, y=RQAOA24, col sep=comma] {data/recursion_bisection.csv};
            \addplot[cqaoa, mark=none] table[x=step, y=CQAOA24, col sep=comma] {data/recursion_bisection.csv};
            \addplot[qaoa, mark=none] table[x=step, y=QAOA24, col sep=comma] {data/recursion_bisection.csv};
        \end{groupplot}
        \foreach \c/\n in {1/8, 2/12, 3/16, 4/20, 5/24} {
            \node[font=\footnotesize, anchor=south] at ([yshift=2pt]group c\c r1.north) {$n = \n$};
        }
        \node[font=\footnotesize\bfseries, anchor=north east] at ([xshift=-0.75cm]group c1r1.north west) {(a)};
        \node[font=\footnotesize\bfseries, anchor=north east] at ([xshift=-0.75cm]group c1r2.north west) {(b)};
        \node[font=\footnotesize, anchor=north] at ([yshift=-0.45cm]group c3r2.south) {recursion step};
    \end{tikzpicture}
    \caption{\label{figure:NumericsRecursion}
        Approximation ratio $\braket{f} / f^{*}$ along the recursion for the instances of \autoref{figure:NumericsFinal}, i.e., \textbf{(a)} knapsack and \textbf{(b)} vertex-weighted max-bisection, with markers for CRQAOA and RQAOA at every recursion step and dashed lines for the non-recursive CQAOA and QAOA as reference.
    }
\end{figure*}
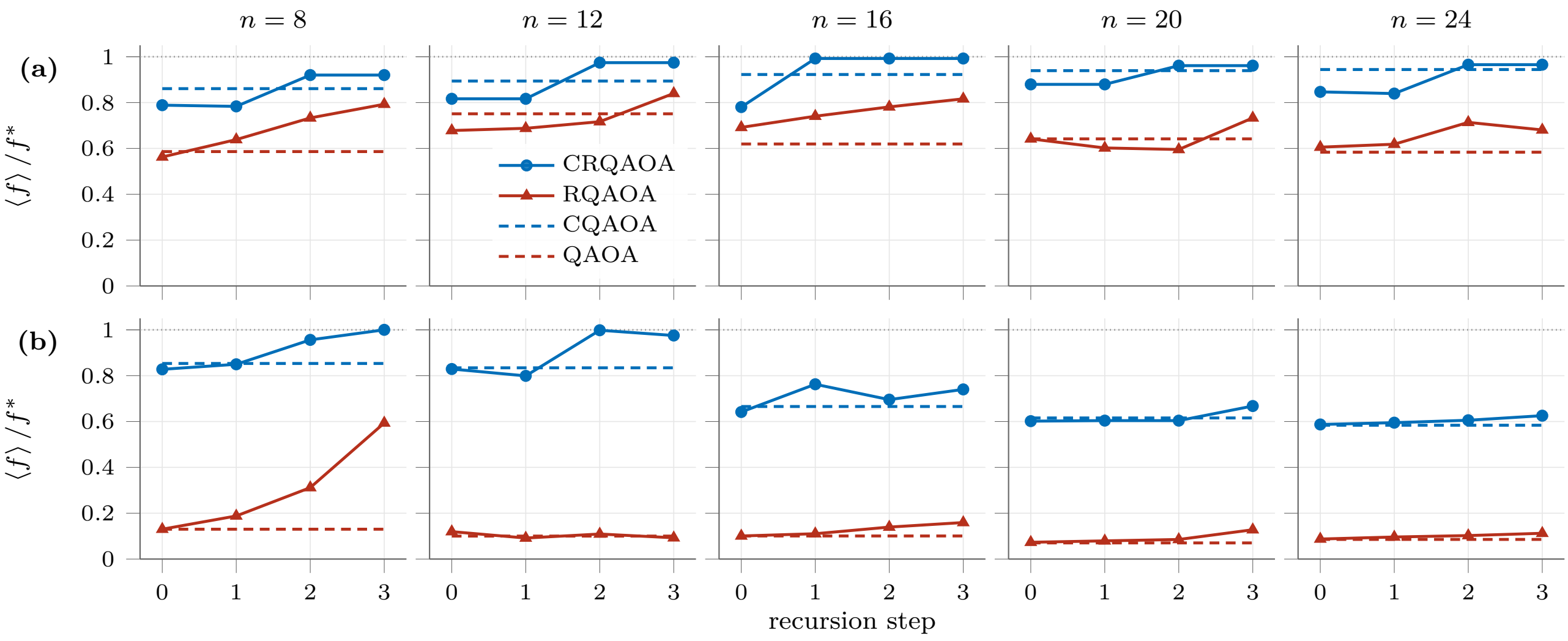

\subsection{\label{subsection:NumericalResults}Results and discussion}

\autoref{figure:NumericsFinal} summarises the final approximation ratios.
CRQAOA achieves the best ratio on all ten instances, followed by CQAOA, while both penalty-based methods fall clearly behind.
On knapsack, CRQAOA attains ratios between $0.92$ and $0.99$ and improves on CQAOA by $0.02$ to $0.08$.
On max-bisection, the improvement amounts to $0.04$ to $0.15$, and CRQAOA even prepares a superposition of optimal solutions for $n = 8$.
However, both constrained methods deteriorate with growing $n$ on max-bisection to about $0.6$, which is to be expected for a fixed circuit depth $p = 2$ and a feasible subspace growing exponentially with $n$.

The penalty-based methods suffer most on max-bisection, where QAOA and RQAOA attain ratios of at most $0.16$ for $n \geq 12$.
Their final states carry only $11\,\%$ to $24\,\%$ of their weight on feasible solutions, the sole exception being RQAOA for $n = 8$ with $69\,\%$.
On knapsack, in contrast, $80\,\%$ to $100\,\%$ of the weight is feasible, and the penalty-based methods lag behind CQAOA by $0.05$ to $0.36$ instead.
This suggests that the penalty succeeds in suppressing infeasible solutions on knapsack, but at the expense of a phase separator dominated by the penalty rather than by the objective function.

\autoref{figure:NumericsRecursion} resolves the recursive methods' performances into their individual recursion steps.
The first circuit of CRQAOA is fitted from a single start and therefore falls short of the best of CQAOA's four starts by $0.06$ to $0.14$ on knapsack, while both are on par on max-bisection.
The subsequent substitutions more than compensate for this:
CRQAOA improves by $0.04$ to $0.21$ over its first circuit and overtakes CQAOA on every instance.
On knapsack, the improvement is concentrated in a single recursion step, after which the ratio stagnates.
RQAOA improves on QAOA for every knapsack instance by $0.09$ to $0.21$, whereas on max-bisection, it stays close to QAOA except for $n = 8$.

The selected substitutions illustrate the necessity of the generalised framework.
On knapsack, CRQAOA exclusively fixes items, whereas RQAOA also enforces anti-correlations, which may render coefficients of the capacity constraint negative by \autoref{proposition:LinearInequalityReduction}.
On max-bisection, $5$ of CRQAOA's $15$ substitutions are binary, i.e., $x_{i} = x_{j}$ or $x_{i} = \neg x_{j}$, which merge vertex weights into sums or differences.
The resulting subproblems are not vertex-weighted max-bisection instances anymore, but they remain linear equality constrained with bounded weight extent, so that \autoref{algorithm:ConnectingFamily} readily fabricates their connecting families and feasible initial states.

These results should be read with the limitations of a proof of principle in mind.
Every problem size is represented by a single instance, all circuits have depth $p = 2$, and the simulations are exact and noiseless, i.e., expectation values are not estimated from samples.
Moreover, we compare the methods per circuit fit rather than per gate:
The threshold-controlled bit flips and the block exchanges of the connecting families are considerably more expensive than the transversal mixer~\eqref{equation:StandardQAOAMixer}, so that a resource-aware comparison remains open.
Nonetheless, the experiments demonstrate that the circuit families of \autoref{section:CatalogueOfRQAOACircuitFamilies} can be compiled in every recursion step, including subproblems outside the root problem's class, and that combining hard constraints with recursion consistently improves on both ingredients alone.

\section{\label{section:Conclusion}Conclusion and outlook}

We have generalised the recursive quantum approximate optimisation algorithm to a recursive quantum alternating operator ansatz, which handles hard constraints intrinsically in every recursion step.
By \autoref{lemma:LocalConsistency}, RQAOA-driven substitutions never render a subproblem infeasible as long as the state preparation is feasibility-preserving, so that the remaining task is to supply feasible initial states and feasibility-preserving, transition-providing mixers for every subproblem that may arise.
Our central design principle is to abandon the original, narrow problem class in favour of a broader class which is closed under all four substitutions $x_{i} \in \{0, 1, x_{j}, \neg x_{j}\}$, and to design mixer and state preparation circuits for this entire class, as summarised in \autoref{table:Catalogue} and \autoref{figure:ClassInclusion}.

Two mechanisms carry the catalogue.
For unate feasibility predicates, i.e., those which are monotone in every binary variable, single-qubit rotations controlled on the feasibility of the unsafe value connect every feasible solution to the safe bitstring; this subsumes the MIS mixer and, for single linear inequality constraints, which are closed under all four substitutions, generalises the quantum tree generator, originally proposed for the 0--1 knapsack problem.
For linear equality constraints, block exchanges between uniform blocks of equal total weight preserve feasibility, and their mixing properties are captured by the connectivity of the fibre graph, which reduces to the connectivity of the grading quotient.
A dynamic program over attainable subset sums constructs connecting families of polynomial size, together with a feasible initial state, whenever the weight extent is polynomially bounded.
Run in both directions, the same dynamic program also identifies all variables forced by the constraint, generalising the multiple-variable elimination for unate feasibility predicates.
This guarantees the efficient construction of mixers and state preparation circuits for all potential subproblems, since the weight extent can never increase along a chain of RQAOA substitutions.
By realising that the dynamic programming machinery is mostly oblivious to the right-hand side of the equality constraint(s) and viewing general constraints on linear forms as multi-valued equality constraints, we generalise the design even further to this broader class.
Here, a star of coherently controlled block exchanges connects the individual feasible fibres.
For the special case of a single linear inequality constraint on top of a system of linear equality constraints, we obtain a refined, threshold-controlled construction, which remains efficient even if the inequality constraint has unbounded coefficients.
The hardness results delimit the scope of these constructions: general weights, polynomially many equality constraints, or polynomially many inequality constraints rule out efficient feasible initial state preparation unless $\mathrm{NP} \subseteq \mathrm{BQP}$, and, for general weights, even the existence or construction of small connecting families fails.

While the mechanical compilation rules of \autoref{section:GeneralisedRecursiveQAOA} suffice for controlled bit flips, \autoref{appendix:InferenceFromParent} shows that inferring block exchanges from a fixed family can be exponentially expensive even for plain Hamming weight constraints, although fabrication from scratch remains cheap.
This has direct consequences for settings in which the admissible operations are fixed in advance, e.g., to a hardware-native gate set.

The numerical experiments provided in \autoref{section:NumericalExperiments} on knapsack and vertex-weighted max-bisection instances confirm that these constructions are viable in practice.
The resulting recursive scheme compiles valid circuits in every recursion step, also for subproblems outside the root problem's class, and outperforms both the non-recursive quantum alternating operator ansatz and penalty-based QAOA and RQAOA on every instance, the latter typically placing most of their weight on infeasible solutions for max-bisection.

Several directions remain open.
First, problems whose number of constraints grows with the problem size, such as the travelling salesperson problem with its $2 n$ overlapping Hamming weight constraints, fall outside our efficiency guarantees, since the number of attainable values grows exponentially with the number of linear forms.
Their specific structure, e.g., permutation constraints, may nonetheless admit tailored closed classes, for instance building upon symmetry-based mixers~\cite{Binkowski2025SymmetryBasedQuantumAlgorithmsForOpenShopSchedulingWithHardConstraints}.
The MIS exemplifies the resulting trade-off: within the unate framework, binary substitutions have to be restricted to admissible pairs, whereas the view as constraints on linear forms lifts this restriction at the expense of one linear form per edge.
Second, it remains to quantify the circuit resources of the presented mixers, in particular of the multi-controlled block exchanges and of the coherent threshold controls, and to assess how they evolve along typical substitution chains, which would also enable a resource-aware comparison beyond our per-fit budgets.
Third, our mixers satisfy Hadfield~\textit{et~al.}'s criteria of feasibility preservation and transition provision, but whether they additionally admit convergence guarantees for $p \to \infty$, e.g., via the conditions of~\cite{Binkowski2024ElementaryProofOfQaoaConvergence}, is open; this would require initial states which are ground states of the respective mixer Hamiltonians rather than single feasible computational basis states.
Fourth, the presented circuit families can be directly embedded into the QIRO framework~\cite{Finzgar2024QuantumInformedRecursiveOptimizationAlgorithms}, including its backtracking, and combined with warm-starting~\cite{Egger2021WarmStartingQuantumOptimization}; extending our proof-of-principle experiments to larger instances, noisy simulations, and quantum hardware, as well as comparisons with knapsack-tailored QAOA variants, are natural next steps.
Lastly, on the complexity-theoretic side, we are not aware of an explicit family of linear equality constraints, derandomising \autoref{proposition:ExponentialConnectingFamilies}.

\bibliographystyle{quantum}
\bibliography{bibliography}

\onecolumngrid

\clearpage

\appendix

\section{\label{appendix:QuantumTreeGenerator}Relation to the quantum tree generator}

In this appendix, we relate the unate mixer~\eqref{equation:UnateMixer} for the 0--1 knapsack constraint $t_{\tau, \bm{a}}^{(n)}$ with item weights $a_{m} \geq 0$ and capacity $\tau \geq 0$ to the quantum tree generator~(QTG) introduced in~\cite{Wilkening2025AQuantumAlgorithmForSolving01KnapsackProblems}.
Besides the $n$-qubit path register, the QTG acts on a capacity register, initialised in the binary representation $\ket{\tau}$ of the capacity, and a profit register, initialised in $\ket{0}$.
It consists of $n$ layers $U_{n} \cdots U_{1}$, where the $m$th layer
\begin{itemize}
    \item[(a)] applies a rotation $R_{\texttt{Y}}(\theta_{m}) = e^{-i \theta_{m} \texttt{Y} / 2}$ to the $m$th path qubit, controlled on the capacity register's state being at least $a_{m}$,
    \item[(b)] subtracts $a_{m}$ from the capacity register, controlled on the $m$th path qubit, and
    \item[(c)] adds the profit of item $m$ to the profit register, controlled on the $m$th path qubit.
\end{itemize}
The angles $\theta_{m}$ are fixed in advance as functions of a bias hyperparameter and the $m$th bit of a reference solution~\cite{Wilkening2025AQuantumAlgorithmForSolving01KnapsackProblems}.
In the following, we write $\bm{p} \in \R^{n}$ for the profits and $V : \ket{\bm{x}} \mapsto \ket{\bm{x}} \ket{\tau - \bm{a} \cdot \bm{x}} \ket{\bm{p} \cdot \bm{x}}$ for the isometry attaching the remaining capacity and the total profit to a path.

\begin{proposition}\label{proposition:QTGAsUnateMixer}
    Let $\bm{a} \in \N^{n}$, $\tau \in \N$, and let $G$ denote the QTG with angles $\theta_{1}, \ldots, \theta_{n}$.
    Furthermore, let $\bm{b} = \bm{1}$ be the unsafe values of $t_{\tau, \bm{a}}^{(n)}$, $D \coloneqq \bigotimes_{m = 1}^{n} \texttt{S}_{m}$ with the phase gate $\texttt{S} = \ket{0}\hspace*{-3pt}\bra{0} + i \ket{1}\hspace*{-3pt}\bra{1}$, and
    \begin{equation*}
        U \coloneqq \Lambda_{t\restriction_{x_{n} = 1}}\big(e^{-i \theta_{n} \texttt{X}_{n} / 2}\big) \cdots \Lambda_{t\restriction_{x_{1} = 1}}\big(e^{-i \theta_{1} \texttt{X}_{1} / 2}\big).
    \end{equation*}
    Then, $G \ket{\bm{0}} \ket{\tau} \ket{0} = V D U \ket{\bm{0}}$.
\end{proposition}

\begin{proof}
    First, we show that the QTG acts on the path register as the product $U_{\texttt{Y}}$ of the partial mixers $\Lambda_{t\restriction_{x_{m} = 1}}(R_{\texttt{Y}}(\theta_{m}))$, $m = 1, \ldots, n$, in this order.
    We argue inductively that, after the first $m - 1$ layers, the joint state is of the form $V \ket{\psi_{m - 1}}$, where $\ket{\psi_{m - 1}}$ is supported on bitstrings $\bm{x}$ with $x_{m} = \cdots = x_{n} = 0$ and $\ket{\psi_{0}} = \ket{\bm{0}}$.
    For such $\bm{x}$, the capacity register holds $\tau - \sum_{j < m} a_{j} x_{j}$, so that the control in step (a) is satisfied if and only if
    \begin{equation*}
        \sum_{j < m} a_{j} x_{j} + a_{m} = \sum_{j \neq m} a_{j} x_{j} + a_{m} \leq \tau \iff t_{\tau, \bm{a}}^{(n)}\hspace*{-3pt}\restriction_{x_{m} = 1}(\bm{x}) = 1,
    \end{equation*}
    where we have used that $x_{j} = 0$ for all $j > m$.
    Hence, step (a) maps $V \ket{\psi_{m - 1}}$ to $(\Lambda_{t\restriction_{x_{m} = 1}}(R_{\texttt{Y}}(\theta_{m})) \otimes \one) V \ket{\psi_{m - 1}}$, where the control is applied as a function of the path register only.
    Since $x_{m} = 0$ before step (a), steps (b) and (c) subtract $a_{m} x_{m}$ and add $p_{m} x_{m}$ for the new value of $x_{m}$, which restores the form $V \ket{\psi_{m}}$ with $\ket{\psi_{m}} \coloneqq \Lambda_{t\restriction_{x_{m} = 1}}(R_{\texttt{Y}}(\theta_{m})) \ket{\psi_{m - 1}}$, supported on bitstrings with $x_{m + 1} = \cdots = x_{n} = 0$.
    After $n$ layers, this yields $G \ket{\bm{0}} \ket{\tau} \ket{0} = V U_{\texttt{Y}} \ket{\bm{0}}$.
    Second, $R_{\texttt{Y}}(\theta) = \texttt{S}\, e^{-i \theta \texttt{X} / 2}\, \texttt{S}^{\dagger}$, since $\texttt{S} \texttt{X} \texttt{S}^{\dagger} = \texttt{Y}$.
    As $\texttt{S}_{m}$ is diagonal, it commutes with every control pattern, and it commutes with all rotations acting on qubits other than $m$.
    Therefore, $U_{\texttt{Y}} = D U D^{\dagger}$, and the claim follows from $D^{\dagger} \ket{\bm{0}} = \ket{\bm{0}}$.
\end{proof}

In other words, the path register of the QTG carries exactly the state produced by one application of the unate mixer with individual angles per qubit to the safe state $\ket{\neg \bm{b}} = \ket{\bm{0}}$, up to the diagonal phases $D$, which do not affect measurement statistics in the computational basis.
The capacity and profit registers remain entangled with the path register, as required for the subsequent amplitude amplification in~\cite{Wilkening2025AQuantumAlgorithmForSolving01KnapsackProblems}.

Note, however, that the QTG's incremental feasibility check relies on the path qubits $m + 1, \ldots, n$ still being in state $\ket{0}$ when item $m$ is processed.
On a general feasible input state, e.g., after a phase separator in a QAOA circuit, the check in step (a) would ignore the weights of all items $j > m$ already included, and could thus create infeasible bitstrings.
To use the QTG layout as a mixer on arbitrary inputs, the control has to account for all other items, which can be achieved as follows:
First, compute the remaining capacity $\tau - \bm{a} \cdot \bm{x}$ into an ancilla register.
Then, for each $m = 1, \ldots, n$, add $a_{m}$ controlled on the $m$th path qubit, so that the register holds $\tau - \sum_{j \neq m} a_{j} x_{j}$, apply the rotation controlled on the register's state being at least $a_{m}$, and subtract $a_{m}$ controlled on the (updated) $m$th path qubit.
Finally, uncompute the register.
This realises exactly the partial mixers $\Lambda_{t\restriction_{x_{m} = 1}}(e^{-i \beta \texttt{X}_{m}})$ of~\eqref{equation:UnateMixer} on the path register, for any input state.
The same construction applies to the subproblems of \autoref{proposition:LinearInequalityReduction} with coefficients of arbitrary sign, where the control checks $\tau' - \sum_{j \neq m} a'_{j} x_{j} \geq a'_{m} b_{m}$, i.e., $t\!\restriction_{x_{m} = b_{m}}(\bm{x}) = 1$.

\section{\label{appendix:ProblematicInstances}Problematic instances for general weights}

By \autoref{theorem:ConnectingMoves}, a polynomially bounded weight extent guarantees both a polynomial-sized connecting family $\mathcal{E}'$ and an efficient routine for constructing it.
In this appendix, we show that neither guarantee extends to general weights.
First, there are COPs for which every connecting family is exponentially large (\autoref{proposition:ExponentialConnectingFamilies}).
Second, there are COPs for which a connecting family consisting of a single move exists, but constructing any polynomial-sized connecting family is NP-hard under randomised reductions (\autoref{proposition:HardConnectingFamilies}).

Throughout, let $S$ denote the set of feasible solutions of the COP $(n, f, h_{d, \bm{w}}^{(n)})$.
Both results rest on the observation that each move $(K, L) \in \mathcal{E}$ changes every bitstring it acts on non-trivially by the same vector $\pm (\bm{1}_{K} - \bm{1}_{L})$, where $\bm{1}_{K} \in \bit^{n}$ denotes the indicator vector of $K \subseteq [n]$.

\subsection{\label{appendix:ExponentialConnectingFamilies}Exponentially large connecting families}

We call four bitstrings $\bm{x}, \bm{x}', \bm{y}, \bm{y}' \in S$ a \emph{parallelogram} in $S$ if $\bm{x} - \bm{x}' = \bm{y} - \bm{y}' \neq \bm{0}$ and $\{\bm{x}, \bm{x}'\} \neq \{\bm{y}, \bm{y}'\}$.

\begin{lemma}\label{lemma:ParallelogramFree}
    Let $\bm{w} \in \Z^{n}$, $d \in \Z$, and $(n, f, h_{d, \bm{w}}^{(n)})$ be a COP whose set of feasible solutions $S$ contains no parallelogram.
    Then, every $\mathcal{E}' \subseteq \mathcal{E}$ so that $F_{\mathcal{E}'}$ is connected satisfies $\abs{\mathcal{E}'} \geq \abs{S} - 1$.
\end{lemma}

\begin{proof}
    Let $(K, L) \in \mathcal{E}'$ and let $\{\bm{x}, \bm{x}'\}$ be an edge of $F_{\mathcal{E}'}$ constituted by $(K, L)$, labelled so that $\bm{x}_{K} = \bm{1}$ and $\bm{x}_{L} = \bm{0}$.
    Then, $\bm{x}' = \bm{x} - \bm{1}_{K} + \bm{1}_{L}$, i.e., $\bm{x} - \bm{x}' = \bm{1}_{K} - \bm{1}_{L}$.
    If $(K, L)$ constituted a second edge $\{\bm{y}, \bm{y}'\} \neq \{\bm{x}, \bm{x}'\}$, labelled analogously, then $\bm{x} - \bm{x}' = \bm{y} - \bm{y}'$ would yield a parallelogram in $S$.
    Hence, every element of $\mathcal{E}'$ constitutes at most one edge, and since a connected graph on $\abs{S}$ vertices has at least $\abs{S} - 1$ edges, the claim follows.
\end{proof}

It thus suffices to exhibit exponentially large, parallelogram-free sets of feasible solutions.
We do so via a probabilistic construction with positive weights.

\begin{proposition}\label{proposition:ExponentialConnectingFamilies}
    For every sufficiently large $n \in \N$, let $A \coloneqq \lceil 6 \cdot 6^{n / 3} \rceil$.
    Then, there exist $\bm{w} \in [A]^{n}$ and $d \in \Z$ so that the COP $(n, f, h_{d, \bm{w}}^{(n)})$ has
    \begin{equation*}
        \abs{S} \geq \frac{2^{n}}{2 (n A + 1)} = \Omega\Big(\frac{1}{n} \Big(\frac{2}{6^{1 / 3}}\Big)^{n}\Big)
    \end{equation*}
    feasible solutions and every $\mathcal{E}' \subseteq \mathcal{E}$ so that $F_{\mathcal{E}'}$ is connected satisfies $\abs{\mathcal{E}'} \geq \abs{S} - 1$.
\end{proposition}

\begin{proof}
    Draw every weight $w_{i}$, $i \in [n]$, independently and uniformly at random from $[A]$, and write $S_{d}$ for the set of feasible solutions for the right-hand side $d \in \Z$.
    Since $0 \leq \bm{w} \cdot \bm{x} \leq n A$ for all $\bm{x} \in \bit^{n}$, at most $n A + 1$ of the sets $S_{d}$ are non-empty, and they partition $\bit^{n}$.
    We will use the following elementary fact about the random weights:
    For $r \in [n]$ and linearly independent $\bm{z}^{(1)}, \ldots, \bm{z}^{(r)} \in \{-1, 0, 1\}^{n}$, it is
    \begin{equation}\label{equation:ProbabilityLinearConditions}
        \Pr\big[\bm{w} \cdot \bm{z}^{(1)} = \cdots = \bm{w} \cdot \bm{z}^{(r)} = 0\big] \leq A^{-r}.
    \end{equation}
    To see this, let $Z \in \Z^{r \times n}$ be the matrix with rows $\bm{z}^{(1)}, \ldots, \bm{z}^{(r)}$, so that the event in~\eqref{equation:ProbabilityLinearConditions} reads $Z \bm{w} = \bm{0}$.
    Since $Z$ has rank $r$, it has $r$ linearly independent columns, say indexed by $R \subseteq [n]$ with $\abs{R} = r$, and the corresponding square submatrix $Z_{R} \in \Z^{r \times r}$ is invertible.
    Writing $\bar{R} \coloneqq [n] \setminus R$, the linear system splits into $Z \bm{w} = Z_{R} \bm{w}_{R} + Z_{\bar{R}} \bm{w}_{\bar{R}}$.
    Now condition on any fixed value $\bm{\omega} \in [A]^{n - r}$ of $\bm{w}_{\bar{R}}$.
    Then, $Z \bm{w} = \bm{0}$ holds if and only if $Z_{R} \bm{w}_{R} = - Z_{\bar{R}} \bm{\omega}$, which, by invertibility of $Z_{R}$, is satisfied by exactly one $\bm{w}_{R} \in \R^{r}$.
    Since the weights are independent, $\bm{w}_{R}$ is, even conditioned on $\bm{w}_{\bar{R}} = \bm{\omega}$, uniformly distributed on the $A^{r}$ elements of $[A]^{r}$, so that it attains this single solution with probability at most $A^{-r}$.
    By the law of total probability,
    \begin{equation*}
        \Pr[Z \bm{w} = \bm{0}] = \sum_{\bm{\omega} \scriptin [A]^{n - r}} \Pr[\bm{w}_{\bar{R}} = \bm{\omega}]\, \Pr[Z \bm{w} = \bm{0} \mid \bm{w}_{\bar{R}} = \bm{\omega}] \leq A^{-r} \sum_{\bm{\omega} \scriptin [A]^{n - r}} \Pr[\bm{w}_{\bar{R}} = \bm{\omega}] = A^{-r}.
    \end{equation*}
    For $r = 1$, linear independence merely means $\bm{z}^{(1)} \neq \bm{0}$.

    Let $P$ denote the number of tuples $(\bm{x}, \bm{x}', \bm{y}, \bm{y}') \in (\bit^{n})^{4}$ forming a parallelogram in some $S_{d}$.
    Since the Boolean equation $x_{i} - x'_{i} = y_{i} - y'_{i}$ admits exactly six bit patterns $(x_{i}, x'_{i}, y_{i}, y'_{i})$ per coordinate $i \in [n]$, there are at most $6^{n}$ candidate tuples with $\bm{x} - \bm{x}' = \bm{y} - \bm{y}'$.
    By definition, such a candidate tuple forms a parallelogram in some $S_{d}$ if and only if $\bm{z} \coloneqq \bm{x} - \bm{x}' \neq \bm{0}$ and $\{\bm{x}, \bm{x}'\} \neq \{\bm{y}, \bm{y}'\}$.
    Since $\bm{x}$, $\bm{x}'$, $\bm{y}$, and $\bm{y}'$ are all in $S_{d}$, it must in particular hold that $\bm{w} \cdot \bm{z} = 0$ and $\bm{w} \cdot \bm{u} = 0$ for $\bm{u} \coloneqq \bm{x} - \bm{y}$.
    Here, $\bm{z}, \bm{u} \in \{-1, 0, 1\}^{n}$ are linearly independent:
    Since $\bm{z} \neq \bm{0}$, a linear dependence reads $\bm{u} = \lambda \bm{z}$ with $\lambda \in \{-1, 0, 1\}$.
    For $\lambda = 0$, this yields $\bm{y} = \bm{x}$ and hence $\bm{y}' = \bm{x}'$.
    For $\lambda = 1$, it yields $\bm{y} = \bm{x}'$ and $\bm{y}' = \bm{x} - 2 \bm{z}$, and for $\lambda = -1$, it yields $\bm{y}' = \bm{x}$ and $\bm{y} = \bm{x} + \bm{z}$; in both cases, some coordinate $q$ with $z_{q} \neq 0$ leaves the bit range.
    Hence, by~\eqref{equation:ProbabilityLinearConditions} with $r = 2$, $\mathbb{E}[P] \leq 6^{n} / A^{2}$.

    Let $\rho \coloneqq 2^{n} / (2 (n A + 1))$ and $H \coloneqq \{d \in \Z \defcolon \abs{S_{d}} \geq \rho\}$.
    The sets $S_{d}$ with $d \notin H$ contain fewer than $(n A + 1) \rho = 2^{n - 1}$ bitstrings in total, so that $\sum_{d \scriptin H} \abs{S_{d}} \geq 2^{n - 1}$.
    Moreover, $\sum_{d} \abs{S_{d}}^{2}$ counts the pairs $(\bm{x}, \bm{y})$ with $\bm{w} \cdot \bm{x} = \bm{w} \cdot \bm{y}$, so that~\eqref{equation:ProbabilityLinearConditions} with $r = 1$ and $\bm{z}^{(1)} = \bm{x} - \bm{y}$ for $\bm{x} \neq \bm{y}$ yields $\mathbb{E}[\sum_{d} \abs{S_{d}}^{2}] \leq 2^{n} + 4^{n} / A \leq 2 \cdot 4^{n} / A$, using $A \leq 2^{n}$ for sufficiently large $n$.
    By Markov's inequality, each of the two random variables $P$ and $\sum_{d} \abs{S_{d}}^{2}$ exceeds four times the above bound on its expectation with probability at most $1 / 4$.
    By the union bound, at least one of these two events occurs with probability at most $1 / 2$, so that with probability at least $1 / 2$, both $P < 4 \cdot 6^{n} / A^{2}$ and $\sum_{d} \abs{S_{d}}^{2} < 8 \cdot 4^{n} / A$.
    In particular, there exist weights satisfying both bounds.
    Fix such weights.
    By the Cauchy--Schwarz inequality,
    \begin{equation*}
        4^{n - 1} \leq \Big(\sum_{d \scriptin H} \abs{S_{d}}\Big)^{2} \leq \abs{H} \sum_{d \scriptin H} \abs{S_{d}}^{2} < \abs{H} \cdot 8 \cdot 4^{n} / A,
    \end{equation*}
    i.e., $\abs{H} > A / 32$.
    Every parallelogram lies in a single fibre, so at most $P < 4 \cdot 6^{n} / A^{2}$ fibres contain one.
    Since $A^{3} \geq 216 \cdot 6^{n} > 128 \cdot 6^{n}$, it is $A / 32 > 4 \cdot 6^{n} / A^{2}$, so that some $d \in H$ has a parallelogram-free fibre $S_{d}$ with $\abs{S_{d}} \geq \rho$.
    \autoref{lemma:ParallelogramFree} completes the proof.
\end{proof}

The weights constructed in \autoref{proposition:ExponentialConnectingFamilies} are exponentially large, in accordance with \autoref{corollary:PolyConnectingMoves}; their weight extent is $W \leq n A = O(n \cdot 6^{n / 3})$.
Expressed in terms of $W$, the lower bound reads $\abs{\mathcal{E}'} = \Omega(W^{\gamma} / \mathrm{poly}(n))$ with $\gamma = \log(2 / 6^{1 / 3}) / \log(6^{1 / 3}) \approx 0.16$, so that no bound of the form $\abs{\mathcal{E}'} \leq \mathrm{poly}(n) \cdot W^{o(1)}$ can hold in general; some polynomial dependence on $W$, as in \autoref{theorem:ConnectingMoves}, is unavoidable.
We are not aware of an explicit, i.e., deterministic, family of COPs with parallelogram-free fibres of exponential size.

\subsection{\label{appendix:HardConnectingFamilies}Connecting families that are hard to construct}

Even where a small connecting family exists, constructing one can be intractable.
We reduce from the \emph{unique subset sum problem}: given $\bm{a} \in \N^{m}$ and $t \in \N$ with the promise that at most one $\bm{x} \in \bit^{m}$ satisfies $\bm{a} \cdot \bm{x} = t$, decide whether such an $\bm{x}$ exists.
Its hardness follows from three ingredients.
First, counting the exact covers of a set system by $3$-element subsets is \#P-complete under parsimonious, i.e., solution-count-preserving, reductions~\cite{Hunt1998TheComplexityOfPlanarCountingProblems}; in particular, every satisfiability instance maps to an exact cover instance with the same number of solutions.
Second, Karp's reduction from exact cover to subset sum~\cite{Karp1972ReducibilityAmongCombinatorialProblems} encodes every subset as a number whose digits indicate the covered elements, in a base exceeding the number of subsets; since no carries occur, the subsets summing to the target correspond one-to-one to the exact covers, so that this reduction is parsimonious as well (see also~\cite{Faliszewski2009TheComplexityOfPowerIndexComparison}).
Third, the Valiant--Vazirani theorem~\cite{Valiant1986NpIsAsEasyAsDetectingUniqueSolutions} provides a randomised polynomial-time reduction from satisfiability to its restriction to instances with at most one solution: it adds random linear constraints (over $\mathbb{F}_{2}$) to a formula, which keeps an unsatisfiable formula unsatisfiable and leaves a satisfiable one with exactly one solution with probability $\Omega(1 / n)$.
Hence, a polynomial-time algorithm for satisfiability instances with at most one solution yields a randomised polynomial-time algorithm for satisfiability with bounded error, i.e., implies $\mathrm{NP} \subseteq \mathrm{BPP}$, where $\mathrm{BPP}$ denotes the class of decision problems solvable in polynomial time by a randomised algorithm erring with probability at most $1 / 3$ on every input.
Since the two parsimonious reductions map instances with at most one solution to such instances, the same conclusion holds for unique subset sum.
The inclusion $\mathrm{NP} \subseteq \mathrm{BPP}$ is widely believed to be false; among other consequences, it would imply that $\mathrm{NP}$ has polynomial-size circuits~\cite{Adleman1978TwoTheoremsOnRandomPolynomialTime} and hence collapse the polynomial hierarchy~\cite{Karp1980SomeConnectionsBetweenNonuniformAndUniformComplexityClasses}.
Unique subset sum is thus intractable unless $\mathrm{NP} \subseteq \mathrm{BPP}$.

\begin{proposition}\label{proposition:HardConnectingFamilies}
    Suppose there is an algorithm $\mathcal{A}$ which, given $\bm{w} \in \N^{n}$ and $d \in \N$ so that the COP $(n, f, h_{d, \bm{w}}^{(n)})$ has $\abs{S} \leq 2$, outputs in polynomial time some $\mathcal{E}' \subseteq \mathcal{E}$ so that $F_{\mathcal{E}'}$ is connected.
    Then, $\mathrm{NP} \subseteq \mathrm{BPP}$.
    At the same time, each of these COPs admits a connecting family with $\abs{\mathcal{E}'} \leq 1$.
\end{proposition}

\begin{proof}
    Let $(\bm{a}, t)$ be an instance of unique subset sum and assume w.l.o.g.~$a_{i} \geq 1$ for all $i \in [m]$ and $t \geq 1$.
    Set $n \coloneqq m + 1$, $\bm{w} \coloneqq (a_{1}, \ldots, a_{m}, t)$, and $d \coloneqq t$.
    A bitstring $(\bm{x}, x_{n})$ is feasible for $h_{d, \bm{w}}^{(n)}$ if and only if either $x_{n} = 1$ and $\bm{a} \cdot \bm{x} = 0$, i.e., $\bm{x} = \bm{0}$ by positivity, or $x_{n} = 0$ and $\bm{a} \cdot \bm{x} = t$.
    Hence, $S = \{\bm{e}_{n}\} \cup \{(\bm{x}, 0) \defcolon \bm{a} \cdot \bm{x} = t\}$ and $\abs{S} \leq 2$ by the promise.
    Run $\mathcal{A}$ on $(\bm{w}, d)$; its output $\mathcal{E}'$ has polynomial size, since it is written in polynomial time.
    For each $(K, L) \in \mathcal{E}'$, compute the image of $\bm{e}_{n}$ under the corresponding block exchange and check whether it is a feasible bitstring different from $\bm{e}_{n}$; accept if and only if some move passes this test.
    If $(\bm{a}, t)$ has a solution $\bm{x}^{*}$, then $S = \{\bm{e}_{n}, (\bm{x}^{*}, 0)\}$ and connectivity of $F_{\mathcal{E}'}$ forces some $(K, L) \in \mathcal{E}'$ to constitute the edge $\{\bm{e}_{n}, (\bm{x}^{*}, 0)\}$, so that we accept.
    Otherwise, $S = \{\bm{e}_{n}\}$ and no move can pass the test, so that we reject.
    This decides unique subset sum in polynomial time, hence $\mathrm{NP} \subseteq \mathrm{BPP}$.
    Lastly, for $S = \{\bm{x}, \bm{y}\}$ with $\bm{x} \neq \bm{y}$, the single move $(K, L)$ with $K \coloneqq \{i \in [n] \defcolon x_{i} = 1 \land y_{i} = 0\}$ and $L \coloneqq \{i \in [n] \defcolon x_{i} = 0 \land y_{i} = 1\}$ connects $S$ (cf.~the proof of \autoref{lemma:ExistenceOfConnectedFibreGraph}), and for $\abs{S} \leq 1$, $\mathcal{E}' = \emptyset$ suffices.
\end{proof}

Note that the COPs in \autoref{proposition:HardConnectingFamilies} come with the trivially feasible solution $\bm{e}_{n}$.
For general weights, however, even preparing some feasible initial state is at least as hard as subset sum; the intractability thus already arises before the mixer design.
Under a polynomially bounded weight extent, instead, the dynamic program in the proof of \autoref{theorem:ConnectingMoves} also yields a feasible solution, namely the indicator vector of the witness $C_{c}(d)$, whenever $d \in \Sigma_{c}$, i.e., whenever a feasible solution exists.

\section{\label{appendix:InferenceFromParent}Inferring connecting families from the current problem}

In \autoref{subsection:HammingWeightConstraints}, we fabricate the connecting family of every subproblem from scratch, using only the subproblem's weights.
A seemingly more economical alternative would be to infer it from the connecting family of the current problem, in the spirit of the compilation rules discussed in \autoref{section:GeneralisedRecursiveQAOA}.
In this appendix, we show that this approach is not viable in general---not even for the Hamming-weight root $\bm{w} = \bm{1}$, whose weight extent is $W = n$.
For simplicity, we only consider unary substitutions $x_{i} = b$, $b \in \bit$.
The feasible solutions of the resulting subproblem are then precisely the feasible solutions $\bm{x} \in S$ of the current problem with $x_{i} = b$, with the $i$th bit removed.
Accordingly, we identify operations on the subproblem with pairs $(K, L)$ of index subsets of $[n] \setminus \{i\}$.
For every pair $(K, L)$, we further write $\bm{v}_{K, L} \coloneqq \bm{1}_{K} - \bm{1}_{L}$ for its \emph{move vector}, i.e., the vector by which $\texttt{s}_{K}^{L}$ changes (up to sign) every bitstring it acts on non-trivially.

Now let $\mathcal{E}' \subseteq \mathcal{E}$ so that $F_{\mathcal{E}'}$ is connected.
The most straightforward attempt is to discard all $(K, L) \in \mathcal{E}'$ with $i \in K \cup L$, since whenever such an operation acts non-trivially on a feasible solution with $x_{i} = b$, it flips the $i$th bit and thus leaves the subproblem's feasible subspace.
The remaining operations are again feasibility-preserving for the subproblem, but, in general, their fibre graph is no longer connected.
Consider, e.g., the Hamming-weight root $\bm{w} = \bm{1}$ with $1 \leq d \leq n - 2$, the adjacency swaps $\mathcal{T}_{1}$ defined in~\eqref{equation:AdjacencySwap}, and the substitution $x_{i} = 0$ for some $1 < i < n$.
After discarding the two swaps involving $i$, the remaining swaps act either within $[i - 1]$ or within $\{i + 1, \ldots, n\}$ and hence preserve the number of ones in $[i - 1]$, which, however, varies among the subproblem's feasible solutions.
In this example, connectivity is easily restored by adding the swap $(\{i - 1\}, \{i + 1\})$.
The question is whether this is always possible at low cost.

In principle, missing operations can always be assembled from $\mathcal{E}'$.
Since $F_{\mathcal{E}'}$ is connected, any two $\bm{x}, \bm{y} \in S$ are joined by a path in $F_{\mathcal{E}'}$, so that
\begin{equation*}
    \bm{y} - \bm{x} = \sum_{(K, L) \scriptin \mathcal{E}'} c_{K, L}\, \bm{v}_{K, L}
\end{equation*}
for some $c_{K, L} \in \Z$.
Every operation constituting an edge in the subproblem's fibre graph connects two feasible solutions with $x_{i} = b$, so its move vector is of the form $\pm (\bm{y} - \bm{x})$ and thus an integer combination of the move vectors of $\mathcal{E}'$.
To assess the cost of such an assembly, we count the operations from $\mathcal{E}'$ it involves, including repetitions.
This covers the two natural ways of inference.
First, we may compile the subproblem's mixer $U$ directly from the gates of the current mixer, i.e., as a product of exponentials $e^{-i \beta \texttt{s}_{K}^{L}}$, $(K, L) \in \mathcal{E}'$, in arbitrary order and with arbitrary repetitions and angles.
As we have seen in the proof of \autoref{proposition:FibreGraphConnectivity}, each such exponential maps a computational basis state $\ket{\bm{x}}$ to a superposition of $\ket{\bm{x}}$ and, if $\texttt{s}_{K}^{L}$ acts non-trivially on $\bm{x}$, the computational basis state $\ket{\bm{x} \pm \bm{v}_{K, L}}$.
Expanding the product, $\braket{\bm{y} \vert U \vert \bm{x}}$ becomes a sum over sequences of computational basis states, along which every exponential either acts as the identity or applies its exchange once.
Hence, if $\braket{\bm{y} \vert U \vert \bm{x}} \neq 0$, there is such a sequence with $\bm{y} - \bm{x} = \sum_{(K, L)} c_{K, L} \bm{v}_{K, L}$, where $c_{K, L}$ denotes the net number of exchanges applied by $(K, L)$, and $U$ contains at least $\sum_{(K, L)} \abs{c_{K, L}}$ exponentials.
Note that the intermediate bitstrings in this sequence need not satisfy $x_{i} = b$.
Second, we may derive the subproblem's operations symbolically by iteratively adding up move vectors of $\mathcal{E}'$.
Representing such a derivation as a binary tree whose leaves are move vectors of $\mathcal{E}'$, the tree has at least $\sum_{(K, L)} \abs{c_{K, L}}$ leaves.
In both cases, the coefficients $c_{K, L}$ are uniquely determined if the move vectors of $\mathcal{E}'$ are linearly independent.
The following result shows that they can nonetheless be exponentially large.

\begin{figure*}[t]
    \centering
    \tikzset{
        locked/.style ={circle,draw=black,very thick,fill=black,minimum size=2.4mm,inner sep=0pt},
        free/.style   ={circle,draw=RoyalBlue!65!black,thick,fill=RoyalBlue!10,minimum size=2.2mm,inner sep=0pt},
        clsbox/.style ={rounded corners=2.2pt,draw=gray!55,thick,fill=gray!6},
        gapbox/.style ={rounded corners=2.2pt,draw=gray!45,thick,
            pattern={north east lines},pattern color=gray!45},
        staredge/.style={black!48,line width=0.85pt},
        g1edge/.style ={BrickRed!80!black,line width=1.5pt},
        g2edge/.style ={RoyalBlue!70!black,line width=1.0pt,dash pattern=on 3pt off 2pt},
        g3edge/.style ={ForestGreen!45!black,line width=1.0pt},
        axisline/.style={-{Latex[length=2mm]},gray!70,thick},
        crossmove/.style={-{Latex[length=2.6mm]},BrickRed!80!black,dash pattern=on 4pt off 2.4pt,line width=1.3pt},
    }
    \newcommand{\bitlab}[1]{{\tiny\texttt{#1}}}
    \begin{tikzpicture}[font=\small]

        \def\dx{1.9}     % horizontal spacing per unit of val
        \def\bw{0.55}    % half-width of a 3-state class box
        \def\bh{0.86}    % half-height of a 3-state class box
        \def\sw{0.4}    % half-width/height of a 1-state class box

        % =======================================================================
        % PANEL (a): ROOT FIBRE GRAPH  (all 8 classes, val = 0,...,7)
        % =======================================================================
        \begin{scope}[shift={(0,0)}]

            \node[font=\bfseries] at (-1.55,1.35) {(a)};

            % ---- class boxes -------------------------------------------------------
            \draw[clsbox] (0-\sw,-\sw) rectangle (0+\sw,\sw);
            \foreach \v in {1,2,3,4,5,6} {
                \draw[clsbox] (\v*\dx-\bw,-\bh) rectangle (\v*\dx+\bw,\bh);
            }
            \draw[clsbox] (7*\dx-\sw,-\sw) rectangle (7*\dx+\sw,\sw);

            % ---- nodes --------------------------------------------------------------
            % val 0
            \node[locked,label={below:\bitlab{010101}}] (n0) at (0,0) {};
            % val 1: hub=110001 (free), leaf-locked=011001, leaf=010011
            \node[free] (n1h) at (1*\dx, 0) {};
            \node[locked,label={above:\bitlab{010110}}] (n1L) at (1*\dx, 0.42) {};
            \node[free] (n1o) at (1*\dx,-0.42) {};
            % val 2: hub=110100 (free), leaf-locked=010110, leaf=011100
            \node[free] (n2h) at (2*\dx, 0) {};
            \node[locked,label={above:\bitlab{011001}}] (n2L) at (2*\dx, 0.42) {};
            \node[free] (n2o) at (2*\dx,-0.42) {};
            % val 3: hub=locked=011010, leaf=110010, leaf=111000
            \node[locked] (n3h) at (3*\dx, 0) {};
            \node[free] (n3a) at (3*\dx, 0.42) {};
            \node[free] (n3b) at (3*\dx,-0.42) {};
            \node[font=\tiny,align=center] at (3*\dx,-\bh-0.16) {\bitlab{011010}};
            % val 4: hub=locked=100101, leaf=001101, leaf=000111
            \node[locked] (n4h) at (4*\dx, 0) {};
            \node[free] (n4a) at (4*\dx, 0.42) {};
            \node[free] (n4b) at (4*\dx,-0.42) {};
            \node[font=\tiny,align=center] at (4*\dx,-\bh-0.16) {\bitlab{100101}};
            % val 5: hub=001011 (free), leaf-locked=101001, leaf=100011
            \node[free] (n5h) at (5*\dx, 0) {};
            \node[locked,label={above:\bitlab{100110}}] (n5L) at (5*\dx, 0.42) {};
            \node[free] (n5o) at (5*\dx,-0.42) {};
            % val 6: hub=001110 (free), leaf-locked=100110, leaf=101100
            \node[free] (n6h) at (6*\dx, 0) {};
            \node[locked,label={above:\bitlab{101001}}] (n6L) at (6*\dx, 0.42) {};
            \node[free] (n6o) at (6*\dx,-0.42) {};
            % val 7
            \node[locked,label={below:\bitlab{101010}}] (n7) at (7*\dx,0) {};

            % ---- star edges (within classes) -----------------------------------------
            \draw[staredge] (n1h)--(n1L); \draw[staredge] (n1h)--(n1o);
            \draw[staredge] (n2h)--(n2L); \draw[staredge] (n2h)--(n2o);
            \draw[staredge] (n3h)--(n3a); \draw[staredge] (n3h)--(n3b);
            \draw[staredge] (n4h)--(n4a); \draw[staredge] (n4h)--(n4b);
            \draw[staredge] (n5h)--(n5L); \draw[staredge] (n5h)--(n5o);
            \draw[staredge] (n6h)--(n6L); \draw[staredge] (n6h)--(n6o);

            % ---- g3 edges (6) ---------------------------------------------------------
            \draw[g3edge] (n0) -- (n1L);
            \draw[g3edge] (n2L) -- (n3h);
            \draw[g3edge] (n3b) -- (n2h);
            \draw[g3edge] (n4b) -- (n5h);
            \draw[g3edge] (n4h) -- (n5L);
            \draw[g3edge] (n6L) -- (n7);

            % ---- g2 edges (2) -----------------------------------------------------------
            \draw[g2edge] (n1L) -- (n2L);
            \draw[g2edge] (n5L) -- (n6L);

            % ---- g1 edge (1) -------------------------------------------------------------
            \draw[g1edge] (n3h) -- (n4h);

            % ---- val axis -----------------------------------------------------------------
            \draw[axisline] (-0.9,-1.35) -- (7*\dx+0.65,-1.35) node[right,gray!70,font=\footnotesize]{$\mathrm{val}$};
            \foreach \v in {0,...,7} {
                \draw[gray!70] (\v*\dx,-1.30) -- (\v*\dx,-1.40);
                \node[gray!40!black,font=\scriptsize] at (\v*\dx,-1.62) {\v};
            }

        \end{scope}

        % =======================================================================
        % PANEL (b): SUBPROBLEM x_t = 0  (classes val in {2,3,6,7} survive)
        % =======================================================================
        \begin{scope}[shift={(0,-4.55)}]

            \node[font=\bfseries] at (-1.55,1.35) {(b)};

            % ---- absent / gap boxes: val = 0,1 (absent) and val = 4,5 (the gap) ------
            \draw[gapbox] (0-\sw,-\sw) rectangle (0+\sw,\sw);
            \draw[gapbox] (1*\dx-\bw,-\bh) rectangle (1*\dx+\bw,\bh);
            \draw[gapbox] (4*\dx-\bw,-\bh) rectangle (4*\dx+\bw,\bh);
            \draw[gapbox] (5*\dx-\bw,-\bh) rectangle (5*\dx+\bw,\bh);

            % ---- surviving class boxes: val = 2,3,6,7 --------------------------------
            \draw[clsbox] (2*\dx-\bw,-\bh) rectangle (2*\dx+\bw,\bh);
            \draw[clsbox] (3*\dx-\bw,-\bh) rectangle (3*\dx+\bw,\bh);
            \draw[clsbox] (6*\dx-\bw,-\bh) rectangle (6*\dx+\bw,\bh);
            \draw[clsbox] (7*\dx-\sw,-\sw) rectangle (7*\dx+\sw,\sw);

            % ---- nodes (2^- deleted from bitstrings) -----------------------------------
            % val 2: hub=11010 (free), leaf-locked=01011, leaf=01110
            \node[free] (m2h) at (2*\dx, 0) {};
            \node[locked,label={above:\bitlab{01101}}] (m2L) at (2*\dx, 0.42) {};
            \node[free] (m2o) at (2*\dx,-0.42) {};
            % val 3: hub=locked=01101, leaf=11001, leaf=11100
            \node[locked] (m3h) at (3*\dx, 0) {};
            \node[free] (m3a) at (3*\dx, 0.42) {};
            \node[free] (m3b) at (3*\dx,-0.42) {};
            \node[font=\tiny,align=center] at (3*\dx,-\bh-0.16) {\bitlab{01110}};
            % val 6: hub=00111 (free), leaf-locked=10011, leaf=10110
            \node[free] (m6h) at (6*\dx, 0) {};
            \node[locked,label={above:\bitlab{10101}}] (m6L) at (6*\dx, 0.42) {};
            \node[free] (m6o) at (6*\dx,-0.42) {};
            % val 7
            \node[locked,label={below:\bitlab{10110}}] (m7) at (7*\dx,0) {};

            % ---- surviving star edges -------------------------------------------------
            \draw[staredge] (m2h)--(m2L); \draw[staredge] (m2h)--(m2o);
            \draw[staredge] (m3h)--(m3a); \draw[staredge] (m3h)--(m3b);
            \draw[staredge] (m6h)--(m6L); \draw[staredge] (m6h)--(m6o);

            % ---- surviving g3 edges -----------------------------------------------------
            \draw[g3edge] (m2L) -- (m3h);
            \draw[g3edge] (m3b) -- (m2h);
            \draw[g3edge] (m6L) -- (m7);

            % ---- required crossing move (dashed arrow over the gap) -------------------
            \draw[crossmove] (m3h) to[bend left=32] node[above,midway,yshift=1mm,font=\scriptsize,align=center]
            {required operation\\
                $\abs{\Delta\,\mathrm{val}} \geq 3$} (m6L);

            % ---- val axis (aligned with panel (a)) -------------------------------------
            \draw[axisline] (-0.9,-1.35) -- (7*\dx+0.65,-1.35) node[right,gray!70,font=\footnotesize]{$\mathrm{val}$};
            \foreach \v in {0,...,7} {
                \draw[gray!70] (\v*\dx,-1.30) -- (\v*\dx,-1.40);
                \node[gray!40!black,font=\scriptsize] at (\v*\dx,-1.62) {\v};
            }

        \end{scope}

        % =======================================================================
        % SHARED LEGEND
        % =======================================================================
        \begin{scope}[shift={(-1.55,-7.3)},font=\footnotesize]
            \node[locked] (lg1) at (0,0) {};
            \node[right=1.1mm of lg1] (lb1) {locked solution};
            \node[free,right=1.4mm of lb1] (lg2) {};
            \node[right=1.1mm of lg2] (lb2) {other solution};
            \node[right=1.4mm of lb2,minimum width=4mm,inner sep=0pt] (lg3a) {};
            \draw[staredge] ([xshift=-2mm]lg3a.center) -- ([xshift=2mm]lg3a.center);
            \node[right=1.1mm of lg3a] (lb3) {transposition};
            \node[right=1.4mm of lb3,minimum width=4mm,inner sep=0pt] (lg4a) {};
            \draw[g1edge] ([xshift=-2mm]lg4a.center) -- ([xshift=2mm]lg4a.center);
            \node[right=1.1mm of lg4a] (lb4) {$\bm{r}_{1}$};
            \node[right=1.4mm of lb4,minimum width=4mm,inner sep=0pt] (lg5a) {};
            \draw[g2edge] ([xshift=-2mm]lg5a.center) -- ([xshift=2mm]lg5a.center);
            \node[right=1.1mm of lg5a] (lb5) {$\bm{r}_{2}$};
            \node[right=1.4mm of lb5,minimum width=4mm,inner sep=0pt] (lg6a) {};
            \draw[g3edge] ([xshift=-2mm]lg6a.center) -- ([xshift=2mm]lg6a.center);
            \node[right=1.1mm of lg6a] (lb6) {$\bm{r}_{3}$};
            \node[right=1.4mm of lb6,minimum width=3.2mm,minimum height=3.2mm,inner sep=0pt] (lg7a) {};
            \draw[gapbox] ([xshift=-1.6mm,yshift=-1.6mm]lg7a.center) rectangle ([xshift=1.6mm,yshift=1.6mm]lg7a.center);
            \node[right=1.1mm of lg7a] {not attained for $x_{2^{-}} = 0$};
        \end{scope}

    \end{tikzpicture}
    \caption{\label{figure:StaircaseFibre}
        Fibre graphs from the proof of \autoref{proposition:ExponentialInference} for $k = 3$, i.e., $n = 6$ variables with coordinates $1^{+}, 1^{-}, 2^{+}, 2^{-}, 3^{+}, 3^{-}$ and Hamming weight $d = 3$.
        Node labels show the locked feasible solutions as bitstrings in this coordinate order.
        \textbf{(a)} Fibre graph $F_{\Sigma}$ of the current problem with the feasible solutions arranged by their counter value $\mathrm{val}$.
        The two transpositions centred at $1^{+}$ (grey) connect all feasible solutions with the same counter value, while $\bm{r}_{1}, \bm{r}_{2}, \bm{r}_{3}$ (coloured) connect the locked feasible solutions (filled) along the path $0 - 1 - \cdots - 7$.
        \textbf{(b)} Fibre graph of the subproblem arising from the substitution $x_{2^{-}} = 0$ (with the bit of $2^{-}$ removed), restricted to the operations in $\Sigma$ not involving $2^{-}$.
        Only the counter values in $R_{1} = \{2, 3\}$ and $R_{2} = \{6, 7\}$ remain, and the remaining operations do not connect them.
        Hence, every connecting family of the subproblem requires an operation changing the counter value by at least $2^{k - 2} + 1 = 3$.
    }
\end{figure*}
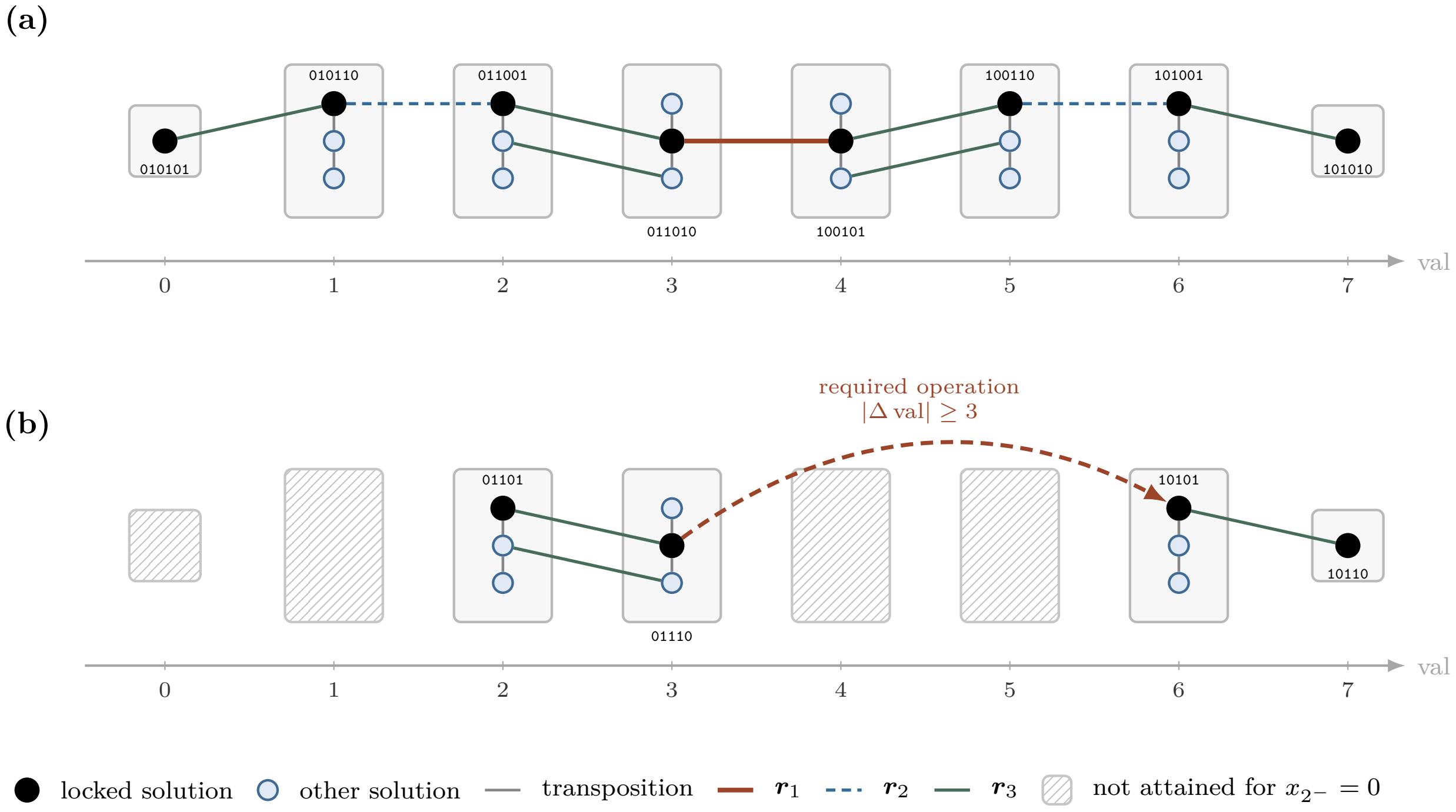

\begin{proposition}\label{proposition:ExponentialInference}
    Let $k \geq 2$, $n = 2 k$, and $(n, f, h_{k, \bm{1}}^{(n)})$ be a COP.
    Then, there exist an index $t \in [n]$ and $\Sigma \subseteq \mathcal{E}$ with $\abs{\Sigma} = 2 k - 1$ so that $F_{\Sigma}$ is connected and the move vectors of $\Sigma$ form a basis of $\ker(\bm{1}^{\top}) \subseteq \R^{n}$, with the following property:
    For the subproblem arising from the substitution $x_{t} = 0$, every family $\mathcal{E}''$ of operations with connected fibre graph contains some $(K, L) \in \mathcal{E}''$ whose move vector's unique expansion $\bm{v}_{K, L} = \sum_{\sigma \scriptin \Sigma} c_{\sigma} \bm{v}_{\sigma}$ satisfies
    \begin{equation*}
        \sum_{\sigma \scriptin \Sigma} \abs{c_{\sigma}} \geq 2^{k - 2} + 1.
    \end{equation*}
\end{proposition}

\begin{proof}
    Label the $n = 2 k$ coordinates by $j^{+}, j^{-}$, $j \in [k]$, write $\bm{e}_{a}$ for the unit vector of coordinate $a$, and set $\bm{p}_{j} \coloneqq \bm{e}_{j^{+}} - \bm{e}_{j^{-}}$.
    The family $\Sigma$ consists of the $k$ operations with move vectors
    \begin{equation*}
        \bm{r}_{i} \coloneqq \bm{p}_{i} - \sum_{j = i + 1}^{k} \bm{p}_{j}, \qquad i \in [k],
    \end{equation*}
    and the $k - 1$ transpositions with move vectors $\bm{e}_{1^{+}} - \bm{e}_{j^{+}}$, $2 \leq j \leq k$.
    Each of these vectors contains as many entries $1$ as $-1$, so that all operations in $\Sigma$ are feasibility-preserving by \autoref{lemma:FeasibilityPreservingGeneralisedXY}.

    Let $A \coloneqq \{j^{+} \defcolon j \in [k]\}$ and $B \coloneqq \{j^{-} \defcolon j \in [k]\}$, which partition the coordinates with $\abs{A} = \abs{B} = k$.
    For $\bm{x} \in \bit^{n}$, define the counter value
    \begin{equation*}
        \mathrm{val}(\bm{x}) \coloneqq \sum_{j = 1}^{k} 2^{k - j} (1 - x_{j^{-}}) \in \{0, \ldots, 2^{k} - 1\},
    \end{equation*}
    i.e., the binary number formed by the complemented bits of $B$.
    It satisfies $\mathrm{val}(\bm{y}) - \mathrm{val}(\bm{x}) = \lambda(\bm{y} - \bm{x})$ for the linear functional $\lambda(\bm{v}) \coloneqq -\sum_{j = 1}^{k} 2^{k - j} v_{j^{-}}$, where $v_{j^{-}}$ denotes the entry of $\bm{v} \in \R^{n}$ at coordinate $j^{-}$.
    For all $i \in [k]$, it is $\lambda(\bm{r}_{i}) = 2^{k - i} - \sum_{j = i + 1}^{k} 2^{k - j} = 1$.
    Since the transpositions are supported on $A$, they satisfy $\lambda = 0$.

    First, the transpositions form a star on $A$ with centre $1^{+}$ and hence connect all feasible solutions sharing the same restriction $\bm{x}_{B}$.
    Every such class is non-empty, since $\bm{x}_{A}$ has to contain $k - \abs{\bm{x}_{B}} \in \{0, \ldots, k\}$ ones, and it contains exactly one feasible solution, which we call \emph{locked}, with $x_{j^{+}} = 1 - x_{j^{-}}$ for all $j \in [k]$.
    The map $\mathrm{val}$ is a bijection between the $2^{k}$ locked solutions and $\{0, \ldots, 2^{k} - 1\}$.
    Second, the operations $\bm{r}_{i}$ connect the locked solutions:
    On a locked solution, the bits $\beta_{j} \coloneqq x_{j^{+}}$, $j \in [k]$, are the binary digits of $\mathrm{val}(\bm{x}) = \sum_{j = 1}^{k} 2^{k - j} \beta_{j}$.
    The move $\bm{r}_{i}$ fires on a locked solution if and only if $\beta_{i} = 0$ and $\beta_{j} = 1$ for all $j > i$ (or vice versa), and then maps it to the locked solution with these digits flipped, i.e., increments (or decrements) $\mathrm{val}$ by $1$.
    Hence, the locked solutions form a path $0 - 1 - \cdots - (2^{k} - 1)$ in $F_{\Sigma}$, and $F_{\Sigma}$ is connected (see \autoref{figure:StaircaseFibre} for $k = 3$).

    Every transposition $\bm{e}_{a} - \bm{e}_{b}$, $a \neq b$, is the difference of two feasible solutions, since $0 < k < n$.
    As $F_{\Sigma}$ is connected, these two feasible solutions are joined by a path in $F_{\Sigma}$, so that $\bm{e}_{a} - \bm{e}_{b}$ lies in the integer span of the move vectors of $\Sigma$.
    These transpositions span $\ker(\bm{1}^{\top})$, which has dimension $2 k - 1 = \abs{\Sigma}$, so the move vectors of $\Sigma$ form a basis.

    Now let $t \coloneqq 2^{-}$ and fix $x_{t} = 0$, i.e., the digit of weight $2^{k - 2}$ in $\mathrm{val}$ is $1$.
    By the same counting argument as above, now with $x_{t} = 0$ fixed, the counter values attained by the subproblem's feasible solutions are precisely those in $R_{1} \cup R_{2}$, where
    \begin{equation*}
        R_{1} \coloneqq \{2^{k - 2}, \ldots, 2^{k - 1} - 1\} \quad \text{and} \quad R_{2} \coloneqq \{3 \cdot 2^{k - 2}, \ldots, 2^{k} - 1\}.
    \end{equation*}
    Let $\mathcal{E}''$ be a family of operations for the subproblem with connected fibre graph.
    Its fibre graph is connected and contains solutions with counter values in $R_{1}$ and in $R_{2}$, so some $(K, L) \in \mathcal{E}''$ constitutes an edge $\{\bm{x}, \bm{y}\}$ with $\mathrm{val}(\bm{x}) \in R_{1}$ and $\mathrm{val}(\bm{y}) \in R_{2}$.
    Its move vector $\bm{v}_{K, L} = \pm (\bm{y} - \bm{x})$ lies in $\ker(\bm{1}^{\top})$ and has a unique expansion $\sum_{\sigma \scriptin \Sigma} c_{\sigma} \bm{v}_{\sigma}$.
    Since $\lambda$ takes the values $1$ and $0$ on $\Sigma$, we conclude
    \begin{equation*}
        \sum_{\sigma \scriptin \Sigma} \abs{c_{\sigma}} \geq \Big|\sum_{\sigma \scriptin \Sigma} c_{\sigma} \lambda(\bm{v}_{\sigma})\Big| = \abs{\lambda(\bm{y} - \bm{x})} = \abs{\mathrm{val}(\bm{y}) - \mathrm{val}(\bm{x})} \geq 3 \cdot 2^{k - 2} - (2^{k - 1} - 1) = 2^{k - 2} + 1,
    \end{equation*}
    concluding the proof.
\end{proof}

Consequently, inferring a connecting family for the subproblem from $\Sigma$ is exponentially expensive in both of the above senses.
On the one hand, every subproblem mixer compiled from the gates of $\Sigma$ requires at least $2^{k - 2} + 1$ exponentials, counted over all mixer layers, to provide transitions between feasible solutions with counter values in $R_{1}$ and $R_{2}$.
In fact, this bound does not even rely on the basis property, since each exponential changes the counter value by at most $1$.
On the other hand, every derivation tree for the operation $(K, L)$ from \autoref{proposition:ExponentialInference} has at least $2^{k - 2} + 1$ leaves.

Let us point out two limitations of this argument.
First, a symbolic derivation may reuse intermediate results, i.e., form a directed acyclic graph instead of a tree.
Since every addition at most doubles the $\ell_{1}$-norm of the coefficients, such a derivation still requires at least $k - 1$ consecutive additions, but its total number of additions is no longer bounded from below.
Second, one may bypass $\Sigma$ altogether, e.g., by computing the required move vector as the difference of two feasible solutions of the subproblem.
This, however, relies on the subproblem's weights rather than on $\Sigma$ and is thus a rudimentary form of fabrication.

Most importantly, the subproblem in \autoref{proposition:ExponentialInference} is again a plain Hamming weight constraint on $n - 1$ variables, for which \autoref{theorem:ConnectingMoves} immediately yields the $n - 2$ adjacency swaps $\mathcal{T}_{1}$.
The exponential cost is thus caused solely by the choice of $\Sigma$, not by the subproblem itself.
This is precisely why we fabricate the connecting family of every subproblem from its weights alone.
If, instead, the admissible operations are fixed in advance, e.g., to a hardware-native gate set, this freedom is lost and the cost exhibited above cannot be avoided in general.

\end{document}